\documentclass[aps,prc,twocolumn,superscriptaddress]{revtex4-1}

\usepackage{graphicx}
\usepackage{threeparttable}
\usepackage{booktabs}
\usepackage{xcolor}

\begin{document}

% Use the \preprint command to place your local institutional report
% number in the upper righthand corner of the title page in preprint mode.
% Multiple \preprint commands are allowed.
% Use the 'preprintnumbers' class option to override journal defaults
% to display numbers if necessary
%\preprint{}

%Title of paper
\title{The decay of quadrupole-octupole $1^-$ states in $^{40}$Ca and $^{140}$Ce}

% repeat the \author .. \affiliation  etc. as needed
% \email, \thanks, \homepage, \altaffiliation all apply to the current
% author. Explanatory text should go in the []'s, actual e-mail
% address or url should go in the {}'s for \email and \homepage.
% Please use the appropriate macro foreach each type of information

% \affiliation command applies to all authors since the last
% \affiliation command. The \affiliation command should follow the
% other information
% \affiliation can be followed by \email, \homepage, \thanks as well.
\author{V.~Derya}
\email[]{derya@ikp.uni-koeln.de}
%\homepage[]{Your web page}
%\thanks{}
%\altaffiliation{}
\affiliation{Institut f\"ur Kernphysik, Universit\"at zu K\"oln, 50937 K\"oln, Germany}
\author{N.~Tsoneva}
\affiliation{Frankfurt Institute for Advanced Studies (FIAS), 60438 Frankfurt am Main, Germany}
\affiliation{Institut f\"ur Theoretische Physik, Universit\"at Gie\ss en, 35392 Gie\ss en, Germany}
\affiliation{Institute for Nuclear Research and Nuclear Energy, 1784 Sofia, Bulgaria}
\author{T.~Aumann}
\affiliation{Institut f\"ur Kernphysik, Technische Universit\"at Darmstadt, 64289 Darmstadt, Germany}
\author{M.~Bhike}
\affiliation{Department of Physics, Duke University, Durham, North Carolina 27708, USA}
\author{J.~Endres}
\affiliation{Institut f\"ur Kernphysik, Universit\"at zu K\"oln, 50937 K\"oln, Germany}
\author{M.~Gooden}
\affiliation{Department of Physics, Duke University, Durham, North Carolina 27708, USA}
\author{A.~Hennig}
\affiliation{Institut f\"ur Kernphysik, Universit\"at zu K\"oln, 50937 K\"oln, Germany}
\author{J.~Isaak}
\affiliation{Frankfurt Institute for Advanced Studies (FIAS), 60438 Frankfurt am Main, Germany}
\affiliation{GSI Helmholtzzentrum f\"ur Schwerionenforschung GmbH, 64291 Darmstadt, Germany}
%\author{J.H.~Kelley}
%\affiliation{Triangle Universities Nuclear Laboratory, Durham, North Carolina 27708, USA}
%\affiliation{Department of Physics, North Carolina State University, Raleigh, North Carolina 27695, USA}
\author{H.~Lenske}
\affiliation{Institut f\"ur Theoretische Physik, Universit\"at Gie\ss en, 35392 Gie\ss en, Germany}
\author{B.~L\"oher}
\affiliation{GSI Helmholtzzentrum f\"ur Schwerionenforschung GmbH, 64291 Darmstadt, Germany}
\author{N.~Pietralla}
\affiliation{Institut f\"ur Kernphysik, Technische Universit\"at Darmstadt, 64289 Darmstadt, Germany}
%\author{R.~Raut}
%\affiliation{Department of Physics, Duke University, Durham, North Carolina 27708, USA}
%\author{G.~Rusev}
%\altaffiliation{Present address: Chemistry Division, Los Alamos National Laboratory, Los Alamos, New Mexico 87545, USA.}
%\affiliation{Department of Physics, Duke University, Durham, North Carolina 27708, USA}
%\affiliation{Triangle Universities Nuclear Laboratory, Durham, North Carolina 27708, USA}
\author{D.~Savran}
\affiliation{GSI Helmholtzzentrum f\"ur Schwerionenforschung GmbH, 64291 Darmstadt, Germany}
%\author{A.~Tonchev}
%\altaffiliation{Present address: Physics Division, Lawrence Livermore National Laboratory, Livermore, California 94550, USA.}
%\affiliation{Department of Physics, Duke University, Durham, North Carolina 27708, USA}
%\affiliation{Triangle Universities Nuclear Laboratory, Durham, North Carolina 27708, USA}
\author{W.~Tornow}
\affiliation{Department of Physics, Duke University, Durham, North Carolina 27708, USA}
\affiliation{Triangle Universities Nuclear Laboratory, Durham, North Carolina 27708, USA}
%\author{H.~Weller}
%\affiliation{Department of Physics, Duke University, Durham, North Carolina 27708, USA}
\author{V.~Werner}
\affiliation{Institut f\"ur Kernphysik, Technische Universit\"at Darmstadt, 64289 Darmstadt, Germany}
\author{A.~Zilges}
\affiliation{Institut f\"ur Kernphysik, Universit\"at zu K\"oln, 50937 K\"oln, Germany}

%Collaboration name if desired (requires use of superscriptaddress
%option in \documentclass). \noaffiliation is required (may also be
%used with the \author command).
%\collaboration can be followed by \email, \homepage, \thanks as well.
%\collaboration{}
%\noaffiliation

\date{\today}
\begin{abstract}

\noindent\textbf{Background:} Two-phonon excitations originating from the coupling of two collective one-phonon states are of great interest in nuclear structure physics. One possibility to generate low-lying $E1$ excitations is the coupling of quadrupole and octupole phonons.

\noindent\textbf{Purpose:} In this work, the $\gamma$-decay behavior of candidates for the $(2_1^+\otimes 3_1^-)_{1^-}$ state in the doubly-magic nucleus $^{40}$Ca and in the heavier and semi-magic nucleus $^{140}$Ce is investigated.

\noindent\textbf{Methods:} $(\vec{\gamma},\gamma')$ experiments have been carried out at the High Intensity $\gamma$-ray Source (HI${\gamma}$S) facility in combination with the high-efficiency $\gamma$-ray spectroscopy setup $\gamma^3$ consisting of HPGe and LaBr$_3$ detectors. The setup enables the acquisition of $\gamma$-$\gamma$ coincidence data and, hence, the detection of direct decay paths.

\noindent\textbf{Results:} In addition to the known ground-state decays, for $^{40}$Ca the decay into the $3^-_1$ state was observed, while for $^{140}$Ce the direct decays into the $2^+_1$ and the $0^+_2$ state were detected. The experimentally deduced transition strengths and excitation energies are compared to theoretical calculations in the framework of EDF theory plus QPM approach and systematically analyzed for $N=82$ isotones. In addition, negative parities for two $J=1$ states in $^{44}$Ca were deduced simultaneously.

\noindent\textbf{Conclusions:} The experimental findings together with the theoretical calculations support the two-phonon character of the $1^-_1$ excitation in the light-to-medium-mass nucleus $^{40}$Ca as well as in the stable even-even $N=82$ nuclei.
\end{abstract}

% insert suggested PACS numbers in braces on next line
\pacs{1}
% insert suggested keywords - APS authors don't need to do this

%\maketitle must follow title, authors, abstract, \pacs, and \keywords
\maketitle

% body of paper here - Use proper section commands
% References should be done using the \cite, \ref, and \label commands
\section{Introduction}

The interaction of the atomic nucleus with an electromagnetic field gives rise to the excitation of various modes of different spin and parity which provide useful information on the nuclear structure. Among them of special importance is the electric dipole ($E1$) response which is generally dominated by a strong, collective isovector nuclear vibration, the isovector giant dipole resonance (IVGDR) \cite{harakeh}. The IVGDR is classically described by a Lorentzian shape \cite{Ber75}. Recently, in nuclei with neutron excess an additional dipole strength component below and around the neutron threshold was found on top of the low-energy tail of the IVGDR \cite{Bart61,Herz99b,Zilg02,Savr13}.
This mode of excitation is usually denoted as pygmy dipole resonance (PDR) because it resembles a resonance-like accumulation of close-lying $J^\pi=1^-$ states with similar spectroscopic features \cite{Savr13}. In a simple macroscopic picture, a displacement of center-of-mass and center-of-charge of the nucleus generates a vibrational motion trying to restore the proton-neutron symmetry.
Nowadays, the rapidly increasing number of experiments using different probes and techniques allow for systematic studies of the PDR over isotopic and isotonic chains from different mass regions \cite{Herz99b,Brys99,Volz06,Sav08,Ton10,Tof11,Schw13,Cre13,Dery14,Brac15,Kru15,Kri15}. A close connection between the total PDR strength and the amount of the neutron excess of neutron-rich nuclei which on the other hand is correlated with the neutron skin thickness was proposed \cite{Savr13,Tso04,Tso08,Klim07}. Furthermore, experiments with complementary probes like $\alpha$-particles at intermediate energy, indicate an isospin splitting of the low-lying $1^-$ states \cite{Savr06}. Similar to the experimental findings, theoretical models show that at lower energies the $E1$ strength is predominantly of isoscalar character which gradually becomes more isovector with increasing excitation energy toward the IVGDR \cite{Tso04,Tso08,Papa14,Repk13,Vret12,Lanz14,Roca12}. 

Various theoretical explanations of the $E1$ strength below and around the particle-emission thresholds exist. These include the low-energy tail of the IVGDR, PDR \cite{Tso04,Tso08,Pie06,Rei13}, multi-phonon excitations \cite{Tso08,Tson15}, toroidal modes \cite{Repk13} and $\alpha$-cluster vibrations \cite{Iach85,Spie15}. Furthermore, in recent studies it has been pointed out that the interaction between quasiparticles and phonons is important for a correct theoretical description of the low-lying $E1$ strength because it can influence its fragmentation and mixing with the core polarization and the IVGDR \cite{Pon98,Tso04,Tso08,Arse12,Litv13,Tson15,Savr11}. This affects strongly the electromagnetic strength distribution, which can have further consequences on the dipole polarizability and nucleosynthesis processes \cite{Gori98,Litv13,Tson15}.

Of particular interest are low-energy two-phonon states related to the coupling of collective quadrupole and octupole core vibrations. 
The collective quadrupole and octupole excitations of electric character are usually among the lowest-lying excitations in nuclei in the vicinity of shell closures. They are interpreted as surface oscillations and theoretically treated as phonons with the possibility to couple to multi-phonon states, like for example double-quadrupole or double-octupole states \cite{bohrmottelson, Apra84, Yeh96, Mukh08}. The mixed harmonic coupling of quadrupole and octupole collective phonons $(2_1^+\otimes 3_1^-)_{J^{\pi}}$ results in a quintuplet of $J^\pi=1^-- 5^-$ states which are located at an excitation energy equal to the sum of the excitation energies of the corresponding $2^{+}_{1}$ and $3^{-}_{1}$ one-phonon states. Anharmonicities in the phonon-phonon interaction can affect the excitation energies and break the degeneracy of the multiplet states. Nevertheless, due to the different nature of the two phonons Pauli blocking is small compared to e.g. $(2^+\otimes 2^+)$ or $(3^-\otimes 3^-)$ states.

Detailed theoretical descriptions of two-phonon states related to members of quadrupole-quadrupole and quadrupole-octupole multiplets \cite{Grin94,Pon98,bohrmottelson} are obtained in the framework of the quasiparticle-phonon model (QPM) \cite{QPM,Grin94,Pon98}. Another model which has been intensively applied in studies of multi-phonon states \cite{Kern95,Smir00} is the interacting boson model (IBM) \cite{IBM}. Recently, the $spdf$ IBM has been applied in systematical studies of low-lying
$J^{π}$=$1^{−}$ states in the Nd isotopes and other rare-earth nuclei \cite{Spie15}.

The first step in identifying two-phonon $1^-$ states is to determine spin, parity, and $B(E1,1^-\rightarrow 0^+_1)$ strength for possible candidates. A widely used experimental tool for the investigation of $J=1$ states is nuclear resonance fluorescence (NRF) \cite{Knei96}. In the last years, the $B(E1)$ strength distributions of many nuclei were measured using this method. The evaluated data serve as a systematic basis for the discussion of two-phonon $E1$ excitations like, e.g., in the Sn isotopes \cite{Brys99,Pysm06}, for $N=82$ isotones \cite{Herz95} and in the compilation of Andrejtscheff \textit{et al.} \cite{Andr01} for $A=48-148$ nuclei. An alternative way to determine $B(E1)$ strengths in particular for states of rare isotopes, for which NRF measurements are difficult, are lifetime measurements using the Doppler-shift attenuation method (DSAM) in particle-$\gamma$ coincidence measurements \cite{Henn15}. Since several years, the DSAM technique is applied in inelastic neutron-scattering at the University of Kentucky \cite{Belg93,Belg96}. Furthermore, direct access to the ground-state decay width $\Gamma_0$ can be obtained using the self-absorption method \cite{Romi15} or inelastic proton-scattering experiments \cite{Tami11} for some cases.

Once a candidate is found, it is desirable to study also its decay behavior to test the two-phonon structure more thoroughly since this information is one of the key signatures in addition to the excitation energy and correlations of transition strengths \cite{Piet99,Jolo04}. In the case of harmonic phonon coupling, the lowest-lying 1$^-$ state is a two-phonon excitation and the corresponding $B(E3)$ strengths for the $1_1^- \rightarrow 2_1^+$ and $3_1^-\rightarrow 0_1^+$ transitions as well as the $B(E2)$ strengths for the $1_1^- \rightarrow 3_1^-$ and $2_1^+\rightarrow 0_1^+$ transitions are equal. Such a direct proof of the two-phonon character of the $1_1^-$ state via its decay behavior was found some years ago only for the two $N=82$ nuclei $^{142}$Nd and $^{144}$Sm in inelastic proton-scattering experiments \cite{Wilh96,Wilh98,Robi94}. It is the aim of the present work to further test the two-phonon quadrupole-octupole $1^{-}$ states in the $N=82$ isotones by extending the knowledge about the decay behavior of the two-phonon $1^-$ candidate at 3.6\,MeV in $^{140}$Ce. In addition, the decay behavior of the two-phonon $1^-$ candidate at 5.9\,MeV in the significantly lighter nucleus $^{40}$Ca is investigated to study the existence of this collective excitation mode in a different mass region.

The experimental method and data analysis tools are introduced in Secs.~\ref{sec:experiments} and \ref{sec:data}. The new experimental results for $^{140}$Ce and $^{40}$Ca are presented and discussed in Secs.~\ref{sec:140ce} and \ref{sec:40ca}, respectively. A systematic theoretical description of two-phonon $1^-_1$ states and corresponding transitions in $N=82$ nuclei is discussed in comparison with data in Sec.~\ref{sec:140ce}.

\section{Experiments}\label{sec:experiments}
Real-photon scattering $(\vec{\gamma},\gamma)$ experiments were performed to study the $\gamma$-decay behavior of possible two-phonon $J^\pi=1^-$ states in $^{40}$Ca and $^{140}$Ce. The states of interest were populated by the quasi-monochromatic, linearly polarized, and intense beam of real photons provided at the High Intensity $\gamma$-ray Source (HI${\gamma}$S) facility \cite{Carm95,Well09} at the Triangle Universities Nuclear Laboratory (TUNL) in Durham, NC, USA. The excitation is selective to low spins (mainly $J=1$) and excitation-energy regions (due to the narrow bandwidth of the beam) and, therefore, well-suited for the study of specific $J^\pi=1^-$ states. The intense $\gamma$-ray source in the entrance channel is combined with the newly installed high-efficiency $\gamma$-$\gamma$ coincidence setup $\gamma^3$ \cite{Loeh13} for the detection of de-exciting $\gamma$-rays in the outgoing channel. For the present experiments the setup was used in a configuration with four $3 '' \times 3''$ LaBr$_3$:Ce scintillation detectors at $\theta=90^\circ$ and four 60\% high-purity Germanium (HPGe) semi-conductor detectors at $\theta=135^\circ$ with respect to the beam axis. The LaBr$_3$ detectors were placed symmetrically at azimuthal angles of $\phi=45^\circ,135^\circ,225^\circ,$ and $315^\circ$ relative to the horizontal polarization axis, whereas two HPGe detectors were placed parallel ($\phi=0^\circ,180^\circ$) and two perpendicular ($\phi=90^\circ,270^\circ$) to the polarization axis. Using this detector configuration and distances of 5 to 10~cm between detector end-cap and target for the LaBr$_3$ and HPGe detectors, respectively, results in a total photopeak efficiency of about 6\% at 1.3\,MeV. Data was acquired in parallel by two data acquisition (DAQ) systems. One is the analog so-called Genie DAQ which was used to store singles spectra of the HPGe detectors. The second DAQ system is the digital MBS DAQ which acquires event-by-event list-mode data for HPGe and LaBr$_3$ detectors. Customized trigger conditions allow to generate, \textit{e.g.}, singles and coincidence triggers and are adjusted individually. More details on the $\gamma^3$ setup can be found in Ref.~\cite{Loeh13}.

\begin{figure*}[tbp]
\centering
\includegraphics[width=1\textwidth]{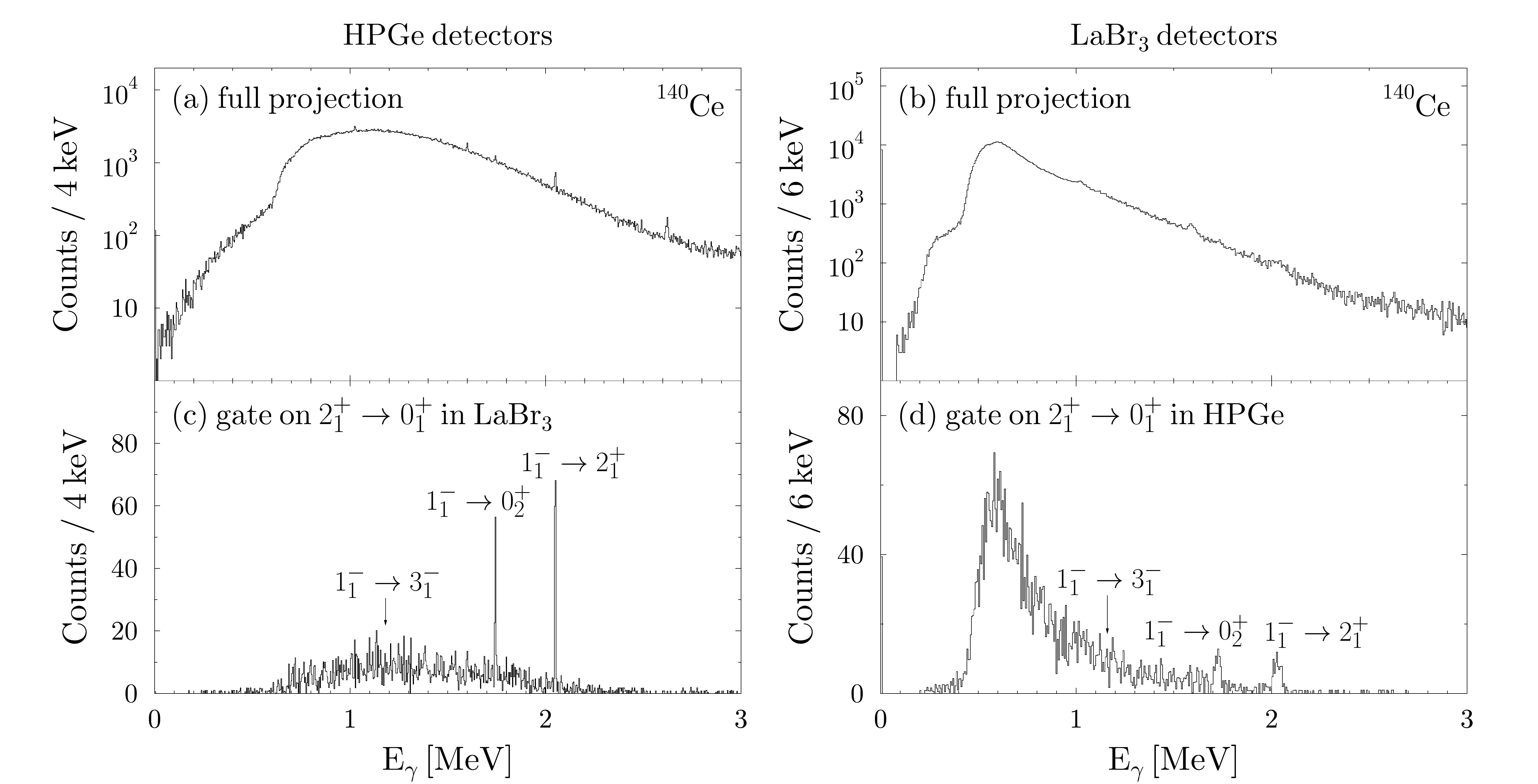}

\caption{Projected $\gamma$-ray spectra of (a) the HPGe detectors with HPGe-LaBr$_3$ coincidence condition, (b) the LaBr$_3$ detectors with HPGe-LaBr$_3$ coincidence condition, (c) the HPGe detectors with HPGe-LaBr$_3$ coincidence condition and energy gate on the $2^+_1 \rightarrow 0^+_1$ transition in the LaBr$_3$ detectors, and (d) the LaBr$_3$ detectors with HPGe-LaBr$_3$ coincidence condition and energy gate on the $2^+_1 \rightarrow 0^+_1$ transition in the HPGe detectors after background subtraction. Arrows indicate a hypothetical $1^-_1 \rightarrow 3^-_1$ transition. The data was taken in the measurement on $^{140}$Ce with a $\gamma$-beam energy of 3.6\,MeV.}
\label{fig:projspecs}
\end{figure*}

Photon energy settings of 3.6~MeV and 5.9~MeV were used in the experiments on $^{140}$Ce and $^{40}$Ca, respectively, to cover the excitation energies of the corresponding two-phonon candidates. The beam-energy profile of the incoming photon beam is monitored by an additional 123\% HPGe detector which can be moved into the beam. In the present experiments the bandwidth of the photon beam amounted to 4\%. The $^{140}$Ce target was composed of 2\,g highly enriched (99.72\%) plus 7.5\,g natural cerium-oxide powder, whereas for the $^{40}$Ca experiment an 11.2\,g natural calcium-carbonate target was used. Both measurements were carried out for about 23\,h, each.

\section{Data analysis}\label{sec:data}
In general, a number of quantities are directly accessible in nuclear resonance fluorescence (NRF) experiments such as spin, parity, excitation energy, and transition strengths. For a transition of electromagnetic character $\sigma$ and multipolarity $L$ without multipole mixing, the reduced transition strength, $B(\sigma L)$, and the partial decay width to a specific final state, $\Gamma_f$, are related via:

\begin{equation}\label{eq:BsigmaL}
B(\sigma L,J_i \rightarrow J_f)=\frac{L[(2L+1)!!]^2}{8\pi(L+1)}\left(\frac{\hbar c}{E_{\gamma}}\right)^{2L+1}g\Gamma_f,
\end{equation}
where $E_\gamma$ is the transition energy and $g=\frac{2J_f+1}{2J_i+1}$ is the spin factor. In the present cases, the cross section, $I_{r,f}$, for the resonant excitation of the $1^-$ states decaying back to the ground state has been measured in previous NRF experiments \cite{Herz99b,Hart02,Volz06} via

\begin{equation}\label{eq:Irf_Gamma}
I_{r,f}=\pi^2 \left(\frac{\hbar c}{E} \right)^2 g\frac{\Gamma_0 \Gamma_f}{\Gamma}.
\end{equation}
In the present analysis, the ratio of partial and total decay widths can be deduced from the peak area in the singles $\gamma$-ray spectra:
\begin{equation}\label{eq:asingle}
A^{single}_{i,f}=g \pi^2 \left(\frac{\hbar c}{E} \right)^2 \frac{\Gamma_0 \Gamma_f}{\Gamma}N_t  N_{\gamma} \Delta_{live,i}\epsilon_i(E - E_f) W_{i,f},
\end{equation}
%\begin{equation}
%A_{f}=g \pi^2 \left(\frac{\hbar c}{E} \right)^2 \frac{\Gamma_0 \Gamma_f}{\Gamma}N_t  N_{\gamma} \Delta_{live}\epsilon(E - E_f) W_{f},
%\end{equation}
where $N_t$ is the number of target nuclei, $N_{\gamma}$ is the photon flux at the resonance energy, $\Delta_{live,i}$ is the relative live-time of detector $i$, $\epsilon_i(E - E_f)$ is the absolute photopeak efficiency of detector $i$ at the transition energy, and $W_{i,f}$ is the angular distribution of the scattered photons at the position of detector $i$.

Using Eq.~(\ref{eq:asingle}), the branching ratio relative to the ground state, $\Gamma_f/\Gamma_0$, can be derived from
\begin{equation}
\frac{\Gamma_f}{\Gamma_0}=\frac{A^{single}_{f}\sum_i\Delta_{live,i}\epsilon_i(E)W_{i,0}}{A^{single}_{0}\sum_i \Delta_{live,i}\epsilon_i(E - E_f) W_{i,f}}
\end{equation}
after summing over all detectors $i$. For the coincidence data, two $\gamma$-rays from the de-exciting $\gamma$ cascade are detected. This leads to an additional experimental access to the relative branching ratio:

\begin{equation}
\frac{\Gamma_f}{\Gamma_0}=\frac{A^{coinc}_{f}\sum_i\Delta_{live,i}\epsilon_i(E)W_{i,0}}{A^{single}_{0}\sum_{ij}\Delta_{live,ij}\epsilon_i(E - E_f)\epsilon_j(E_{\gamma_2}) W_{ij,f}},
\end{equation}

where $A^{coinc}_{f}$ is the peak area in the energy-gated coincidence spectrum summed for all detector combinations, $\Delta_{live,ij}$ is the relative live-time of detector $i$ and $j$, $W_{ij,f}$ is the angular distribution of the scattered photons at the position of detector $i$ and $j$, and $\gamma_2$ denotes the second $\gamma$-ray that is detected in addition to the $1^-\rightarrow J_f$ transition.

\begin{figure}[tbp]
\centering
\includegraphics[width=0.5\textwidth]{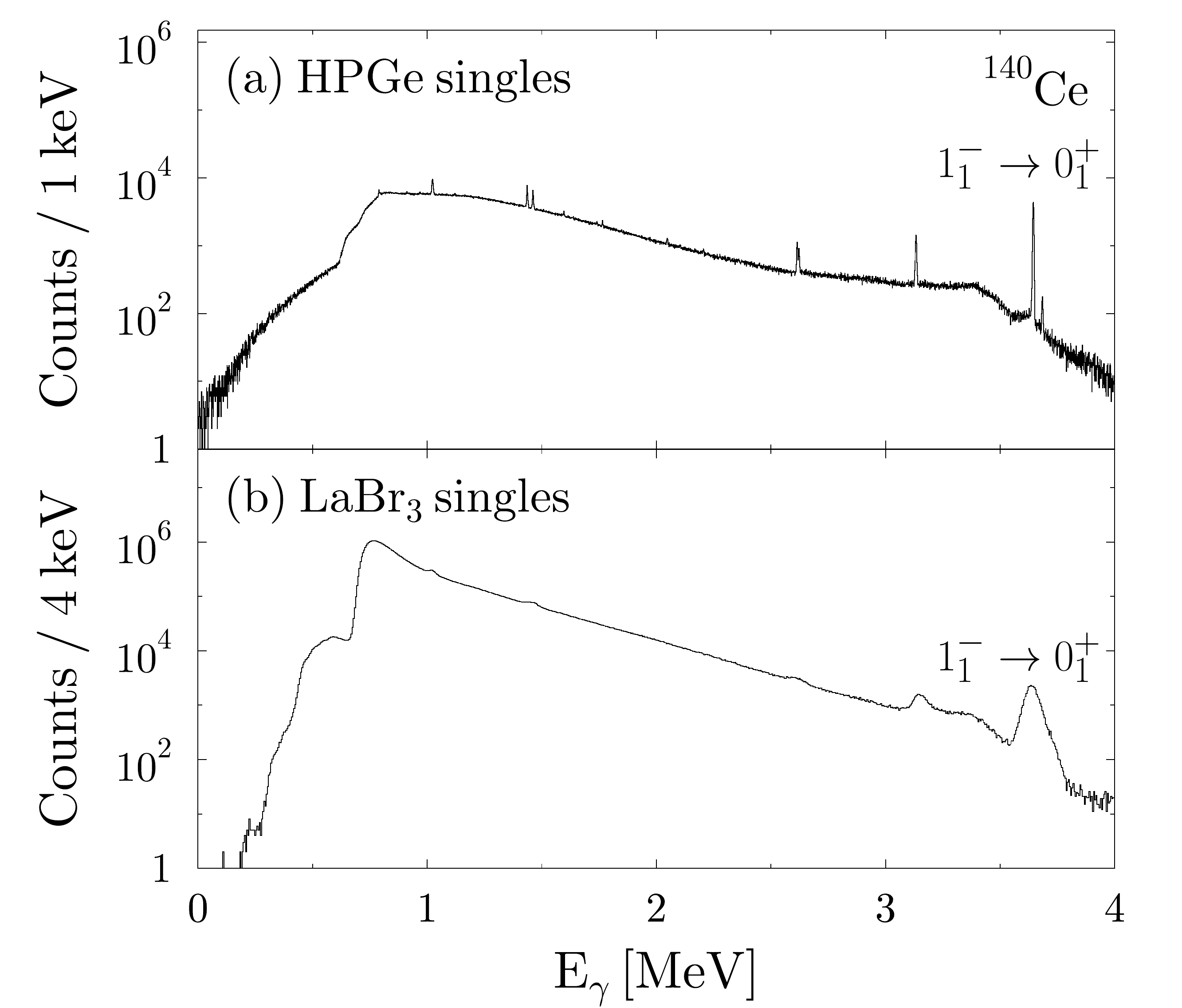}

\caption{Singles $\gamma$-ray spectra of (a) HPGe detectors and (b) LaBr$_3$ detectors. The data was taken in the measurement on $^{140}$Ce with a $\gamma$-beam energy of 3.6\,MeV.}
\label{fig:140ce_singles}
\end{figure}

The focus in the present work lies on the determination of relative branching ratios $\Gamma_f/\Gamma_0$ which gives access to $\Gamma_f$ for known $\Gamma_0$ and can be transfered into reduced transition strengths using Eq.~(\ref{eq:BsigmaL}). In principle, both, singles and coincidence data, can be used for the determination of $\Gamma_f/\Gamma_0$ as shown above. With the coincidence data the selectivity is improved, however, the intensity in the $\gamma$-ray spectra is reduced. Two-dimensional $\gamma$-$\gamma$ coincidence matrices filled with the $\gamma$-ray energies measured by HPGe and LaBr$_3$ detectors are used to generate projected $\gamma$-ray spectra as shown in Fig.~\ref{fig:projspecs} for the measurement on $^{140}$Ce. The upper panel shows the full projections of the HPGe-LaBr$_3$ coincidence data. A large background in particular at lower energies is visible in the full projections which mainly stems from non-resonant scattering processes in the target itself. The lower panel of Fig.~\ref{fig:projspecs} shows the projected $\gamma$-ray spectra after applying an energy gate ($E_\gamma\approx 1596$\,keV) on the secondary $2^+_1 \rightarrow 0^+_1$ transition of $^{140}$Ce. The primary transitions $1^-_1 \rightarrow 2^+_1$ and $1^-_1 \rightarrow 0^+_2$ are clearly visible in the gated $\gamma$-ray spectra obtained with the HPGe and LaBr$_3$ detectors. Their peak areas can be used to determine branching ratios for the different decay channels relative to the ground-state decay. The singles $\gamma$-ray spectra of HPGe and LaBr$_3$ detectors are shown in Fig.~\ref{fig:140ce_singles}.

In addition, the setup allows for parity measurements via the polarization information carried by the angular distribution, $W(\theta,\phi)$, of the de-exciting $\gamma$ rays. The analyzing power for a fixed scattering angle $\theta$ is defined as

\begin{equation}
\Sigma=\frac{W(\theta,0^\circ)-W(\theta,90^\circ)}{W(\theta,0^\circ)+W(\theta,90^\circ)}.
\end{equation}

The position of the HPGe detectors differed from the usual parity measurements where the analyzing power is maximized \cite{Piet02}.
The detectors at $\theta=135^\circ$ give analyzing powers of $\Sigma=\pm 1/3 \textrm{ for } J=1^\pm$ states and $\Sigma=\mp 1 \textrm{ for } J=2^\pm$ states. The experimentally accessible observable is the asymmetry

\begin{equation}
\epsilon=\frac{I_\parallel-I_\perp}{I_\parallel-I_\perp}=q\Sigma,
\end{equation}

where $I_\parallel$ and $I_\perp$ are the efficiency-corrected photon intensities in the horizontal ($\parallel$) and vertical ($\perp$) detectors with respect to the horizontal polarization axis. The experimental sensitivity $q\approx 0.9$ accounts for the finite opening angle of the detectors.

\section{Results for $^{140}$Ce}\label{sec:140ce}
The two-phonon candidate in $^{140}$Ce, which is investigated in the present work, is the $1_1^-$ state at 3.6\,MeV with a $B(E1,1_1^-\rightarrow 0^+_1)$ transition strength of 4.1(6)\,mW.u. \cite{Volz06}. Its decays to the first $2^+_1$ and to the second $0^+_2$ states are clearly visible in the projected $\gamma$-ray spectrum of the HPGe detectors with a gate on the ground-state transition of the first $2^+_1$ state (see Fig.~\ref{fig:projspecs}). The transition strengths can be deduced using these primary $\gamma$-ray transition from the excited $1^-_1$ state into the corresponding excited state (seen in the coincidence $\gamma$-ray spectra shown in Fig.~\ref{fig:projspecs}) and ground state (seen in the singles $\gamma$-ray spectra shown in Fig.~\ref{fig:140ce_singles}) as well as the known $B(E1,1_1^-\rightarrow 0^+_1)$ transition strength. The results are 0.54(3) and 0.75(6)\,m.W.u. for the $B(E1,1_1^-\rightarrow 2^+_1)$ and $B(E1,1_1^-\rightarrow 0^+_2)$ transition, respectively. The decay of the first $1^-$ state to the $3^-_1$ state is not visible on top of a pronounced background. However, for the $B(E2,1_1^-\rightarrow 3^-_1)$ transition strength an upper limit of 28\,W.u. was deduced by analyzing the background in the $\gamma$-ray spectrum. In the harmonic model a $1^-_1\rightarrow 2^+_1$ $E3$ transition would be expected, but a measurement of this transition is difficult because $E1$ radiation dominates over $E3$ radiation. We assumed that the observed $1^-_1\rightarrow 2^+_1$ transition is of $E1$ character. The observation of $1^-_1\rightarrow 2^+_1$ and $1^-_1\rightarrow 0^+_2$ $E1$ transitions cannot be explained in the simple harmonic picture but needs further explanation which will be discussed in the following paragraphs.

Already some years ago, the QPM was applied to study two-phonon structures including the quadrupole-octupole coupled $1^-$ state in stable $N=82$ nuclei \cite{Grin94}. Lowest-lying 1$_1^-$ states with a large $(2^+_1\otimes 3^-_1)_{1^{-}}$ content and excitation energies close to the sum energy of the first 2$^+$ and 3$^-$ states were calculated and interpreted as two-phonon excitations. However, the previous calculations in the $N=82$ isotones do not discuss the excited $0^+_2$ state. Thus, also the $B(E1,1_1^-\rightarrow 0^+_2)$ transition strength which we measured for the first time could not be compared to available theoretical predictions within a consistent framework. For this reason we performed new calculations for the $N=82$ nuclei $^{138}$Ba, $^{140}$Ce, $^{142}$Nd, and $^{144}$Sm in the framework of a more advanced microscopic nuclear structure approach based on the self-consistent energy-density functional (EDF) theory and QPM including up to three-phonon configurations\cite{Tso04,Tso08}. The theoretical method has been widely tested in systematic studies of electric and magnetic excitations from different energy and mass regions \cite{Herz99b,Volz06,Ton10,Rus13,Schw13,Kru15,Kri15} and also in predictions of new modes of nuclear excitations related to the pygmy quadrupole resonance (PQR) \cite{Tso11,Pel15,Spie16}. A further advantage of the three-phonon EDF+QPM calculations is that we consider explicitly all one-phonon configurations up to the neutron threshold including explicitly the PDR. Additional dynamical dipole core polarization contributions are accounted for by the isovector interaction strength which is fitted to reproduce the properties of the GDR. Differently from Ref.~\cite{Grin94} no additional effective charges are needed. 

In Table~\ref{tab:n82syst}, the experimental and theoretical QPM results for excitation energies, wave function structures and transition strengths are summarized. 
The QPM wave functions of the $2^{+}_{1}$ and $3^{-}_{1}$ excited states are dominated by one-phonon components related to the collective $2^{+}_{1}$ (about 93$\%$) and $3^{-}_{1}$ (about 90$\%$) QRPA one-phonon states, respectively. The main contributions to the $2^{+}_{1}$ QRPA state vectors in $N=82$ nuclei come from $[2d_{5/2}]^{2}_{p}$, $[1h_{11/2}]^{2}_{p}$, $[1g_{7/2}]^{2}_{p}$, and $[1g_{7/2}2d_{5/2}]_{p}$ two-quasiparticle proton configurations located close to the Fermi surface. This is related to the fact that the $[1g_{7/2}]_p$ level is the proton Fermi-level in $^{138}$Ba and $^{140}$Ce and the $[2d_{5/2}]_p$ level is the proton Fermi-level in $^{142}$Nd and $^{144}$Sm. Because of the pairing interaction the two-quasiparticle states situated close to the Fermi surface could spend part of the time below or above the Fermi surface. The major configuration reaches from a fraction of about 38$\%$ in $^{140}$Ce up to about 47$\%$ in $^{138}$Ba. The neutron contribution is related mainly to the $[1h_{11/2}2f_{7/2}]_{n}$ two-quasiparticle neutron configuration and varies between $\approx 3-5 \%$.
The $B(E2)$ transition probabilities follow closely the amount of collectivity of the $2^{+}_{1}$ QRPA states and consequently the largest $B(E2)$ value is obtained for the $^{140}$Ce nucleus as it is shown in Table~\ref{tab:n82syst} both from theory and experiment. 

In the case of the $3^{-}_{1}$ QRPA states there are two major competing contributions to the state vectors due to the $[2d_{5/2}1h_{11/2}]_{p}$ and $[1g_{7/2}1h_{11/2}]_{p}$ two-quasiparticle proton configurations. The $[2d_{5/2}1h_{11/2}]_{p}$ proton component contributes from about 63$\%$ in $^{138}$Ba up to about 76$\%$ in $^{144}$Sm. The $[2d_{5/2}1h_{11/2}]_{p}$ proton component also contributes dominantly to the $B(E3)$ transition matrix elements to the ground state.
The neutron contribution to the $3^{-}_{1}$ QRPA states in $N=82$ nuclei is related mainly to the $[1h_{11/2}1i_{13/2}]_{n}$ two-quasiparticle neutron component and varies between $\approx 3-6 \%$. 
The experimentally observed general trend of decreasing energy of the $3^-$ excited states with the increase of the proton number in $N=82$ nuclei is reproduced well in our calculations with smooth changes of the residual interaction model parameters.

\begin{figure}[tbp]
\centering
\includegraphics[width=0.5\textwidth]{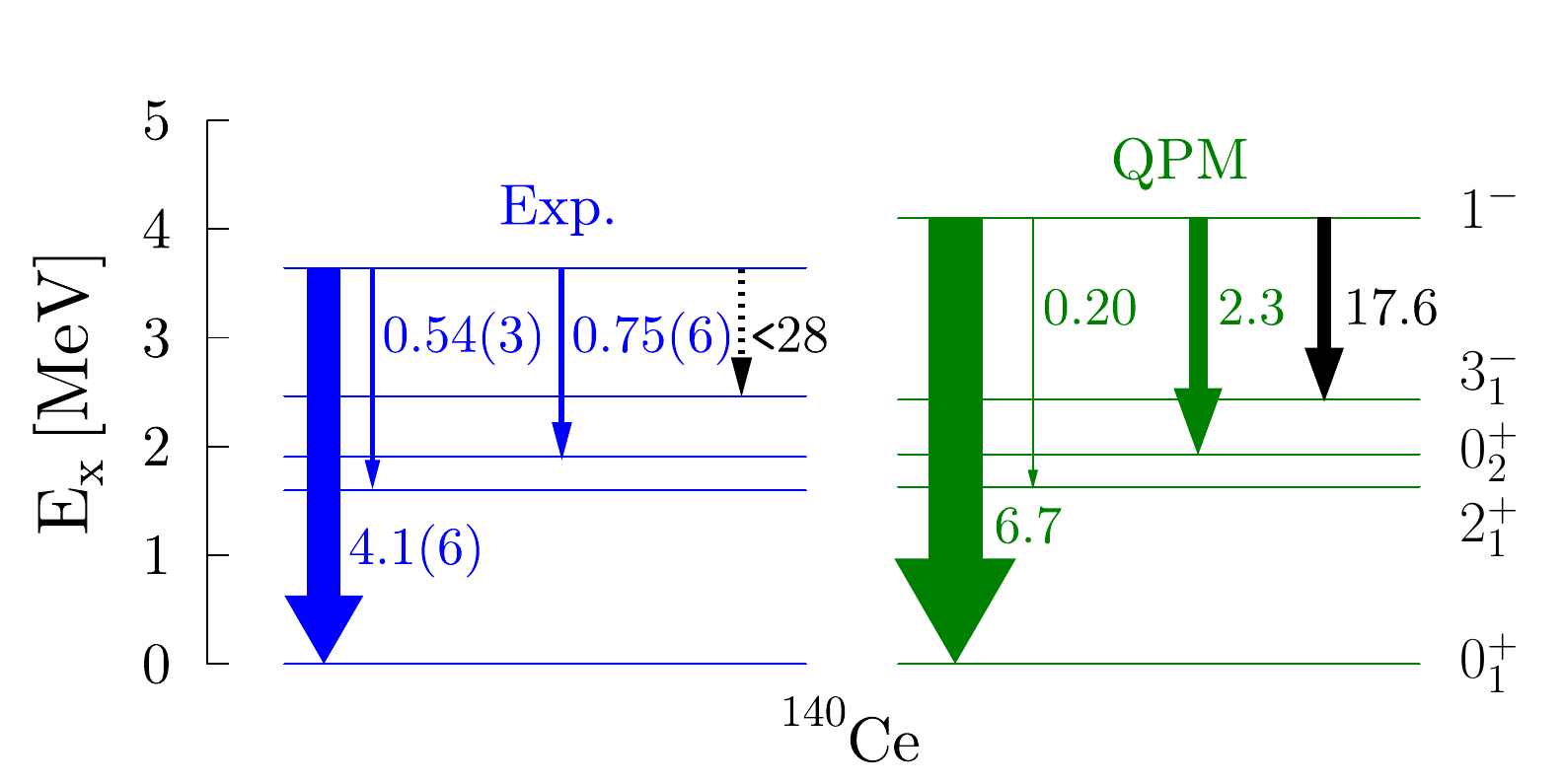}

\caption{(Color online) Experimental (left) and theoretical (right) decay pattern of the $1^-_1$ state in $^{140}$Ce. The numbers indicate the transition strengths in mW.u. for $E1$ transitions (blue and green) and W.u. for $E2$ transitions (black).}
\label{fig:exp_qpm_140ce}
\end{figure}

The theoretical properties of the $2^{+}_{1}$ and $3^{-}_{1}$ QRPA phonons can be further examined in studies of low-energy two-phonon states related to the quadrupole-octupole multiplet. The QPM $1^-_1$ state has a major two-phonon $(2^{+}_{1}\otimes3^{-}_{1})_{1^{-}}$ content of more than 93$\%$. However, in all nuclei a contribution to the state wave function of higher-lying one-phonon PDR states of larger than 1$\%$ was found. With increasing proton number toward $^{144}$Sm, the excitation energy of the $1^-_1$ state decreases following the decrease of the excitation energy of the $3^-_1$ state, which, on the other hand, reduces the coupling with PDR and IVGDR phonons. Three-phonon contributions are found of minor importance for the wave function and transition properties of the $1_1^-$ states in the considered $N=82$ nuclei. The decay pattern of the $1^-_1$ state in $^{140}$Ce is illustrated in Fig.~\ref{fig:exp_qpm_140ce} in terms of the transition strengths, indicated by the arrow thicknesses. The general agreement between experiment and QPM calculations is reasonably good.

\begin{table*}[tbp]
\centering
\caption[QPM results]{Comparison of experimental data with QPM results for stable even-even $N=82$ isotones.}
\vskip 6pt
\begin{threeparttable}
\begin{tabular}{p{3.8cm}p{3cm}p{3cm}p{3cm}p{3cm}p{1.3cm}}
\hline
\hline
 &   $^{138}$Ba   &   $^{140}$Ce   &   $^{142}$Nd   &   $^{144}$Sm    &\\
  \hline
 $E_x(2^+_1)$~[MeV]  & 1.436 &   1.596  & 1.576   & 1.660  & Exp.\\
\hline 
 $E_x(2^+_1)$~[MeV]  & 1.415 & 1.550  & 1.547 & 1.670 & QPM\\
Structure & 97.3$\%$ $2^+_1$   &  96.0$\%$ $2^+_1$   & 92.7$\%$ $2^+_1$    & 94.2$\%$ $2^+_1$    & \\
          &    &  + 1.9$\%$ $(3^-_1 \otimes 3^-_1)_{2^+}$  & + 3.4$\%$ $(3^-_1 \otimes 3^-_1)_{2^+}$  & + 3.5$\%$ $(3^-_1 \otimes 3^-_1)_{2^+}$    & \\

\hline
$E_x(0^+_2)$~[MeV] & 2.340 & 1.903  &  2.217 & 2.477 &Exp.\\ 

\hline
 $E_x(0^+_2)$~[MeV] & 2.400 & 1.901 &  2.170 & 2.220 &QPM\\
Structure & 97.6$\%$ $0^+_2$ &  64.2$\%$ $0^+_2$   & 60.3$\%$ $0^+_2$   & 61.1$\%$ $0^+_2$  & \\
          &     + 1.3$\%$ $(2^+_1 \otimes 2^+_1)_{0^+}$              &  + 15$\%$ $0^+_3$   & + 20.8$\%$ $0^+_3$   & + 14.7$\%$ $0^+_3$  & \\
          &   &  + 14.3$\%$ $(3^-_1 \otimes 3^-_1)_{0^+}$  & + 14.8$\%$ $(3^-_1 \otimes 3^-_1)_{0^+}$  & + 22.6$\%$ $(3^-_1 \otimes 3^-_1)_{0^+}$    & \\

\hline					
 $E_x(3^-_1)$~[MeV]    &  2.881 &  2.464 & 2.084  & 1.810 &Exp.\\
\hline
 $E_x(3^-_1)$~[MeV] & 2.845 & 2.390 & 2.030    &1.730 &QPM\\
Structure & 92.7$\%$ $3^-_1$  &  88.9$\%$ $3^-_1$   &  91.2$\%$ $3^-_1$  & 92.3$\%$ $3^-_1$ & \\
          & + 7.2$\%$ $(2^+_1 \otimes 3^-_1)_{3^-}$  &  + 8.4$\%$ $(2^+_1 \otimes 3^-_1)_{3^-}$  & + 6.6$\%$ $(2^+_1 \otimes 3^-_1)_{3^-}$  & + 3.6$\%$ $(2^+_1 \otimes 3^-_1)_{3^-}$     & \\

\hline					
 $E_x(1^-_1)$~[MeV] & 4.026 & 3.643 & 3.424  & 3.225 &Exp. \\

\hline
  $E_x(1^-_1)$~[MeV] & 4.350 & 4.140 & 3.850    &3.589 &QPM\\
Structure & 94.6$\%$ $(2^+_1 \otimes 3^-_1)_{1^-}$  &  93.1$\%$ $(2^+_1 \otimes 3^-_1)_{1^-} $   & 93.2$\%$ $(2^+_1 \otimes 3^-_1)_{1^-} $   &  93.1$\%$ $(2^+_1 \otimes 3^-_1)_{1^-}$  & \\
          & + 1.9$\%$ $1^-_5 $   &  + 1.8$\%$ $1^-_5 $  & + 1.8$\%$ $1^-_4$   &  + 1.7$\%$ $1^-_4$   & \\
					& + 2.6$\%$  & + 3.5$\%$     & + $2.2$\%     &    &\\
					& $(2^+_1 \otimes 2^+_1\otimes 3^-_1)_{1^-} $      & $(2^+_1 \otimes 2^+_1\otimes 3^-_1)_{1^-} $    &  
					$(2^+_1 \otimes 2^+_1\otimes 3^-_1)_{1^-}$  &    &\\

\hline
\hline
$B(E1,1^-_1\rightarrow 0^+_1)$~[mW.u.]& 5.6(3)\tnote{e}& 4.1(6)\tnote{c} & 3.3(7)\tnote{a} & 3.7(5)\tnote{d}  & Exp.\\
& 7.8 & 6.7 & 5.3 & 4.9 & QPM \\
\hline
$B(E1,1^-_1\rightarrow 2^+_1)$~[mW.u.]& 0.48(12)\tnote{e} & 0.54(3)\tnote{b} & 0.77(16)\tnote{a}  & 0.61(13)\tnote{d} &Exp.\\
 & 0.21 & 0.20 & 0.33    & 0.30 &QPM\\
\hline
$B(E1,1^-_1\rightarrow 0^+_2)$~[mW.u.] &  - & 0.75(6)\tnote{b}  & -  & - &Exp. \\ 
& 0.5 &  2.3 &2.1& 2.3&QPM\\
\hline
$B(E2,1^-_1\rightarrow 3^-_1)$~[W.u.] &- & $<$ 28\tnote{b}&15.7(33)\tnote{a} & 16.6(40)\tnote{d}&Exp. \\
& 14.2 & 17.6 &16.1 & 16.0& QPM\\
\hline

$B(E2,2^+_1\rightarrow 0^+_1)$~[W.u.]& 10.7(4)\tnote{f} & 13.7(3)\tnote{f} & 12.3(4)\tnote{f} & 11.9(4)\tnote{f} &Exp. \\
 & 10.6 & 13.2 & 12.1& 12.0 & QPM\\
 \hline

$B(E3,3^-_1\rightarrow 0^+_1)$~[W.u.]& 16.8(1.6)\tnote{g} & 26(3)\tnote{h} & 29(5)\tnote{i} & 38(3)\tnote{j} & Exp. \\
 & 16.0 & 19.5 & 24.6 & 29.5 & QPM\\

\hline
\hline 
\end{tabular}
\label{tab:n82syst}
\begin{tablenotes}
    \item[a] adopted from Ref.~\cite{Wilh98}
    \item[b] this work
    \item[c] adopted from Ref.~\cite{Volz06}    
    \item[d] adopted from Ref.~\cite{Wilh96}
    \item[e] adopted from Ref.~\cite{Herz99}
    \item[f] adopted from Ref.~\cite{Rama87}
    \item[g] adopted from Ref.~\cite{Burn85}
    \item[h] adopted from Ref.~\cite{Pitt70}
    \item[i] adopted from Ref.~\cite{Mads71}
    \item[j] adopted from Ref.~\cite{Barf89}

\end{tablenotes}
\end{threeparttable}
\end{table*}

\begin{figure}[tbp]
\centering
\includegraphics[width=0.5\textwidth]{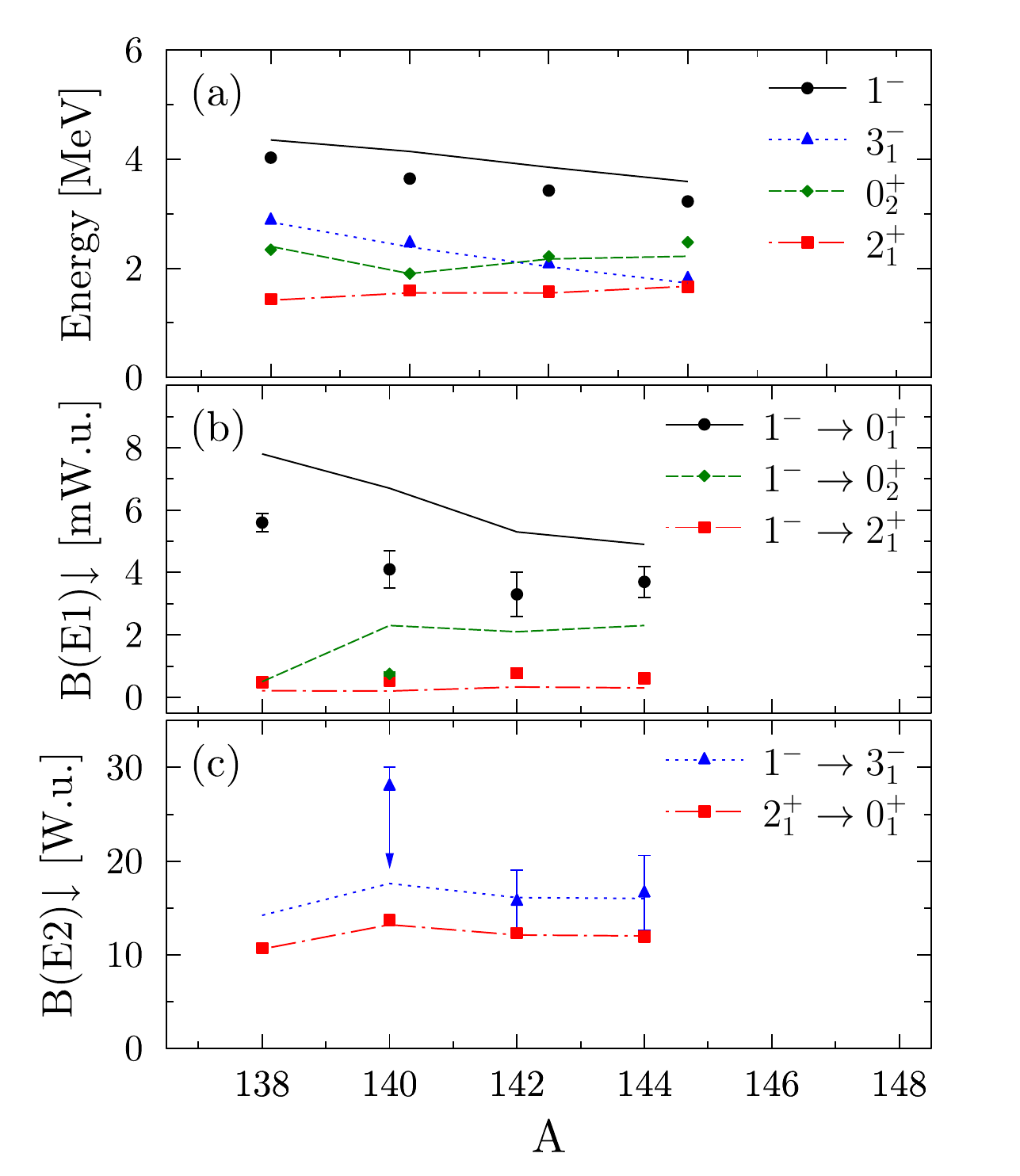}

\caption{(Color online) Compilation of experimental (markers) and QPM (lines) data in $N=82$ isotones: (a) Excitation energies, (b) $E1$ transition strengths of $1^- \rightarrow J_f^{\pi_f}$ transitions, (c) $E2$ transition strengths.}
\label{fig:n82syst}
\end{figure}

Now we would like to discuss the results within the systematics of the two-phonon $E1$ excitation mode in the $N=82$ isotones. For this purpose the compiled experimental data and the QPM results are shown in Fig.~\ref{fig:n82syst}. The upper panel shows the energy trend of the excitation energies with increasing proton number. The excitation energy of the $3^-_1$ state decreases steeper than the energy of the $2^+_1$ state increases. This leads to a decrease of the excitation energy of the $1^-_1$ state. The proton number dependence of the excitation energy of the $0^+_2$ state shows a different behavior with a minimum for cerium. The energy trends for all states are well reproduced by the QPM. The experimentally observed excitation energies of the $1^-_1$ states are typically lower than compared to the sum energy of the constituent phonons which is a known feature for two-phonon $1^-_1$ states \cite{Andr01}.

In the following, some theoretical details on the structure of the QPM $0^+_2$ state in $N=82$ nuclei are given (see Table~\ref{tab:n82syst}). The QPM $0^{+}_{2}$ excited states are dominated by one-phonon components related to the $0^{+}_{2}$ QRPA phonon which has the largest contribution of 97.6$\%$ in $^{138}$Ba. For comparison the same component gives 64.2$\%$ in $^{140}$Ce, 60.3$\%$ in $^{142}$Nd, and 61.1$\%$ in $^{144}$Sm, respectively. A considerable contribution, due to to the $0^{+}_{3}$ QRPA phonon, of 15$\%$ in $^{140}$Ce, 20.8$\%$ in $^{142}$Nd, and 14.7$\%$ in $^{144}$Sm, is found as well. In addition two-phonon $(2^{+}_{1}\otimes 2^{+}_{1})_{0^{+}}$ and $(3^{-}_{1}\otimes 3^{-}_{1})_{0^{+}}$ configurations contribute to the structure of the $0^+_2$ excited states. In particular, the latter are very important for transitions between two-phonon states. Thus, from the calculations it is found that the $(3^{-}_{1}\otimes 3^{-}_{1})_{0^{+}}$ state has the largest counterpart to the structure of the $0^+_2$ excited state in $^{144}$Sm which corresponds also to one of the largest $B(E1,1^-_1\rightarrow 0^+_2)$ transition probabilities in comparison with the other considered $N=82$ nuclei.
In general, the energy of the $0_2^+$ QRPA state should increase with the total strength of the monopole pairing interaction and the width of the pairing gap $\Delta_p$, which in turn increases with the proton number in the case of the neutron-magic $N=82$ isotones. This means the pairing gap in $^{140}$Ce is larger than that in $^{138}$Ba. However, different effects can lead to a lowering of the energy of the $0_2^+$ state. In particular, the structure of the QRPA $0_2^+$ state is a pure proton excitation resulting from re-coupling processes of two-quasiparticle states from the $[2d_{5/2}]^{2}_{p}$, $[1g_{7/2}]^{2}_{p}$, and $[1h_{11/2}]^{2}_{p}$ proton subshells. The energy of the $[2d_{5/2}]^{2}_{p}$ two-quasiparticle proton configuration, which has the major contribution of 55.9\% to the QRPA $0_2^+$ state in $^{138}$Ba, is higher than that in $^{140}$Ce, where the $[2d_{5/2}]^{2}_{p}$ two-quasiparticle proton configuration is the second of importance with 48\%. Furthermore, the main contribution of the QRPA $0_2^+$ state in $^{140}$Ce is due to the $[1g_{7/2}]^{2}_{p}$ (50.1\%) two-quasiparticle proton configuration whose energy is also lower than the energy of the $[2d_{5/2}]^{2}_{p}$ two-quasiparticle proton configuration in $^{138}$Ba. Consequently, even though the total pairing energy $\Delta_p^2/G_p$, where $G_p$ is the monopole pairing strength constant, is larger in $^{140}$Ce than that in $^{138}$Ba, the mentioned shell effects lead to the lowest energy of the QRPA $0_2^+$ state in $^{140}$Ce in comparison with the other investigated $N=82$ isotones. In addition, the calculated anharmonicity contributions to the QPM $0_2^+$ state are larger than those in the neighboring $^{138}$Ba and $^{142}$Nd nuclei, in the case of $^{140}$Ce which further reduce the excitation energy of the $0^+_2$ state. 

The $B(E1){\downarrow}$ transition strengths for three different decay channels of the $1_1^-$ state are shown in Fig.~\ref{fig:n82syst}(b). The predicted minimum of the $1_1^-\rightarrow 0^+_1$ transition strength for $^{144}$Sm is not seen in the data, but still the trend is consistent within the experimental uncertainties. The theoretical value of this transition strength is strongly correlated with the contribution of the two-phonon matrix element. The latter depends strongly on the amplitude of the two-phonon $(2^{+}_{1} \otimes 3^{-}_{1})_{1^{-}_{1}}$ component which is one of the smallest in $^{144}$Sm (see also Table~\ref{tab:n82syst}) and also on the collectivity of the involved two-phonon states. Furthermore, as discussed above, the presence of PDR and IVGDR counterparts to the wave function of the $1_1^-$ states influences as well their decay rates. In particular, the total amount of one-phonon contributions to the $B(E1,1^-_1 \rightarrow 0^+_1)$ transition probability varies from 7.2$\%$ in $^{140}$Ce up to 29$\%$ in $^{138}$Ba. A relatively constant behavior of the $1_1^-\rightarrow 2^+_1$ transition strengths for $^{140}$Ce, $^{142}$Nd, $^{144}$Sm is found in both, experiment and theory, although the absolute values are slightly underestimated. In the QPM the $1_1^-\rightarrow 2^+_1$ transition strength is determined by the matrix element which couples the two-phonon components of the $1_1^-$ and $2^+_1$ state and depends mainly on the collectivity of the $2^+_1$ state which is an almost pure one-phonon state in the $N=82$ isotones. In this case, the nucleus $^{138}$Ba has the least collective $2^+_1$ state and consequently one of the smallest $1^-\rightarrow 2^+_1$ transition strength. However, one should also note that the $B(E1)$ transition probability is determined by the sum of the matrix elements of all two-phonon contributions which might have different signs and cancel out. This is the case for $^{142}$Nd and $^{144}$Sm. In general, this transition belongs to the so-called boson-forbidden transitions. In particular its value is very small and even minor contributions to the state vectors can affect the transition probability.

The lower panel of Fig.~\ref{fig:n82syst} displays the $B(E2)$ values for the $1^-_1\rightarrow 3^-_1$ and $2^+_1 \rightarrow 0^+_1$ transition, respectively. The agreement between QPM and experimental data is excellent for, both, the $2^+_1 \rightarrow 0^+_1$ and the $1^-_1\rightarrow 3^-_1$ transition strengths. The $B(E2,1^-_1\rightarrow 3^-_1)$ values for $^{142}$Nd and $^{144}$Sm were measured in proton-scattering experiments \cite{Wilh96,Wilh98}. The presently determined upper limit for the $B(E2,1^-_1\rightarrow 3^-_1)$ value of $^{140}$Ce is consistent with the QPM and would also fit into the $N=82$ systematics. More experimental effort is needed to measure this transition strength or further reduce its upper limit.

From the newly observed decays of the $1^-_1$ state into the $2^+_1$ and $0^+_2$ states we find strong evidence for the two-phonon character of the $1^-_1$ state in $^{140}$Ce. This conclusion is fully supported by our new QPM calculations.

\section{Results for $^{40}$Ca and $^{44}$Ca}\label{sec:40ca}
The calcium chain has five stable even-even isotopes in the light-to-medium mass region covering a wide $N/Z$ range. Low-lying $E1$ excitations have been studied systematically in $^{40}$Ca, $^{44}$Ca, and $^{48}$Ca by means of NRF experiments \cite{Hart00,Hart02,Hart04,Isaa11,Dery14}. The doubly-magic $N=Z$ nucleus $^{40}$Ca exhibits almost no low-lying $E1$ strength, whereas $^{44}$Ca and $^{48}$Ca exhaust more and a similar amount of the Thomas-Reiche-Kuhn energy-weighted sum rule \cite{Hart04,Isaa11}.

\begin{figure}[tbp]
\centering
\includegraphics[width=0.5\textwidth]{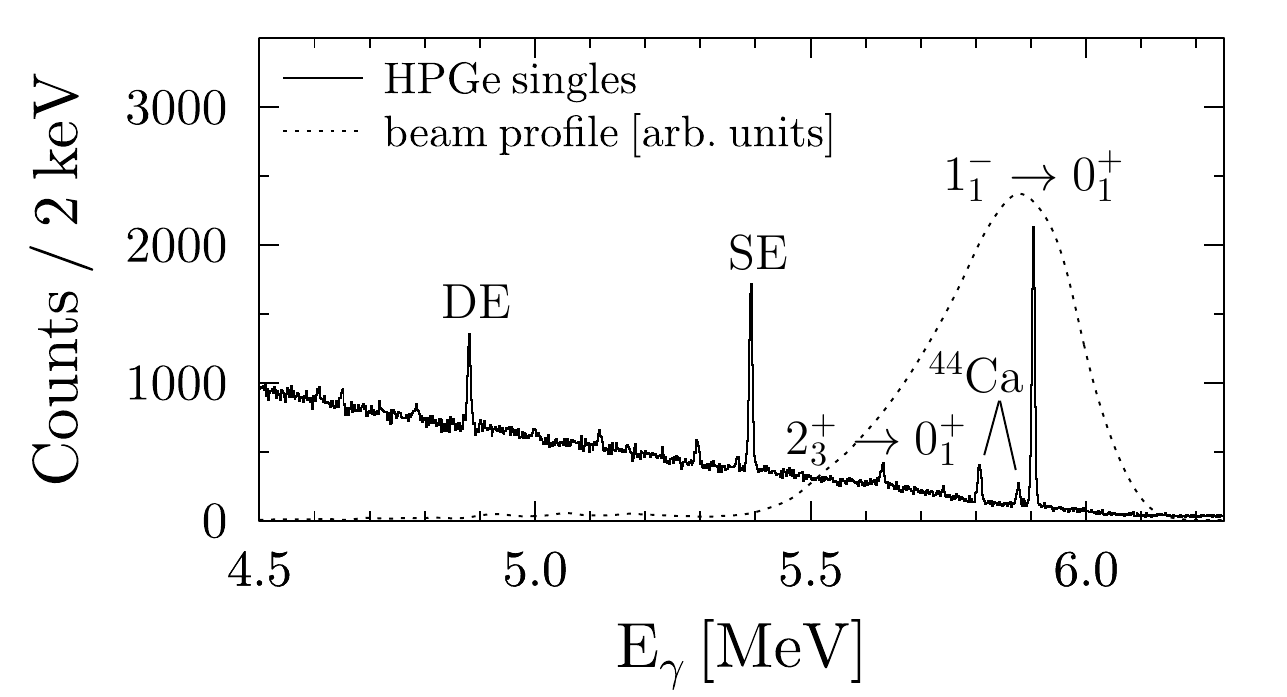}

\caption{Singles $\gamma$-ray spectrum of the HPGe detectors for the measurement on $^{40}$Ca. The transitions within the excitation window defined by the beam profile (dashed curve) are labeled. Single (SE) and double (DE) escape peaks are visible at lower energies.}
\label{fig:40ca_spec}
\end{figure}

The $B(E1)$ strength of $^{40}$Ca below the particle threshold is mainly carried by one excitation at 6.9\,MeV which was also strongly excited in an $(\alpha,\alpha'\gamma)$ experiment \cite{Poel92}. It is interpreted as a pure isoscalar oscillation which is predicted in all Ca isotopes \cite{Papa12} and was experimentally identified in $^{40}$Ca and $^{48}$Ca \cite{Dery14}. The quadrupole-octupole two-phonon candidate which is investigated in the present work, is the $1^-_1$ state at 5.9\,MeV that has a $B(E1,1^-_1\rightarrow 0^+_1)$ strength of 0.20(2)\,mW.u. \cite{Hart02}. In total four ground-state transitions of excited states in $^{40}$Ca and $^{44}$Ca lie within the beam profile as shown in Fig.~\ref{fig:40ca_spec}. Spin and parity of the two excited states in $^{40}$Ca are known from previous studies \cite{Hart02} and are confirmed in the present experiment. For the two $J=1$ states in $^{44}$Ca the parity was unknown. Therefore, the data taken in the present experiment was also used to perform a parity assignment as explained in Sec.~\ref{sec:data}. The results are given in Table~\ref{tab:ca_parities}. On the basis of the measured experimental asymmetries, negative parity can be assigned to both states.

\begin{table}[tbp]
\centering
\caption[Asymmetries of ${J=1}$ and ${J=2}$ states in $^{40,44}$Ca]{Experimental asymmetries of ${J=1}$ and ${J=2}$ states in $^{40,44}$Ca obtained in the present $(\vec{\gamma},\gamma')$ experiment.}
\vskip 6pt
\begin{threeparttable}
\begin{tabular}{p{0.08\textwidth}p{0.065\textwidth}p{0.045\textwidth}p{0.13\textwidth}c}
 \hline
 $E_x$ [keV] & nucleus & $J^\pi$ & asymmetry $\epsilon$  & $J^\pi$ (this work) \\
 \hline
5628.9	& $^{40}$Ca & $2^+$\tnote{a}& -0.8(4) & $2^+$\\
5806.3	& $^{44}$Ca & $1$\tnote{b} & -0.31(4) & $1^-$\\
5875.8	& $^{44}$Ca & $1$\tnote{b} & -0.33(6) & $1^-$\\
5902.5	& $^{40}$Ca & $1^-$\tnote{a} & -0.336(12) & $1^-$\\
\hline
\end{tabular}
\label{tab:ca_parities}
\begin{tablenotes}
    \item[a] Ref. \cite{Hart02} and references therein
    \item[b] Ref. \cite{Isaa11}
   % \item[c] adopted from Ref. \cite{Hart02}
\end{tablenotes}
\end{threeparttable}
\end{table}

\begin{figure}[tbp]
\centering
\includegraphics[width=0.5\textwidth]{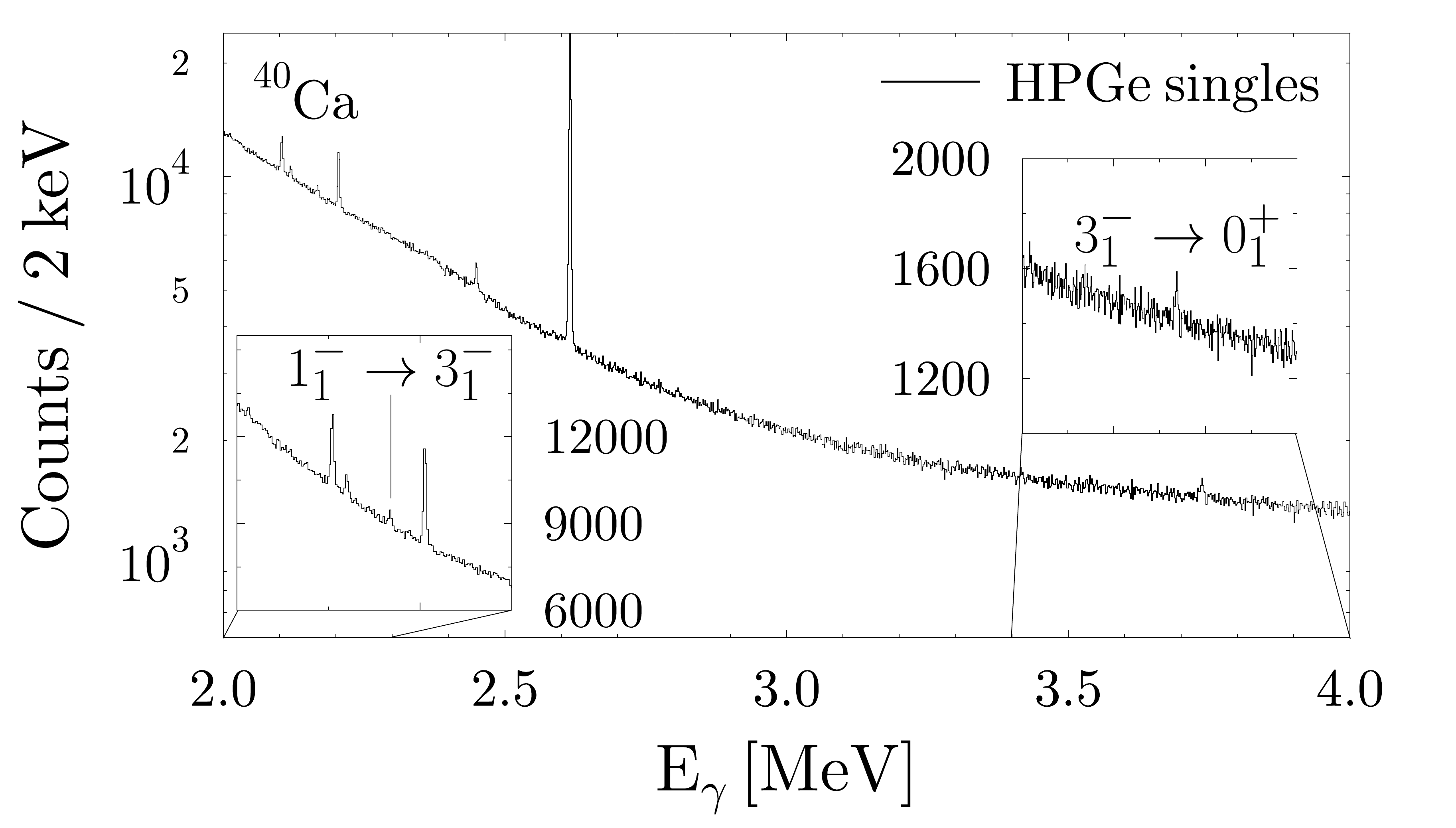}

\caption{Low-energy part of the singles $\gamma$-ray spectrum obtained with the HPGe detectors. The insets show the energy regions where the $1^-_1\rightarrow 3^-_1$ (left) and $3^-_1 \rightarrow 0^+_1$ (right) transitions are located.}
\label{fig:40ca_spec_decay}
\end{figure}

Concerning the decay behavior of the $1^-_1$ state in $^{40}$Ca, the coincidence data suffered from low statistics. Therefore, the $\gamma$-ray singles spectra were taken into account in the further analysis. Compared to the much heavier $^{140}$Ce the non-resonant background at low energies is strongly reduced in $^{40}$Ca. The decay into the first $3^-_1$ state at 3.7\,MeV is observed in the singles $\gamma$-ray spectrum in terms of the primary $1^-_1\rightarrow 3^-_1$ transition as well as the secondary $3^-_1\rightarrow 0^+_1$ transition. These transitions are visible in the $\gamma$-ray spectrum shown in Fig.~\ref{fig:40ca_spec_decay}. A decay into the higher-lying first $2^+_1$ state at 3.9\,MeV is not observed. Note that the $3^-_1$ state is the lowest-lying excited state in $^{40}$Ca. The reduced transition strengths which were determined in previous experiments and in this work are summarized in Table~\ref{tab:ca_bes}. The $B(E2)$ values for the $1^-_1\rightarrow 3^-_1$ and the $2^+_1\rightarrow 0^+_1$ transitions agree within the error bars. This means the first $1^-_1$ state in $^{40}$Ca is supported as a candidate for the two-phonon $1^-$ state. Hence, the possibility of a collective phonon mode exists also in light nuclei like $^{40}$Ca.

\begin{table}[tbp]
\centering
\caption{Experimental results for the $\gamma$-decay behavior of the $1^-_1$ state in $^{40}$Ca.}
\vskip 6pt
\begin{threeparttable}
\begin{tabular}{ll}
 \hline
transition strength &  \\
 \hline
$B(E1,1^-_1\rightarrow 0^+_1)$ [mW.u.] & 0.20(2)\tnote{a}\\
$B(E2,1^-_1\rightarrow 3^-_1)$ [W.u.] & 4.2(12) \\
$B(E2,2^+_1\rightarrow 0^+_1)$ [W.u.] & 2.7(7)\tnote{a}\\
\hline
\end{tabular}
\label{tab:ca_bes}
\begin{tablenotes}
    \item[a] taken from Ref.~\cite{Hart02}
\end{tablenotes}
\end{threeparttable}
\end{table}

\section{Summary}
We investigated the decay pattern of two-phonon $1^-$ candidates in $^{40}$Ca and $^{140}$Ce by means of $(\vec{\gamma},\gamma')$ experiments at the HI$\gamma$S facility. The experiments were performed using the $\gamma$-$\gamma$ coincidence setup $\gamma^3$. For both nuclei new decay paths were found in addition to the known strong ground-state decay. For $^{140}$Ce the $E1$ strength for the $1^-_1\rightarrow 2^+_1$ transition was determined. The deduced value fits into the $N=82$ systematics. For the first time in $^{140}$Ce and in the $N=82$ isotones an $E1$ transition of the $1^-_1$ state into the first excited $0^+_2$ state was observed and quantified. Microscopic calculations on the basis of the EDF+QPM approach support the interpretation of a dominant two-phonon character of the $1^-_1$ state. In the future, a measurement of the $1^-_1 \rightarrow 3^-_1$ transition strength or a more stringent upper limit for this observable could serve as an additional test of the model and associated interpretation.

For $^{40}$Ca the direct decay of the $1^-_1$ state into the first $3^-_1$ state was observed. Its transition strength is equal to the $2^+_1\rightarrow 0^+_1$ transition strength within the experimental errors. Thus, it is consistent with the harmonic model and hints to a two-phonon structure of the $1^-_1$ state. A systematic investigation of the decay behavior of two-phonon $1^-$ candidates in other Ca isotopes could help to establish this collective excitation mode in the light-to-medium mass region.

\begin{acknowledgments}
The authors would like to thank K.-O.~Zell and S.~Thiel for their help concerning the target preparation and J.~Jolie for providing LaBr$_3$ detectors. We thank M.~Spieker for valuable discussions. We furthermore highly acknowledge the support of the accelerator staff at HI${\gamma}$S during the beam times.
This work is supported by the DFG (ZI 510/7-1 and SFB 634) and the Alliance Program of the Helmholtz Association (HA216/EMMI). N.T. is supported by the Helmholtz International Center for FAIR within the framework of the LOEWE program. HI${\gamma}$S and the TUNL group is supported by the U.S. Department of Energy, Office of Nuclear Physics, Grant No. DE-FG02-97ER41033.
\end{acknowledgments}

% Create the reference section using BibTeX:
%\bibliography{lit.bib.bib}

\begin{thebibliography}{79}%
\makeatletter
\providecommand \@ifxundefined [1]{%
 \@ifx{#1\undefined}
}%
\providecommand \@ifnum [1]{%
 \ifnum #1\expandafter \@firstoftwo
 \else \expandafter \@secondoftwo
 \fi
}%
\providecommand \@ifx [1]{%
 \ifx #1\expandafter \@firstoftwo
 \else \expandafter \@secondoftwo
 \fi
}%
\providecommand \natexlab [1]{#1}%
\providecommand \enquote  [1]{``#1''}%
\providecommand \bibnamefont  [1]{#1}%
\providecommand \bibfnamefont [1]{#1}%
\providecommand \citenamefont [1]{#1}%
\providecommand \href@noop [0]{\@secondoftwo}%
\providecommand \href [0]{\begingroup \@sanitize@url \@href}%
\providecommand \@href[1]{\@@startlink{#1}\@@href}%
\providecommand \@@href[1]{\endgroup#1\@@endlink}%
\providecommand \@sanitize@url [0]{\catcode `\\12\catcode `\$12\catcode
  `\&12\catcode `\#12\catcode `\^12\catcode `\_12\catcode `\%12\relax}%
\providecommand \@@startlink[1]{}%
\providecommand \@@endlink[0]{}%
\providecommand \url  [0]{\begingroup\@sanitize@url \@url }%
\providecommand \@url [1]{\endgroup\@href {#1}{\urlprefix }}%
\providecommand \urlprefix  [0]{URL }%
\providecommand \Eprint [0]{\href }%
\providecommand \doibase [0]{http://dx.doi.org/}%
\providecommand \selectlanguage [0]{\@gobble}%
\providecommand \bibinfo  [0]{\@secondoftwo}%
\providecommand \bibfield  [0]{\@secondoftwo}%
\providecommand \translation [1]{[#1]}%
\providecommand \BibitemOpen [0]{}%
\providecommand \bibitemStop [0]{}%
\providecommand \bibitemNoStop [0]{.\EOS\space}%
\providecommand \EOS [0]{\spacefactor3000\relax}%
\providecommand \BibitemShut  [1]{\csname bibitem#1\endcsname}%
\let\auto@bib@innerbib\@empty
%</preamble>
\bibitem [{\citenamefont {Harakeh}\ and\ \citenamefont {van~der
  Woude}(2001)}]{harakeh}%
  \BibitemOpen
  \bibfield  {author} {\bibinfo {author} {\bibfnamefont {M.~N.}\ \bibnamefont
  {Harakeh}}\ and\ \bibinfo {author} {\bibfnamefont {A.}~\bibnamefont {van~der
  Woude}},\ }\href@noop {} {\emph {\bibinfo {title} {{Giant Resonances}}}}\
  (\bibinfo  {publisher} {Oxford University Press},\ \bibinfo {address} {New
  York},\ \bibinfo {year} {2001})\BibitemShut {NoStop}%
\bibitem [{\citenamefont {Berman}(1975)}]{Ber75}%
  \BibitemOpen
  \bibfield  {author} {\bibinfo {author} {\bibfnamefont {B.~L.}\ \bibnamefont
  {Berman}},\ }\href@noop {} {\bibfield  {journal} {\bibinfo  {journal} {At.
  Data Nucl. Data Tables}\ }\textbf {\bibinfo {volume} {15}},\ \bibinfo {pages}
  {319} (\bibinfo {year} {1975})}\BibitemShut {NoStop}%
\bibitem [{\citenamefont {Bartholomew}(1961)}]{Bart61}%
  \BibitemOpen
  \bibfield  {author} {\bibinfo {author} {\bibfnamefont {G.~A.}\ \bibnamefont
  {Bartholomew}},\ }\href@noop {} {\bibfield  {journal} {\bibinfo  {journal}
  {Ann. Rev. Nucl. Sc.}\ }\textbf {\bibinfo {volume} {11}},\ \bibinfo {pages}
  {259} (\bibinfo {year} {1961})}\BibitemShut {NoStop}%
\bibitem [{\citenamefont {Herzberg}\ \emph {et~al.}(1997)\citenamefont
  {Herzberg}, \citenamefont {von Brentano}, \citenamefont {Eberth},
  \citenamefont {Enders}, \citenamefont {Fischer}, \citenamefont {Huxel},
  \citenamefont {Klemme}, \citenamefont {von Neumann-Cosel}, \citenamefont
  {Nicolay}, \citenamefont {Pietralla}, \citenamefont {Ponomarev},
  \citenamefont {Reif}, \citenamefont {Richter}, \citenamefont {Schlegel},
  \citenamefont {Schwengner}, \citenamefont {Skoda}, \citenamefont {Thomas},
  \citenamefont {Wiedenhöver}, \citenamefont {Winter},\ and\ \citenamefont
  {Zilges}}]{Herz99b}%
  \BibitemOpen
  \bibfield  {author} {\bibinfo {author} {\bibfnamefont {R.-D.}\ \bibnamefont
  {Herzberg}}, \bibinfo {author} {\bibfnamefont {P.}~\bibnamefont {von
  Brentano}}, \bibinfo {author} {\bibfnamefont {J.}~\bibnamefont {Eberth}},
  \bibinfo {author} {\bibfnamefont {J.}~\bibnamefont {Enders}}, \bibinfo
  {author} {\bibfnamefont {R.}~\bibnamefont {Fischer}}, \bibinfo {author}
  {\bibfnamefont {N.}~\bibnamefont {Huxel}}, \bibinfo {author} {\bibfnamefont
  {T.}~\bibnamefont {Klemme}}, \bibinfo {author} {\bibfnamefont
  {P.}~\bibnamefont {von Neumann-Cosel}}, \bibinfo {author} {\bibfnamefont
  {N.}~\bibnamefont {Nicolay}}, \bibinfo {author} {\bibfnamefont
  {N.}~\bibnamefont {Pietralla}}, \bibinfo {author} {\bibfnamefont
  {V.~{\BIBYu}.}\ \bibnamefont {Ponomarev}}, \bibinfo {author} {\bibfnamefont
  {J.}~\bibnamefont {Reif}}, \bibinfo {author} {\bibfnamefont {A.}~\bibnamefont
  {Richter}}, \bibinfo {author} {\bibfnamefont {C.}~\bibnamefont {Schlegel}},
  \bibinfo {author} {\bibfnamefont {R.}~\bibnamefont {Schwengner}}, \bibinfo
  {author} {\bibfnamefont {S.}~\bibnamefont {Skoda}}, \bibinfo {author}
  {\bibfnamefont {H.~G.}\ \bibnamefont {Thomas}}, \bibinfo {author}
  {\bibfnamefont {I.}~\bibnamefont {Wiedenhöver}}, \bibinfo {author}
  {\bibfnamefont {G.}~\bibnamefont {Winter}}, \ and\ \bibinfo {author}
  {\bibfnamefont {A.}~\bibnamefont {Zilges}},\ }\href@noop {} {\bibfield
  {journal} {\bibinfo  {journal} {Phys. Lett. B}\ }\textbf {\bibinfo {volume}
  {390}},\ \bibinfo {pages} {49} (\bibinfo {year} {1997})}\BibitemShut
  {NoStop}%
\bibitem [{\citenamefont {Zilges}\ \emph {et~al.}(2002)\citenamefont {Zilges},
  \citenamefont {Volz}, \citenamefont {Babilon}, \citenamefont {Hartmann},
  \citenamefont {Mohr},\ and\ \citenamefont {Vogt}}]{Zilg02}%
  \BibitemOpen
  \bibfield  {author} {\bibinfo {author} {\bibfnamefont {A.}~\bibnamefont
  {Zilges}}, \bibinfo {author} {\bibfnamefont {S.}~\bibnamefont {Volz}},
  \bibinfo {author} {\bibfnamefont {M.}~\bibnamefont {Babilon}}, \bibinfo
  {author} {\bibfnamefont {T.}~\bibnamefont {Hartmann}}, \bibinfo {author}
  {\bibfnamefont {P.}~\bibnamefont {Mohr}}, \ and\ \bibinfo {author}
  {\bibfnamefont {K.}~\bibnamefont {Vogt}},\ }\href@noop {} {\bibfield
  {journal} {\bibinfo  {journal} {Phys. Lett. B}\ }\textbf {\bibinfo {volume}
  {542}},\ \bibinfo {pages} {43} (\bibinfo {year} {2002})}\BibitemShut
  {NoStop}%
\bibitem [{\citenamefont {Savran}\ \emph {et~al.}(2013)\citenamefont {Savran},
  \citenamefont {Aumann},\ and\ \citenamefont {Zilges}}]{Savr13}%
  \BibitemOpen
  \bibfield  {author} {\bibinfo {author} {\bibfnamefont {D.}~\bibnamefont
  {Savran}}, \bibinfo {author} {\bibfnamefont {T.}~\bibnamefont {Aumann}}, \
  and\ \bibinfo {author} {\bibfnamefont {A.}~\bibnamefont {Zilges}},\
  }\href@noop {} {\bibfield  {journal} {\bibinfo  {journal} {Prog. Part. Nucl.
  Phys.}\ }\textbf {\bibinfo {volume} {70}},\ \bibinfo {pages} {210} (\bibinfo
  {year} {2013})}\BibitemShut {NoStop}%
\bibitem [{\citenamefont {Bryssinck}\ \emph {et~al.}(1999)\citenamefont
  {Bryssinck}, \citenamefont {Govor}, \citenamefont {Belic}, \citenamefont
  {Bauwens}, \citenamefont {Beck}, \citenamefont {von Brentano}, \citenamefont
  {De~Frenne}, \citenamefont {Eckert}, \citenamefont {Fransen}, \citenamefont
  {Govaert}, \citenamefont {Herzberg}, \citenamefont {Jacobs}, \citenamefont
  {Kneissl}, \citenamefont {Maser}, \citenamefont {Nord}, \citenamefont
  {Pietralla}, \citenamefont {Pitz}, \citenamefont {Ponomarev},\ and\
  \citenamefont {Werner}}]{Brys99}%
  \BibitemOpen
  \bibfield  {author} {\bibinfo {author} {\bibfnamefont {J.}~\bibnamefont
  {Bryssinck}}, \bibinfo {author} {\bibfnamefont {L.}~\bibnamefont {Govor}},
  \bibinfo {author} {\bibfnamefont {D.}~\bibnamefont {Belic}}, \bibinfo
  {author} {\bibfnamefont {F.}~\bibnamefont {Bauwens}}, \bibinfo {author}
  {\bibfnamefont {O.}~\bibnamefont {Beck}}, \bibinfo {author} {\bibfnamefont
  {P.}~\bibnamefont {von Brentano}}, \bibinfo {author} {\bibfnamefont
  {D.}~\bibnamefont {De~Frenne}}, \bibinfo {author} {\bibfnamefont
  {T.}~\bibnamefont {Eckert}}, \bibinfo {author} {\bibfnamefont
  {C.}~\bibnamefont {Fransen}}, \bibinfo {author} {\bibfnamefont
  {K.}~\bibnamefont {Govaert}}, \bibinfo {author} {\bibfnamefont {R.-D.}\
  \bibnamefont {Herzberg}}, \bibinfo {author} {\bibfnamefont {E.}~\bibnamefont
  {Jacobs}}, \bibinfo {author} {\bibfnamefont {U.}~\bibnamefont {Kneissl}},
  \bibinfo {author} {\bibfnamefont {H.}~\bibnamefont {Maser}}, \bibinfo
  {author} {\bibfnamefont {A.}~\bibnamefont {Nord}}, \bibinfo {author}
  {\bibfnamefont {N.}~\bibnamefont {Pietralla}}, \bibinfo {author}
  {\bibfnamefont {H.~H.}\ \bibnamefont {Pitz}}, \bibinfo {author}
  {\bibfnamefont {V.~{\BIBYu}.}\ \bibnamefont {Ponomarev}}, \ and\ \bibinfo
  {author} {\bibfnamefont {V.}~\bibnamefont {Werner}},\ }\href@noop {}
  {\bibfield  {journal} {\bibinfo  {journal} {Phys. Rev. C}\ }\textbf {\bibinfo
  {volume} {59}},\ \bibinfo {pages} {1930} (\bibinfo {year}
  {1999})}\BibitemShut {NoStop}%
\bibitem [{\citenamefont {Volz}\ \emph {et~al.}(2006)\citenamefont {Volz},
  \citenamefont {Tsoneva}, \citenamefont {Babilon}, \citenamefont {Elvers},
  \citenamefont {Hasper}, \citenamefont {Herzberg}, \citenamefont {Lenske},
  \citenamefont {Lindenberg}, \citenamefont {Savran},\ and\ \citenamefont
  {Zilges}}]{Volz06}%
  \BibitemOpen
  \bibfield  {author} {\bibinfo {author} {\bibfnamefont {S.}~\bibnamefont
  {Volz}}, \bibinfo {author} {\bibfnamefont {N.}~\bibnamefont {Tsoneva}},
  \bibinfo {author} {\bibfnamefont {M.}~\bibnamefont {Babilon}}, \bibinfo
  {author} {\bibfnamefont {M.}~\bibnamefont {Elvers}}, \bibinfo {author}
  {\bibfnamefont {J.}~\bibnamefont {Hasper}}, \bibinfo {author} {\bibfnamefont
  {R.-D.}\ \bibnamefont {Herzberg}}, \bibinfo {author} {\bibfnamefont
  {H.}~\bibnamefont {Lenske}}, \bibinfo {author} {\bibfnamefont
  {K.}~\bibnamefont {Lindenberg}}, \bibinfo {author} {\bibfnamefont
  {D.}~\bibnamefont {Savran}}, \ and\ \bibinfo {author} {\bibfnamefont
  {A.}~\bibnamefont {Zilges}},\ }\href@noop {} {\bibfield  {journal} {\bibinfo
  {journal} {Nucl. Phys. A}\ }\textbf {\bibinfo {volume} {779}},\ \bibinfo
  {pages} {1} (\bibinfo {year} {2006})}\BibitemShut {NoStop}%
\bibitem [{\citenamefont {Savran}\ \emph {et~al.}(2008)\citenamefont {Savran},
  \citenamefont {Fritzsche}, \citenamefont {Hasper}, \citenamefont
  {Lindenberg}, \citenamefont {M\"uller}, \citenamefont {Ponomarev},
  \citenamefont {Sonnabend},\ and\ \citenamefont {Zilges}}]{Sav08}%
  \BibitemOpen
  \bibfield  {author} {\bibinfo {author} {\bibfnamefont {D.}~\bibnamefont
  {Savran}}, \bibinfo {author} {\bibfnamefont {M.}~\bibnamefont {Fritzsche}},
  \bibinfo {author} {\bibfnamefont {J.}~\bibnamefont {Hasper}}, \bibinfo
  {author} {\bibfnamefont {K.}~\bibnamefont {Lindenberg}}, \bibinfo {author}
  {\bibfnamefont {S.}~\bibnamefont {M\"uller}}, \bibinfo {author}
  {\bibfnamefont {V.~{\BIBYu}.}\ \bibnamefont {Ponomarev}}, \bibinfo {author}
  {\bibfnamefont {K.}~\bibnamefont {Sonnabend}}, \ and\ \bibinfo {author}
  {\bibfnamefont {A.}~\bibnamefont {Zilges}},\ }\href@noop {} {\bibfield
  {journal} {\bibinfo  {journal} {Phys. Rev. Lett.}\ }\textbf {\bibinfo
  {volume} {100}},\ \bibinfo {pages} {232501} (\bibinfo {year}
  {2008})}\BibitemShut {NoStop}%
\bibitem [{\citenamefont {Tonchev}\ \emph {et~al.}(2010)\citenamefont
  {Tonchev}, \citenamefont {Hammond}, \citenamefont {Kelley}, \citenamefont
  {Kwan}, \citenamefont {Lenske}, \citenamefont {Rusev}, \citenamefont
  {Tornow},\ and\ \citenamefont {Tsoneva}}]{Ton10}%
  \BibitemOpen
  \bibfield  {author} {\bibinfo {author} {\bibfnamefont {A.~P.}\ \bibnamefont
  {Tonchev}}, \bibinfo {author} {\bibfnamefont {S.~L.}\ \bibnamefont
  {Hammond}}, \bibinfo {author} {\bibfnamefont {J.~H.}\ \bibnamefont {Kelley}},
  \bibinfo {author} {\bibfnamefont {E.}~\bibnamefont {Kwan}}, \bibinfo {author}
  {\bibfnamefont {H.}~\bibnamefont {Lenske}}, \bibinfo {author} {\bibfnamefont
  {G.}~\bibnamefont {Rusev}}, \bibinfo {author} {\bibfnamefont
  {W.}~\bibnamefont {Tornow}}, \ and\ \bibinfo {author} {\bibfnamefont
  {N.}~\bibnamefont {Tsoneva}},\ }\href@noop {} {\bibfield  {journal} {\bibinfo
   {journal} {Phys. Rev. Lett.}\ }\textbf {\bibinfo {volume} {104}},\ \bibinfo
  {pages} {072501} (\bibinfo {year} {2010})}\BibitemShut {NoStop}%
\bibitem [{\citenamefont {Toft}\ \emph {et~al.}(2010)\citenamefont {Toft},
  \citenamefont {Larsen}, \citenamefont {Agvaanluvsan}, \citenamefont
  {B\"urger}, \citenamefont {Guttormsen}, \citenamefont {Mitchell},
  \citenamefont {Nyhus}, \citenamefont {Schiller}, \citenamefont {Siem},
  \citenamefont {Syed},\ and\ \citenamefont {Voinov}}]{Tof11}%
  \BibitemOpen
  \bibfield  {author} {\bibinfo {author} {\bibfnamefont {H.~K.}\ \bibnamefont
  {Toft}}, \bibinfo {author} {\bibfnamefont {A.~C.}\ \bibnamefont {Larsen}},
  \bibinfo {author} {\bibfnamefont {U.}~\bibnamefont {Agvaanluvsan}}, \bibinfo
  {author} {\bibfnamefont {A.}~\bibnamefont {B\"urger}}, \bibinfo {author}
  {\bibfnamefont {M.}~\bibnamefont {Guttormsen}}, \bibinfo {author}
  {\bibfnamefont {G.~E.}\ \bibnamefont {Mitchell}}, \bibinfo {author}
  {\bibfnamefont {H.~T.}\ \bibnamefont {Nyhus}}, \bibinfo {author}
  {\bibfnamefont {A.}~\bibnamefont {Schiller}}, \bibinfo {author}
  {\bibfnamefont {S.}~\bibnamefont {Siem}}, \bibinfo {author} {\bibfnamefont
  {N.~U.~H.}\ \bibnamefont {Syed}}, \ and\ \bibinfo {author} {\bibfnamefont
  {A.}~\bibnamefont {Voinov}},\ }\href@noop {} {\bibfield  {journal} {\bibinfo
  {journal} {Phys. Rev. C}\ }\textbf {\bibinfo {volume} {81}},\ \bibinfo
  {pages} {064311} (\bibinfo {year} {2010})}\BibitemShut {NoStop}%
\bibitem [{\citenamefont {Schwengner}\ \emph {et~al.}(2013)\citenamefont
  {Schwengner}, \citenamefont {Massarczyk}, \citenamefont {Rusev},
  \citenamefont {Tsoneva}, \citenamefont {Bemmerer}, \citenamefont {Beyer},
  \citenamefont {Hannaske}, \citenamefont {Junghans}, \citenamefont {Kelley},
  \citenamefont {Kwan}, \citenamefont {Lenske}, \citenamefont {Marta},
  \citenamefont {Raut}, \citenamefont {Schilling}, \citenamefont {Tonchev},
  \citenamefont {Tornow},\ and\ \citenamefont {Wagner}}]{Schw13}%
  \BibitemOpen
  \bibfield  {author} {\bibinfo {author} {\bibfnamefont {R.}~\bibnamefont
  {Schwengner}}, \bibinfo {author} {\bibfnamefont {R.}~\bibnamefont
  {Massarczyk}}, \bibinfo {author} {\bibfnamefont {G.}~\bibnamefont {Rusev}},
  \bibinfo {author} {\bibfnamefont {N.}~\bibnamefont {Tsoneva}}, \bibinfo
  {author} {\bibfnamefont {D.}~\bibnamefont {Bemmerer}}, \bibinfo {author}
  {\bibfnamefont {R.}~\bibnamefont {Beyer}}, \bibinfo {author} {\bibfnamefont
  {R.}~\bibnamefont {Hannaske}}, \bibinfo {author} {\bibfnamefont {A.~R.}\
  \bibnamefont {Junghans}}, \bibinfo {author} {\bibfnamefont {J.~H.}\
  \bibnamefont {Kelley}}, \bibinfo {author} {\bibfnamefont {E.}~\bibnamefont
  {Kwan}}, \bibinfo {author} {\bibfnamefont {H.}~\bibnamefont {Lenske}},
  \bibinfo {author} {\bibfnamefont {M.}~\bibnamefont {Marta}}, \bibinfo
  {author} {\bibfnamefont {R.}~\bibnamefont {Raut}}, \bibinfo {author}
  {\bibfnamefont {K.~D.}\ \bibnamefont {Schilling}}, \bibinfo {author}
  {\bibfnamefont {A.}~\bibnamefont {Tonchev}}, \bibinfo {author} {\bibfnamefont
  {W.}~\bibnamefont {Tornow}}, \ and\ \bibinfo {author} {\bibfnamefont
  {A.}~\bibnamefont {Wagner}},\ }\href@noop {} {\bibfield  {journal} {\bibinfo
  {journal} {Phys. Rev. C}\ }\textbf {\bibinfo {volume} {87}},\ \bibinfo
  {pages} {024306} (\bibinfo {year} {2013})}\BibitemShut {NoStop}%
\bibitem [{\citenamefont {Crespi}\ \emph {et~al.}(2014)\citenamefont {Crespi},
  \citenamefont {Bracco}, \citenamefont {Nicolini}, \citenamefont {Mengoni},
  \citenamefont {Pellegri}, \citenamefont {Lanza}, \citenamefont {Leoni},
  \citenamefont {Maj}, \citenamefont {Kmiecik}, \citenamefont {Avigo},
  \citenamefont {Benzoni}, \citenamefont {Blasi}, \citenamefont {Boiano},
  \citenamefont {Bottoni}, \citenamefont {Brambilla}, \citenamefont {Camera},
  \citenamefont {Ceruti}, \citenamefont {Giaz}, \citenamefont {Million},
  \citenamefont {Morales}, \citenamefont {Vandone}, \citenamefont {Wieland},
  \citenamefont {Bednarczyk}, \citenamefont {Ciema\l{}a}, \citenamefont
  {Grebosz}, \citenamefont {Krzysiek}, \citenamefont {Mazurek}, \citenamefont
  {Zieblinski}, \citenamefont {Bazzacco}, \citenamefont {Bellato},
  \citenamefont {Birkenbach}, \citenamefont {Bortolato}, \citenamefont
  {Calore}, \citenamefont {Cederwall}, \citenamefont {Charles}, \citenamefont
  {de~Angelis}, \citenamefont {D\'esesquelles}, \citenamefont {Eberth},
  \citenamefont {Farnea}, \citenamefont {Gadea}, \citenamefont {G\"orgen},
  \citenamefont {Gottardo}, \citenamefont {Isocrate}, \citenamefont {Jolie},
  \citenamefont {Jungclaus}, \citenamefont {Karkour}, \citenamefont {Korten},
  \citenamefont {Menegazzo}, \citenamefont {Michelagnoli}, \citenamefont
  {Molini}, \citenamefont {Napoli}, \citenamefont {Pullia}, \citenamefont
  {Recchia}, \citenamefont {Reiter}, \citenamefont {Rosso}, \citenamefont
  {Sahin}, \citenamefont {Salsac}, \citenamefont {Siebeck}, \citenamefont
  {Siem}, \citenamefont {Simpson}, \citenamefont {S\"oderstr\"om},
  \citenamefont {Stezowski}, \citenamefont {Theisen}, \citenamefont {Ur},\ and\
  \citenamefont {Valiente-Dob\'on}}]{Cre13}%
  \BibitemOpen
  \bibfield  {author} {\bibinfo {author} {\bibfnamefont {F.~C.~L.}\
  \bibnamefont {Crespi}}, \bibinfo {author} {\bibfnamefont {A.}~\bibnamefont
  {Bracco}}, \bibinfo {author} {\bibfnamefont {R.}~\bibnamefont {Nicolini}},
  \bibinfo {author} {\bibfnamefont {D.}~\bibnamefont {Mengoni}}, \bibinfo
  {author} {\bibfnamefont {L.}~\bibnamefont {Pellegri}}, \bibinfo {author}
  {\bibfnamefont {E.~G.}\ \bibnamefont {Lanza}}, \bibinfo {author}
  {\bibfnamefont {S.}~\bibnamefont {Leoni}}, \bibinfo {author} {\bibfnamefont
  {A.}~\bibnamefont {Maj}}, \bibinfo {author} {\bibfnamefont {M.}~\bibnamefont
  {Kmiecik}}, \bibinfo {author} {\bibfnamefont {R.}~\bibnamefont {Avigo}},
  \bibinfo {author} {\bibfnamefont {G.}~\bibnamefont {Benzoni}}, \bibinfo
  {author} {\bibfnamefont {N.}~\bibnamefont {Blasi}}, \bibinfo {author}
  {\bibfnamefont {C.}~\bibnamefont {Boiano}}, \bibinfo {author} {\bibfnamefont
  {S.}~\bibnamefont {Bottoni}}, \bibinfo {author} {\bibfnamefont
  {S.}~\bibnamefont {Brambilla}}, \bibinfo {author} {\bibfnamefont
  {F.}~\bibnamefont {Camera}}, \bibinfo {author} {\bibfnamefont
  {S.}~\bibnamefont {Ceruti}}, \bibinfo {author} {\bibfnamefont
  {A.}~\bibnamefont {Giaz}}, \bibinfo {author} {\bibfnamefont {B.}~\bibnamefont
  {Million}}, \bibinfo {author} {\bibfnamefont {A.~I.}\ \bibnamefont
  {Morales}}, \bibinfo {author} {\bibfnamefont {V.}~\bibnamefont {Vandone}},
  \bibinfo {author} {\bibfnamefont {O.}~\bibnamefont {Wieland}}, \bibinfo
  {author} {\bibfnamefont {P.}~\bibnamefont {Bednarczyk}}, \bibinfo {author}
  {\bibfnamefont {M.}~\bibnamefont {Ciema\l{}a}}, \bibinfo {author}
  {\bibfnamefont {J.}~\bibnamefont {Grebosz}}, \bibinfo {author} {\bibfnamefont
  {M.}~\bibnamefont {Krzysiek}}, \bibinfo {author} {\bibfnamefont
  {K.}~\bibnamefont {Mazurek}}, \bibinfo {author} {\bibfnamefont
  {M.}~\bibnamefont {Zieblinski}}, \bibinfo {author} {\bibfnamefont
  {D.}~\bibnamefont {Bazzacco}}, \bibinfo {author} {\bibfnamefont
  {M.}~\bibnamefont {Bellato}}, \bibinfo {author} {\bibfnamefont
  {B.}~\bibnamefont {Birkenbach}}, \bibinfo {author} {\bibfnamefont
  {D.}~\bibnamefont {Bortolato}}, \bibinfo {author} {\bibfnamefont
  {E.}~\bibnamefont {Calore}}, \bibinfo {author} {\bibfnamefont
  {B.}~\bibnamefont {Cederwall}}, \bibinfo {author} {\bibfnamefont
  {L.}~\bibnamefont {Charles}}, \bibinfo {author} {\bibfnamefont
  {G.}~\bibnamefont {de~Angelis}}, \bibinfo {author} {\bibfnamefont
  {P.}~\bibnamefont {D\'esesquelles}}, \bibinfo {author} {\bibfnamefont
  {J.}~\bibnamefont {Eberth}}, \bibinfo {author} {\bibfnamefont
  {E.}~\bibnamefont {Farnea}}, \bibinfo {author} {\bibfnamefont
  {A.}~\bibnamefont {Gadea}}, \bibinfo {author} {\bibfnamefont
  {A.}~\bibnamefont {G\"orgen}}, \bibinfo {author} {\bibfnamefont
  {A.}~\bibnamefont {Gottardo}}, \bibinfo {author} {\bibfnamefont
  {R.}~\bibnamefont {Isocrate}}, \bibinfo {author} {\bibfnamefont
  {J.}~\bibnamefont {Jolie}}, \bibinfo {author} {\bibfnamefont
  {A.}~\bibnamefont {Jungclaus}}, \bibinfo {author} {\bibfnamefont
  {N.}~\bibnamefont {Karkour}}, \bibinfo {author} {\bibfnamefont
  {W.}~\bibnamefont {Korten}}, \bibinfo {author} {\bibfnamefont
  {R.}~\bibnamefont {Menegazzo}}, \bibinfo {author} {\bibfnamefont
  {C.}~\bibnamefont {Michelagnoli}}, \bibinfo {author} {\bibfnamefont
  {P.}~\bibnamefont {Molini}}, \bibinfo {author} {\bibfnamefont {D.~R.}\
  \bibnamefont {Napoli}}, \bibinfo {author} {\bibfnamefont {A.}~\bibnamefont
  {Pullia}}, \bibinfo {author} {\bibfnamefont {F.}~\bibnamefont {Recchia}},
  \bibinfo {author} {\bibfnamefont {P.}~\bibnamefont {Reiter}}, \bibinfo
  {author} {\bibfnamefont {D.}~\bibnamefont {Rosso}}, \bibinfo {author}
  {\bibfnamefont {E.}~\bibnamefont {Sahin}}, \bibinfo {author} {\bibfnamefont
  {M.~D.}\ \bibnamefont {Salsac}}, \bibinfo {author} {\bibfnamefont
  {B.}~\bibnamefont {Siebeck}}, \bibinfo {author} {\bibfnamefont
  {S.}~\bibnamefont {Siem}}, \bibinfo {author} {\bibfnamefont {J.}~\bibnamefont
  {Simpson}}, \bibinfo {author} {\bibfnamefont {P.-A.}\ \bibnamefont
  {S\"oderstr\"om}}, \bibinfo {author} {\bibfnamefont {O.}~\bibnamefont
  {Stezowski}}, \bibinfo {author} {\bibfnamefont {C.}~\bibnamefont {Theisen}},
  \bibinfo {author} {\bibfnamefont {C.}~\bibnamefont {Ur}}, \ and\ \bibinfo
  {author} {\bibfnamefont {J.~J.}\ \bibnamefont {Valiente-Dob\'on}},\
  }\href@noop {} {\bibfield  {journal} {\bibinfo  {journal} {Phys. Rev. Lett.}\
  }\textbf {\bibinfo {volume} {113}},\ \bibinfo {pages} {012501} (\bibinfo
  {year} {2014})}\BibitemShut {NoStop}%
\bibitem [{\citenamefont {Derya}\ \emph {et~al.}(2014)\citenamefont {Derya},
  \citenamefont {Savran}, \citenamefont {Endres}, \citenamefont {Harakeh},
  \citenamefont {Hergert}, \citenamefont {Kelley}, \citenamefont
  {Papakonstantinou}, \citenamefont {Pietralla}, \citenamefont {Ponomarev},
  \citenamefont {Roth}, \citenamefont {Rusev}, \citenamefont {Tonchev},
  \citenamefont {Tornow}, \citenamefont {W\"ortche},\ and\ \citenamefont
  {Zilges}}]{Dery14}%
  \BibitemOpen
  \bibfield  {author} {\bibinfo {author} {\bibfnamefont {V.}~\bibnamefont
  {Derya}}, \bibinfo {author} {\bibfnamefont {D.}~\bibnamefont {Savran}},
  \bibinfo {author} {\bibfnamefont {J.}~\bibnamefont {Endres}}, \bibinfo
  {author} {\bibfnamefont {M.~N.}\ \bibnamefont {Harakeh}}, \bibinfo {author}
  {\bibfnamefont {H.}~\bibnamefont {Hergert}}, \bibinfo {author} {\bibfnamefont
  {J.~H.}\ \bibnamefont {Kelley}}, \bibinfo {author} {\bibfnamefont
  {P.}~\bibnamefont {Papakonstantinou}}, \bibinfo {author} {\bibfnamefont
  {N.}~\bibnamefont {Pietralla}}, \bibinfo {author} {\bibfnamefont
  {V.~{\BIBYu}.}\ \bibnamefont {Ponomarev}}, \bibinfo {author} {\bibfnamefont
  {R.}~\bibnamefont {Roth}}, \bibinfo {author} {\bibfnamefont {G.}~\bibnamefont
  {Rusev}}, \bibinfo {author} {\bibfnamefont {A.~P.}\ \bibnamefont {Tonchev}},
  \bibinfo {author} {\bibfnamefont {W.}~\bibnamefont {Tornow}}, \bibinfo
  {author} {\bibfnamefont {H.~J.}\ \bibnamefont {W\"ortche}}, \ and\ \bibinfo
  {author} {\bibfnamefont {A.}~\bibnamefont {Zilges}},\ }\href@noop {}
  {\bibfield  {journal} {\bibinfo  {journal} {Phys. Lett. B}\ }\textbf
  {\bibinfo {volume} {730}},\ \bibinfo {pages} {288} (\bibinfo {year}
  {2014})}\BibitemShut {NoStop}%
\bibitem [{\citenamefont {Bracco}\ \emph {et~al.}(2015)\citenamefont {Bracco},
  \citenamefont {Crespi},\ and\ \citenamefont {Lanza}}]{Brac15}%
  \BibitemOpen
  \bibfield  {author} {\bibinfo {author} {\bibfnamefont {A.}~\bibnamefont
  {Bracco}}, \bibinfo {author} {\bibfnamefont {F.~C.~L.}\ \bibnamefont
  {Crespi}}, \ and\ \bibinfo {author} {\bibfnamefont {E.~G.}\ \bibnamefont
  {Lanza}},\ }\href@noop {} {\bibfield  {journal} {\bibinfo  {journal} {Eur.
  Phys. J. A}\ }\textbf {\bibinfo {volume} {51}},\ \bibinfo {pages} {99}
  (\bibinfo {year} {2015})}\BibitemShut {NoStop}%
\bibitem [{\citenamefont {Krumbholz}\ \emph {et~al.}(2015)\citenamefont
  {Krumbholz}, \citenamefont {von Neumann-Cosel}, \citenamefont {Hashimoto},
  \citenamefont {Tamii}, \citenamefont {Adachi}, \citenamefont {Bertulani},
  \citenamefont {Fujita}, \citenamefont {Fujita}, \citenamefont {Ganioglu},
  \citenamefont {Hatanaka}, \citenamefont {Iwamoto}, \citenamefont {Kawabata},
  \citenamefont {Khai}, \citenamefont {Krugmann}, \citenamefont {Martin},
  \citenamefont {Matsubara}, \citenamefont {Neveling}, \citenamefont {Okamura},
  \citenamefont {Ong}, \citenamefont {Poltoratska}, \citenamefont {Ponomarev},
  \citenamefont {Richter}, \citenamefont {Sakaguchi}, \citenamefont {Shimbara},
  \citenamefont {Shimizu}, \citenamefont {Simonis}, \citenamefont {Smit},
  \citenamefont {Susoy}, \citenamefont {Thies}, \citenamefont {Suzuki},
  \citenamefont {Yosoi},\ and\ \citenamefont {Zenihiro}}]{Kru15}%
  \BibitemOpen
  \bibfield  {author} {\bibinfo {author} {\bibfnamefont {A.~M.}\ \bibnamefont
  {Krumbholz}}, \bibinfo {author} {\bibfnamefont {P.}~\bibnamefont {von
  Neumann-Cosel}}, \bibinfo {author} {\bibfnamefont {T.}~\bibnamefont
  {Hashimoto}}, \bibinfo {author} {\bibfnamefont {A.}~\bibnamefont {Tamii}},
  \bibinfo {author} {\bibfnamefont {T.}~\bibnamefont {Adachi}}, \bibinfo
  {author} {\bibfnamefont {C.~A.}\ \bibnamefont {Bertulani}}, \bibinfo {author}
  {\bibfnamefont {H.}~\bibnamefont {Fujita}}, \bibinfo {author} {\bibfnamefont
  {Y.}~\bibnamefont {Fujita}}, \bibinfo {author} {\bibfnamefont
  {E.}~\bibnamefont {Ganioglu}}, \bibinfo {author} {\bibfnamefont
  {K.}~\bibnamefont {Hatanaka}}, \bibinfo {author} {\bibfnamefont
  {C.}~\bibnamefont {Iwamoto}}, \bibinfo {author} {\bibfnamefont
  {T.}~\bibnamefont {Kawabata}}, \bibinfo {author} {\bibfnamefont {N.~T.}\
  \bibnamefont {Khai}}, \bibinfo {author} {\bibfnamefont {A.}~\bibnamefont
  {Krugmann}}, \bibinfo {author} {\bibfnamefont {D.}~\bibnamefont {Martin}},
  \bibinfo {author} {\bibfnamefont {H.}~\bibnamefont {Matsubara}}, \bibinfo
  {author} {\bibfnamefont {R.}~\bibnamefont {Neveling}}, \bibinfo {author}
  {\bibfnamefont {H.}~\bibnamefont {Okamura}}, \bibinfo {author} {\bibfnamefont
  {H.~J.}\ \bibnamefont {Ong}}, \bibinfo {author} {\bibfnamefont
  {I.}~\bibnamefont {Poltoratska}}, \bibinfo {author} {\bibfnamefont
  {V.~{\BIBYu}.}\ \bibnamefont {Ponomarev}}, \bibinfo {author} {\bibfnamefont
  {A.}~\bibnamefont {Richter}}, \bibinfo {author} {\bibfnamefont
  {H.}~\bibnamefont {Sakaguchi}}, \bibinfo {author} {\bibfnamefont
  {Y.}~\bibnamefont {Shimbara}}, \bibinfo {author} {\bibfnamefont
  {Y.}~\bibnamefont {Shimizu}}, \bibinfo {author} {\bibfnamefont
  {J.}~\bibnamefont {Simonis}}, \bibinfo {author} {\bibfnamefont {F.~D.}\
  \bibnamefont {Smit}}, \bibinfo {author} {\bibfnamefont {G.}~\bibnamefont
  {Susoy}}, \bibinfo {author} {\bibfnamefont {J.~H.}\ \bibnamefont {Thies}},
  \bibinfo {author} {\bibfnamefont {T.}~\bibnamefont {Suzuki}}, \bibinfo
  {author} {\bibfnamefont {M.}~\bibnamefont {Yosoi}}, \ and\ \bibinfo {author}
  {\bibfnamefont {J.}~\bibnamefont {Zenihiro}},\ }\href@noop {} {\bibfield
  {journal} {\bibinfo  {journal} {Phys. Lett. B}\ }\textbf {\bibinfo {volume}
  {744}},\ \bibinfo {pages} {7} (\bibinfo {year} {2015})}\BibitemShut {NoStop}%
\bibitem [{\citenamefont {Krishichayan}\ \emph {et~al.}(2015)\citenamefont
  {Krishichayan}, \citenamefont {Bhike}, \citenamefont {Tornow}, \citenamefont
  {Rusev}, \citenamefont {Tonchev}, \citenamefont {Tsoneva},\ and\
  \citenamefont {Lenske}}]{Kri15}%
  \BibitemOpen
  \bibfield  {author} {\bibinfo {author} {\bibnamefont {Krishichayan}},
  \bibinfo {author} {\bibfnamefont {M.}~\bibnamefont {Bhike}}, \bibinfo
  {author} {\bibfnamefont {W.}~\bibnamefont {Tornow}}, \bibinfo {author}
  {\bibfnamefont {G.}~\bibnamefont {Rusev}}, \bibinfo {author} {\bibfnamefont
  {A.~P.}\ \bibnamefont {Tonchev}}, \bibinfo {author} {\bibfnamefont
  {N.}~\bibnamefont {Tsoneva}}, \ and\ \bibinfo {author} {\bibfnamefont
  {H.}~\bibnamefont {Lenske}},\ }\href@noop {} {\bibfield  {journal} {\bibinfo
  {journal} {Phys. Rev. C}\ }\textbf {\bibinfo {volume} {91}},\ \bibinfo
  {pages} {044328} (\bibinfo {year} {2015})}\BibitemShut {NoStop}%
\bibitem [{\citenamefont {Tsoneva}\ \emph {et~al.}(2004)\citenamefont
  {Tsoneva}, \citenamefont {Lenske},\ and\ \citenamefont {Stoyanov}}]{Tso04}%
  \BibitemOpen
  \bibfield  {author} {\bibinfo {author} {\bibfnamefont {N.}~\bibnamefont
  {Tsoneva}}, \bibinfo {author} {\bibfnamefont {H.}~\bibnamefont {Lenske}}, \
  and\ \bibinfo {author} {\bibfnamefont {{\BIBCh}.}~\bibnamefont {Stoyanov}},\
  }\href@noop {} {\bibfield  {journal} {\bibinfo  {journal} {Phys. Lett. B}\
  }\textbf {\bibinfo {volume} {586}},\ \bibinfo {pages} {213} (\bibinfo {year}
  {2004})}\BibitemShut {NoStop}%
\bibitem [{\citenamefont {Tsoneva}\ and\ \citenamefont {Lenske}(2008)}]{Tso08}%
  \BibitemOpen
  \bibfield  {author} {\bibinfo {author} {\bibfnamefont {N.}~\bibnamefont
  {Tsoneva}}\ and\ \bibinfo {author} {\bibfnamefont {H.}~\bibnamefont
  {Lenske}},\ }\href@noop {} {\bibfield  {journal} {\bibinfo  {journal} {Phys.
  Rev. C}\ }\textbf {\bibinfo {volume} {77}},\ \bibinfo {pages} {024321}
  (\bibinfo {year} {2008})}\BibitemShut {NoStop}%
\bibitem [{\citenamefont {Klimkiewicz}\ \emph {et~al.}(2007)\citenamefont
  {Klimkiewicz}, \citenamefont {Paar}, \citenamefont {Adrich}, \citenamefont
  {Fallot}, \citenamefont {Boretzky}, \citenamefont {Aumann}, \citenamefont
  {Cortina-Gil}, \citenamefont {Pramanik}, \citenamefont {Elze}, \citenamefont
  {Emling}, \citenamefont {Geissel}, \citenamefont {Hellstr\"om}, \citenamefont
  {Jones}, \citenamefont {Kratz}, \citenamefont {Kulessa}, \citenamefont
  {Nociforo}, \citenamefont {Palit}, \citenamefont {Simon}, \citenamefont
  {Sur\'owka}, \citenamefont {S\"ummerer}, \citenamefont {Vretenar},\ and\
  \citenamefont {Walu\ifmmode~\acute{s}\else \'{s}\fi{}}}]{Klim07}%
  \BibitemOpen
  \bibfield  {author} {\bibinfo {author} {\bibfnamefont {A.}~\bibnamefont
  {Klimkiewicz}}, \bibinfo {author} {\bibfnamefont {N.}~\bibnamefont {Paar}},
  \bibinfo {author} {\bibfnamefont {P.}~\bibnamefont {Adrich}}, \bibinfo
  {author} {\bibfnamefont {M.}~\bibnamefont {Fallot}}, \bibinfo {author}
  {\bibfnamefont {K.}~\bibnamefont {Boretzky}}, \bibinfo {author}
  {\bibfnamefont {T.}~\bibnamefont {Aumann}}, \bibinfo {author} {\bibfnamefont
  {D.}~\bibnamefont {Cortina-Gil}}, \bibinfo {author} {\bibfnamefont {U.~D.}\
  \bibnamefont {Pramanik}}, \bibinfo {author} {\bibfnamefont {{\BIBTh}.~W.}\
  \bibnamefont {Elze}}, \bibinfo {author} {\bibfnamefont {H.}~\bibnamefont
  {Emling}}, \bibinfo {author} {\bibfnamefont {H.}~\bibnamefont {Geissel}},
  \bibinfo {author} {\bibfnamefont {M.}~\bibnamefont {Hellstr\"om}}, \bibinfo
  {author} {\bibfnamefont {K.~L.}\ \bibnamefont {Jones}}, \bibinfo {author}
  {\bibfnamefont {J.~V.}\ \bibnamefont {Kratz}}, \bibinfo {author}
  {\bibfnamefont {R.}~\bibnamefont {Kulessa}}, \bibinfo {author} {\bibfnamefont
  {C.}~\bibnamefont {Nociforo}}, \bibinfo {author} {\bibfnamefont
  {R.}~\bibnamefont {Palit}}, \bibinfo {author} {\bibfnamefont
  {H.}~\bibnamefont {Simon}}, \bibinfo {author} {\bibfnamefont
  {G.}~\bibnamefont {Sur\'owka}}, \bibinfo {author} {\bibfnamefont
  {K.}~\bibnamefont {S\"ummerer}}, \bibinfo {author} {\bibfnamefont
  {D.}~\bibnamefont {Vretenar}}, \ and\ \bibinfo {author} {\bibfnamefont
  {W.}~\bibnamefont {Walu\ifmmode~\acute{s}\else \'{s}\fi{}}},\ }\href@noop {}
  {\bibfield  {journal} {\bibinfo  {journal} {Phys. Rev. C}\ }\textbf {\bibinfo
  {volume} {76}},\ \bibinfo {pages} {051603(R)} (\bibinfo {year}
  {2007})}\BibitemShut {NoStop}%
\bibitem [{\citenamefont {Savran}\ \emph {et~al.}(2006)\citenamefont {Savran},
  \citenamefont {Babilon}, \citenamefont {van~den Berg}, \citenamefont
  {Harakeh}, \citenamefont {Hasper}, \citenamefont {Matic}, \citenamefont
  {W\"ortche},\ and\ \citenamefont {Zilges}}]{Savr06}%
  \BibitemOpen
  \bibfield  {author} {\bibinfo {author} {\bibfnamefont {D.}~\bibnamefont
  {Savran}}, \bibinfo {author} {\bibfnamefont {M.}~\bibnamefont {Babilon}},
  \bibinfo {author} {\bibfnamefont {A.~M.}\ \bibnamefont {van~den Berg}},
  \bibinfo {author} {\bibfnamefont {M.~N.}\ \bibnamefont {Harakeh}}, \bibinfo
  {author} {\bibfnamefont {J.}~\bibnamefont {Hasper}}, \bibinfo {author}
  {\bibfnamefont {A.}~\bibnamefont {Matic}}, \bibinfo {author} {\bibfnamefont
  {H.~J.}\ \bibnamefont {W\"ortche}}, \ and\ \bibinfo {author} {\bibfnamefont
  {A.}~\bibnamefont {Zilges}},\ }\href@noop {} {\bibfield  {journal} {\bibinfo
  {journal} {Phys. Rev. Lett.}\ }\textbf {\bibinfo {volume} {97}},\ \bibinfo
  {pages} {172502} (\bibinfo {year} {2006})}\BibitemShut {NoStop}%
\bibitem [{\citenamefont {Papakonstantinou}\ \emph {et~al.}(2014)\citenamefont
  {Papakonstantinou}, \citenamefont {Hergert}, \citenamefont {Ponomarev},\ and\
  \citenamefont {Roth}}]{Papa14}%
  \BibitemOpen
  \bibfield  {author} {\bibinfo {author} {\bibfnamefont {P.}~\bibnamefont
  {Papakonstantinou}}, \bibinfo {author} {\bibfnamefont {H.}~\bibnamefont
  {Hergert}}, \bibinfo {author} {\bibfnamefont {V.~{\BIBYu}.}\ \bibnamefont
  {Ponomarev}}, \ and\ \bibinfo {author} {\bibfnamefont {R.}~\bibnamefont
  {Roth}},\ }\href@noop {} {\bibfield  {journal} {\bibinfo  {journal} {Phys.
  Rev. C}\ }\textbf {\bibinfo {volume} {89}},\ \bibinfo {pages} {034306}
  (\bibinfo {year} {2014})}\BibitemShut {NoStop}%
\bibitem [{\citenamefont {Repko}\ \emph {et~al.}(2013)\citenamefont {Repko},
  \citenamefont {Reinhard}, \citenamefont {Nesterenko},\ and\ \citenamefont
  {Kvasil}}]{Repk13}%
  \BibitemOpen
  \bibfield  {author} {\bibinfo {author} {\bibfnamefont {A.}~\bibnamefont
  {Repko}}, \bibinfo {author} {\bibfnamefont {P.-G.}\ \bibnamefont {Reinhard}},
  \bibinfo {author} {\bibfnamefont {V.~O.}\ \bibnamefont {Nesterenko}}, \ and\
  \bibinfo {author} {\bibfnamefont {J.}~\bibnamefont {Kvasil}},\ }\href@noop {}
  {\bibfield  {journal} {\bibinfo  {journal} {Phys. Rev. C}\ }\textbf {\bibinfo
  {volume} {87}},\ \bibinfo {pages} {024305} (\bibinfo {year}
  {2013})}\BibitemShut {NoStop}%
\bibitem [{\citenamefont {Vretenar}\ \emph {et~al.}(2012)\citenamefont
  {Vretenar}, \citenamefont {Niu}, \citenamefont {Paar},\ and\ \citenamefont
  {Meng}}]{Vret12}%
  \BibitemOpen
  \bibfield  {author} {\bibinfo {author} {\bibfnamefont {D.}~\bibnamefont
  {Vretenar}}, \bibinfo {author} {\bibfnamefont {Y.~F.}\ \bibnamefont {Niu}},
  \bibinfo {author} {\bibfnamefont {N.}~\bibnamefont {Paar}}, \ and\ \bibinfo
  {author} {\bibfnamefont {J.}~\bibnamefont {Meng}},\ }\href@noop {} {\bibfield
   {journal} {\bibinfo  {journal} {Phys. Rev. C}\ }\textbf {\bibinfo {volume}
  {85}},\ \bibinfo {pages} {044317} (\bibinfo {year} {2012})}\BibitemShut
  {NoStop}%
\bibitem [{\citenamefont {Lanza}\ \emph {et~al.}(2014)\citenamefont {Lanza},
  \citenamefont {Vitturi}, \citenamefont {Litvinova},\ and\ \citenamefont
  {Savran}}]{Lanz14}%
  \BibitemOpen
  \bibfield  {author} {\bibinfo {author} {\bibfnamefont {E.~G.}\ \bibnamefont
  {Lanza}}, \bibinfo {author} {\bibfnamefont {A.}~\bibnamefont {Vitturi}},
  \bibinfo {author} {\bibfnamefont {E.}~\bibnamefont {Litvinova}}, \ and\
  \bibinfo {author} {\bibfnamefont {D.}~\bibnamefont {Savran}},\ }\href@noop {}
  {\bibfield  {journal} {\bibinfo  {journal} {Phys. Rev. C}\ }\textbf {\bibinfo
  {volume} {89}},\ \bibinfo {pages} {041601} (\bibinfo {year}
  {2014})}\BibitemShut {NoStop}%
\bibitem [{\citenamefont {Roca-Maza}\ \emph {et~al.}(2012)\citenamefont
  {Roca-Maza}, \citenamefont {Pozzi}, \citenamefont {Brenna}, \citenamefont
  {Mizuyama},\ and\ \citenamefont {Col\`o}}]{Roca12}%
  \BibitemOpen
  \bibfield  {author} {\bibinfo {author} {\bibfnamefont {X.}~\bibnamefont
  {Roca-Maza}}, \bibinfo {author} {\bibfnamefont {G.}~\bibnamefont {Pozzi}},
  \bibinfo {author} {\bibfnamefont {M.}~\bibnamefont {Brenna}}, \bibinfo
  {author} {\bibfnamefont {K.}~\bibnamefont {Mizuyama}}, \ and\ \bibinfo
  {author} {\bibfnamefont {G.}~\bibnamefont {Col\`o}},\ }\href@noop {}
  {\bibfield  {journal} {\bibinfo  {journal} {Phys. Rev. C}\ }\textbf {\bibinfo
  {volume} {85}},\ \bibinfo {pages} {024601} (\bibinfo {year}
  {2012})}\BibitemShut {NoStop}%
\bibitem [{\citenamefont {Piekarewicz}(2006)}]{Pie06}%
  \BibitemOpen
  \bibfield  {author} {\bibinfo {author} {\bibfnamefont {J.}~\bibnamefont
  {Piekarewicz}},\ }\href@noop {} {\bibfield  {journal} {\bibinfo  {journal}
  {Phys. Rev. C}\ }\textbf {\bibinfo {volume} {73}},\ \bibinfo {pages} {044325}
  (\bibinfo {year} {2006})}\BibitemShut {NoStop}%
\bibitem [{\citenamefont {Reinhard}\ and\ \citenamefont
  {Nazarewicz}(2013)}]{Rei13}%
  \BibitemOpen
  \bibfield  {author} {\bibinfo {author} {\bibfnamefont {P.-G.}\ \bibnamefont
  {Reinhard}}\ and\ \bibinfo {author} {\bibfnamefont {W.}~\bibnamefont
  {Nazarewicz}},\ }\href@noop {} {\bibfield  {journal} {\bibinfo  {journal}
  {Phys. Rev. C}\ }\textbf {\bibinfo {volume} {87}},\ \bibinfo {pages} {014324}
  (\bibinfo {year} {2013})}\BibitemShut {NoStop}%
\bibitem [{\citenamefont {Tsoneva}\ \emph {et~al.}(2015)\citenamefont
  {Tsoneva}, \citenamefont {Goriely}, \citenamefont {Lenske},\ and\
  \citenamefont {Schwengner}}]{Tson15}%
  \BibitemOpen
  \bibfield  {author} {\bibinfo {author} {\bibfnamefont {N.}~\bibnamefont
  {Tsoneva}}, \bibinfo {author} {\bibfnamefont {S.}~\bibnamefont {Goriely}},
  \bibinfo {author} {\bibfnamefont {H.}~\bibnamefont {Lenske}}, \ and\ \bibinfo
  {author} {\bibfnamefont {R.}~\bibnamefont {Schwengner}},\ }\href@noop {}
  {\bibfield  {journal} {\bibinfo  {journal} {Phys. Rev. C}\ }\textbf {\bibinfo
  {volume} {91}},\ \bibinfo {pages} {044318} (\bibinfo {year}
  {2015})}\BibitemShut {NoStop}%
\bibitem [{\citenamefont {Iachello}(1985)}]{Iach85}%
  \BibitemOpen
  \bibfield  {author} {\bibinfo {author} {\bibfnamefont {F.}~\bibnamefont
  {Iachello}},\ }\href@noop {} {\bibfield  {journal} {\bibinfo  {journal}
  {Phys. Lett. B}\ }\textbf {\bibinfo {volume} {160}},\ \bibinfo {pages} {1}
  (\bibinfo {year} {1985})}\BibitemShut {NoStop}%
\bibitem [{\citenamefont {Spieker}\ \emph {et~al.}(2015)\citenamefont
  {Spieker}, \citenamefont {Pascu}, \citenamefont {Zilges},\ and\ \citenamefont
  {Iachello}}]{Spie15}%
  \BibitemOpen
  \bibfield  {author} {\bibinfo {author} {\bibfnamefont {M.}~\bibnamefont
  {Spieker}}, \bibinfo {author} {\bibfnamefont {S.}~\bibnamefont {Pascu}},
  \bibinfo {author} {\bibfnamefont {A.}~\bibnamefont {Zilges}}, \ and\ \bibinfo
  {author} {\bibfnamefont {F.}~\bibnamefont {Iachello}},\ }\href@noop {}
  {\bibfield  {journal} {\bibinfo  {journal} {Phys. Rev. Lett.}\ }\textbf
  {\bibinfo {volume} {114}},\ \bibinfo {pages} {192504} (\bibinfo {year}
  {2015})}\BibitemShut {NoStop}%
\bibitem [{\citenamefont {Ponomarev}\ \emph {et~al.}(1998)\citenamefont
  {Ponomarev}, \citenamefont {Stoyanov}, \citenamefont {Tsoneva},\ and\
  \citenamefont {Grinberg}}]{Pon98}%
  \BibitemOpen
  \bibfield  {author} {\bibinfo {author} {\bibfnamefont {V.~{\BIBYu}.}\
  \bibnamefont {Ponomarev}}, \bibinfo {author} {\bibfnamefont
  {{\BIBCh}.}~\bibnamefont {Stoyanov}}, \bibinfo {author} {\bibfnamefont
  {N.}~\bibnamefont {Tsoneva}}, \ and\ \bibinfo {author} {\bibfnamefont
  {M.}~\bibnamefont {Grinberg}},\ }\href@noop {} {\bibfield  {journal}
  {\bibinfo  {journal} {Nucl. Phys. A}\ }\textbf {\bibinfo {volume} {635}},\
  \bibinfo {pages} {470} (\bibinfo {year} {1998})}\BibitemShut {NoStop}%
\bibitem [{\citenamefont {Arsenyev}\ \emph {et~al.}(2012)\citenamefont
  {Arsenyev}, \citenamefont {Severyukhin}, \citenamefont {Voronov},\ and\
  \citenamefont {Van~Giai}}]{Arse12}%
  \BibitemOpen
  \bibfield  {author} {\bibinfo {author} {\bibfnamefont {N.~N.}\ \bibnamefont
  {Arsenyev}}, \bibinfo {author} {\bibfnamefont {A.~P.}\ \bibnamefont
  {Severyukhin}}, \bibinfo {author} {\bibfnamefont {V.~V.}\ \bibnamefont
  {Voronov}}, \ and\ \bibinfo {author} {\bibfnamefont {N.}~\bibnamefont
  {Van~Giai}},\ }\href@noop {} {\bibfield  {journal} {\bibinfo  {journal} {EPJ
  Conf.}\ }\textbf {\bibinfo {volume} {38}},\ \bibinfo {pages} {17002}
  (\bibinfo {year} {2012})}\BibitemShut {NoStop}%
\bibitem [{\citenamefont {Litvinova}\ \emph {et~al.}(2013)\citenamefont
  {Litvinova}, \citenamefont {Ring},\ and\ \citenamefont {Tselyaev}}]{Litv13}%
  \BibitemOpen
  \bibfield  {author} {\bibinfo {author} {\bibfnamefont {E.}~\bibnamefont
  {Litvinova}}, \bibinfo {author} {\bibfnamefont {P.}~\bibnamefont {Ring}}, \
  and\ \bibinfo {author} {\bibfnamefont {V.}~\bibnamefont {Tselyaev}},\
  }\href@noop {} {\bibfield  {journal} {\bibinfo  {journal} {Phys. Rev. C}\
  }\textbf {\bibinfo {volume} {88}},\ \bibinfo {pages} {044320} (\bibinfo
  {year} {2013})}\BibitemShut {NoStop}%
\bibitem [{\citenamefont {Savran}\ \emph {et~al.}()\citenamefont {Savran},
  \citenamefont {Elvers}, \citenamefont {Endres}, \citenamefont {Fritzsche},
  \citenamefont {L\"oher}, \citenamefont {Pietralla}, \citenamefont
  {Ponomarev}, \citenamefont {Romig}, \citenamefont {Schnorrenberger},
  \citenamefont {Sonnabend},\ and\ \citenamefont {Zilges}}]{Savr11}%
  \BibitemOpen
  \bibfield  {author} {\bibinfo {author} {\bibfnamefont {D.}~\bibnamefont
  {Savran}}, \bibinfo {author} {\bibfnamefont {M.}~\bibnamefont {Elvers}},
  \bibinfo {author} {\bibfnamefont {J.}~\bibnamefont {Endres}}, \bibinfo
  {author} {\bibfnamefont {M.}~\bibnamefont {Fritzsche}}, \bibinfo {author}
  {\bibfnamefont {B.}~\bibnamefont {L\"oher}}, \bibinfo {author} {\bibfnamefont
  {N.}~\bibnamefont {Pietralla}}, \bibinfo {author} {\bibfnamefont
  {V.~{\BIBYu}.}\ \bibnamefont {Ponomarev}}, \bibinfo {author} {\bibfnamefont
  {C.}~\bibnamefont {Romig}}, \bibinfo {author} {\bibfnamefont
  {L.}~\bibnamefont {Schnorrenberger}}, \bibinfo {author} {\bibfnamefont
  {K.}~\bibnamefont {Sonnabend}}, \ and\ \bibinfo {author} {\bibfnamefont
  {A.}~\bibnamefont {Zilges}},\ }\href@noop {} {\bibfield  {journal} {\bibinfo
  {journal} {Phys. Rev. C}\ }\textbf {\bibinfo {volume} {84}},\ \bibinfo
  {pages} {024326}}\BibitemShut {NoStop}%
\bibitem [{\citenamefont {Goriely}(1998)}]{Gori98}%
  \BibitemOpen
  \bibfield  {author} {\bibinfo {author} {\bibfnamefont {S.}~\bibnamefont
  {Goriely}},\ }\href@noop {} {\bibfield  {journal} {\bibinfo  {journal} {Phys.
  Lett. B}\ }\textbf {\bibinfo {volume} {436}},\ \bibinfo {pages} {10}
  (\bibinfo {year} {1998})}\BibitemShut {NoStop}%
\bibitem [{\citenamefont {Bohr}\ and\ \citenamefont
  {Mottelson}(1975)}]{bohrmottelson}%
  \BibitemOpen
  \bibfield  {author} {\bibinfo {author} {\bibfnamefont {A.}~\bibnamefont
  {Bohr}}\ and\ \bibinfo {author} {\bibfnamefont {B.~R.}\ \bibnamefont
  {Mottelson}},\ }\href@noop {} {\emph {\bibinfo {title} {{Nuclear structure.
  Volume II: Nuclear deformations}}}}\ (\bibinfo  {publisher} {W. A. Benjamin,
  Inc.},\ \bibinfo {address} {Reading, Massachussetts},\ \bibinfo {year}
  {1975})\BibitemShut {NoStop}%
\bibitem [{\citenamefont {Aprahamian}\ \emph {et~al.}(1984)\citenamefont
  {Aprahamian}, \citenamefont {Brenner}, \citenamefont {Casten}, \citenamefont
  {Gill}, \citenamefont {Piotrowski},\ and\ \citenamefont {Heyde}}]{Apra84}%
  \BibitemOpen
  \bibfield  {author} {\bibinfo {author} {\bibfnamefont {A.}~\bibnamefont
  {Aprahamian}}, \bibinfo {author} {\bibfnamefont {D.~S.}\ \bibnamefont
  {Brenner}}, \bibinfo {author} {\bibfnamefont {R.~F.}\ \bibnamefont {Casten}},
  \bibinfo {author} {\bibfnamefont {R.~L.}\ \bibnamefont {Gill}}, \bibinfo
  {author} {\bibfnamefont {A.}~\bibnamefont {Piotrowski}}, \ and\ \bibinfo
  {author} {\bibfnamefont {K.}~\bibnamefont {Heyde}},\ }\href@noop {}
  {\bibfield  {journal} {\bibinfo  {journal} {Phys. Lett. B}\ }\textbf
  {\bibinfo {volume} {140}},\ \bibinfo {pages} {22} (\bibinfo {year}
  {1984})}\BibitemShut {NoStop}%
\bibitem [{\citenamefont {Yeh}\ \emph {et~al.}(1996)\citenamefont {Yeh},
  \citenamefont {Garrett}, \citenamefont {McGrath}, \citenamefont {Yates},\
  and\ \citenamefont {Belgya}}]{Yeh96}%
  \BibitemOpen
  \bibfield  {author} {\bibinfo {author} {\bibfnamefont {M.}~\bibnamefont
  {Yeh}}, \bibinfo {author} {\bibfnamefont {P.~E.}\ \bibnamefont {Garrett}},
  \bibinfo {author} {\bibfnamefont {C.~A.}\ \bibnamefont {McGrath}}, \bibinfo
  {author} {\bibfnamefont {S.~W.}\ \bibnamefont {Yates}}, \ and\ \bibinfo
  {author} {\bibfnamefont {T.}~\bibnamefont {Belgya}},\ }\href@noop {}
  {\bibfield  {journal} {\bibinfo  {journal} {Phys. Rev. Lett.}\ }\textbf
  {\bibinfo {volume} {76}},\ \bibinfo {pages} {1208} (\bibinfo {year}
  {1996})}\BibitemShut {NoStop}%
\bibitem [{\citenamefont {Mukhopadhyay}\ \emph {et~al.}(2008)\citenamefont
  {Mukhopadhyay}, \citenamefont {Scheck}, \citenamefont {Crider}, \citenamefont
  {Choudry}, \citenamefont {Elhami}, \citenamefont {Peters}, \citenamefont
  {McEllistrem}, \citenamefont {Orce},\ and\ \citenamefont {Yates}}]{Mukh08}%
  \BibitemOpen
  \bibfield  {author} {\bibinfo {author} {\bibfnamefont {S.}~\bibnamefont
  {Mukhopadhyay}}, \bibinfo {author} {\bibfnamefont {M.}~\bibnamefont
  {Scheck}}, \bibinfo {author} {\bibfnamefont {B.}~\bibnamefont {Crider}},
  \bibinfo {author} {\bibfnamefont {S.~N.}\ \bibnamefont {Choudry}}, \bibinfo
  {author} {\bibfnamefont {E.}~\bibnamefont {Elhami}}, \bibinfo {author}
  {\bibfnamefont {E.}~\bibnamefont {Peters}}, \bibinfo {author} {\bibfnamefont
  {M.~T.}\ \bibnamefont {McEllistrem}}, \bibinfo {author} {\bibfnamefont
  {J.~N.}\ \bibnamefont {Orce}}, \ and\ \bibinfo {author} {\bibfnamefont
  {S.~W.}\ \bibnamefont {Yates}},\ }\href@noop {} {\bibfield  {journal}
  {\bibinfo  {journal} {Phys. Rev. C}\ }\textbf {\bibinfo {volume} {78}},\
  \bibinfo {pages} {034317} (\bibinfo {year} {2008})}\BibitemShut {NoStop}%
\bibitem [{\citenamefont {Grinberg}\ and\ \citenamefont
  {Stoyanov}(1994)}]{Grin94}%
  \BibitemOpen
  \bibfield  {author} {\bibinfo {author} {\bibfnamefont {M.}~\bibnamefont
  {Grinberg}}\ and\ \bibinfo {author} {\bibfnamefont {{\BIBCh}.}~\bibnamefont
  {Stoyanov}},\ }\href@noop {} {\bibfield  {journal} {\bibinfo  {journal}
  {Nucl. Phys. A}\ }\textbf {\bibinfo {volume} {573}},\ \bibinfo {pages} {231}
  (\bibinfo {year} {1994})}\BibitemShut {NoStop}%
\bibitem [{\citenamefont {Soloviev}(1992)}]{QPM}%
  \BibitemOpen
  \bibfield  {author} {\bibinfo {author} {\bibfnamefont {V.~G.}\ \bibnamefont
  {Soloviev}},\ }\href@noop {} {\emph {\bibinfo {title} {{Theory of Atomic
  Nuclei: Quasiparticles and Phonons}}}}\ (\bibinfo  {publisher} {Institute of
  Physics Publishing, Bristol},\ \bibinfo {year} {1992})\BibitemShut {NoStop}%
\bibitem [{\citenamefont {Kern}\ \emph {et~al.}(1995)\citenamefont {Kern},
  \citenamefont {Garrett}, \citenamefont {Jolie},\ and\ \citenamefont
  {Lehmann}}]{Kern95}%
  \BibitemOpen
  \bibfield  {author} {\bibinfo {author} {\bibfnamefont {J.}~\bibnamefont
  {Kern}}, \bibinfo {author} {\bibfnamefont {P.~E.}\ \bibnamefont {Garrett}},
  \bibinfo {author} {\bibfnamefont {J.}~\bibnamefont {Jolie}}, \ and\ \bibinfo
  {author} {\bibfnamefont {H.}~\bibnamefont {Lehmann}},\ }\href@noop {}
  {\bibfield  {journal} {\bibinfo  {journal} {Nucl. Phys. A}\ }\textbf
  {\bibinfo {volume} {593}},\ \bibinfo {pages} {21} (\bibinfo {year}
  {1995})}\BibitemShut {NoStop}%
\bibitem [{\citenamefont {Smirnova}\ \emph {et~al.}(2000)\citenamefont
  {Smirnova}, \citenamefont {Pietralla}, \citenamefont {Mizusaki},\ and\
  \citenamefont {van Isacker}}]{Smir00}%
  \BibitemOpen
  \bibfield  {author} {\bibinfo {author} {\bibfnamefont {N.~A.}\ \bibnamefont
  {Smirnova}}, \bibinfo {author} {\bibfnamefont {N.}~\bibnamefont {Pietralla}},
  \bibinfo {author} {\bibfnamefont {T.}~\bibnamefont {Mizusaki}}, \ and\
  \bibinfo {author} {\bibfnamefont {P.}~\bibnamefont {van Isacker}},\
  }\href@noop {} {\bibfield  {journal} {\bibinfo  {journal} {Nucl. Phys. A}\
  }\textbf {\bibinfo {volume} {678}},\ \bibinfo {pages} {235} (\bibinfo {year}
  {2000})}\BibitemShut {NoStop}%
\bibitem [{\citenamefont {Iachello}\ and\ \citenamefont {Arima}(1987)}]{IBM}%
  \BibitemOpen
  \bibfield  {author} {\bibinfo {author} {\bibfnamefont {F.}~\bibnamefont
  {Iachello}}\ and\ \bibinfo {author} {\bibfnamefont {A.}~\bibnamefont
  {Arima}},\ }\href@noop {} {\emph {\bibinfo {title} {{The Interacting Boson
  Model}}}}\ (\bibinfo  {publisher} {Cambridge University Press, Cambridge},\
  \bibinfo {year} {1987})\BibitemShut {NoStop}%
\bibitem [{\citenamefont {Kneissl}\ \emph {et~al.}(1996)\citenamefont
  {Kneissl}, \citenamefont {Pitz},\ and\ \citenamefont {Zilges}}]{Knei96}%
  \BibitemOpen
  \bibfield  {author} {\bibinfo {author} {\bibfnamefont {U.}~\bibnamefont
  {Kneissl}}, \bibinfo {author} {\bibfnamefont {H.~H.}\ \bibnamefont {Pitz}}, \
  and\ \bibinfo {author} {\bibfnamefont {A.}~\bibnamefont {Zilges}},\
  }\href@noop {} {\bibfield  {journal} {\bibinfo  {journal} {Prog. Part. Nucl.
  Phys.}\ }\textbf {\bibinfo {volume} {37}},\ \bibinfo {pages} {349} (\bibinfo
  {year} {1996})}\BibitemShut {NoStop}%
\bibitem [{\citenamefont {Pysmenetska}\ \emph {et~al.}(2006)\citenamefont
  {Pysmenetska}, \citenamefont {Walter}, \citenamefont {Enders}, \citenamefont
  {Garrel}, \citenamefont {Karg}, \citenamefont {Kneissl}, \citenamefont
  {Kohstall}, \citenamefont {von Neumann-Cosel}, \citenamefont {Pitz},
  \citenamefont {Ponomarev}, \citenamefont {Scheck}, \citenamefont {Stedile},\
  and\ \citenamefont {Volz}}]{Pysm06}%
  \BibitemOpen
  \bibfield  {author} {\bibinfo {author} {\bibfnamefont {I.}~\bibnamefont
  {Pysmenetska}}, \bibinfo {author} {\bibfnamefont {S.}~\bibnamefont {Walter}},
  \bibinfo {author} {\bibfnamefont {J.}~\bibnamefont {Enders}}, \bibinfo
  {author} {\bibfnamefont {H.~v.}\ \bibnamefont {Garrel}}, \bibinfo {author}
  {\bibfnamefont {O.}~\bibnamefont {Karg}}, \bibinfo {author} {\bibfnamefont
  {U.}~\bibnamefont {Kneissl}}, \bibinfo {author} {\bibfnamefont
  {C.}~\bibnamefont {Kohstall}}, \bibinfo {author} {\bibfnamefont
  {P.}~\bibnamefont {von Neumann-Cosel}}, \bibinfo {author} {\bibfnamefont
  {H.~H.}\ \bibnamefont {Pitz}}, \bibinfo {author} {\bibfnamefont
  {V.~{\BIBYu}.}\ \bibnamefont {Ponomarev}}, \bibinfo {author} {\bibfnamefont
  {M.}~\bibnamefont {Scheck}}, \bibinfo {author} {\bibfnamefont
  {F.}~\bibnamefont {Stedile}}, \ and\ \bibinfo {author} {\bibfnamefont
  {S.}~\bibnamefont {Volz}},\ }\href@noop {} {\bibfield  {journal} {\bibinfo
  {journal} {Phys. Rev. C}\ }\textbf {\bibinfo {volume} {73}},\ \bibinfo
  {pages} {017302} (\bibinfo {year} {2006})}\BibitemShut {NoStop}%
\bibitem [{\citenamefont {Herzberg}\ \emph {et~al.}(1995)\citenamefont
  {Herzberg}, \citenamefont {Bauske}, \citenamefont {von Brentano},
  \citenamefont {Eckert}, \citenamefont {Fischer}, \citenamefont {Geiger},
  \citenamefont {Kneissl}, \citenamefont {Margraf}, \citenamefont {Maser},
  \citenamefont {Pietralla}, \citenamefont {Pitz},\ and\ \citenamefont
  {Zilges}}]{Herz95}%
  \BibitemOpen
  \bibfield  {author} {\bibinfo {author} {\bibfnamefont {R.-D.}\ \bibnamefont
  {Herzberg}}, \bibinfo {author} {\bibfnamefont {I.}~\bibnamefont {Bauske}},
  \bibinfo {author} {\bibfnamefont {P.}~\bibnamefont {von Brentano}}, \bibinfo
  {author} {\bibfnamefont {{\BIBTh}.}~\bibnamefont {Eckert}}, \bibinfo {author}
  {\bibfnamefont {R.}~\bibnamefont {Fischer}}, \bibinfo {author} {\bibfnamefont
  {W.}~\bibnamefont {Geiger}}, \bibinfo {author} {\bibfnamefont
  {U.}~\bibnamefont {Kneissl}}, \bibinfo {author} {\bibfnamefont
  {J.}~\bibnamefont {Margraf}}, \bibinfo {author} {\bibfnamefont
  {H.}~\bibnamefont {Maser}}, \bibinfo {author} {\bibfnamefont
  {N.}~\bibnamefont {Pietralla}}, \bibinfo {author} {\bibfnamefont {H.~H.}\
  \bibnamefont {Pitz}}, \ and\ \bibinfo {author} {\bibfnamefont
  {A.}~\bibnamefont {Zilges}},\ }\href@noop {} {\bibfield  {journal} {\bibinfo
  {journal} {Nucl. Phys. A}\ }\textbf {\bibinfo {volume} {592}},\ \bibinfo
  {pages} {211} (\bibinfo {year} {1995})}\BibitemShut {NoStop}%
\bibitem [{\citenamefont {Andrejtscheff}\ \emph {et~al.}(2001)\citenamefont
  {Andrejtscheff}, \citenamefont {Kohstall}, \citenamefont {von Brentano},
  \citenamefont {Fransen}, \citenamefont {Kneissl}, \citenamefont {Pietralla},\
  and\ \citenamefont {Pitz}}]{Andr01}%
  \BibitemOpen
  \bibfield  {author} {\bibinfo {author} {\bibfnamefont {W.}~\bibnamefont
  {Andrejtscheff}}, \bibinfo {author} {\bibfnamefont {C.}~\bibnamefont
  {Kohstall}}, \bibinfo {author} {\bibfnamefont {P.}~\bibnamefont {von
  Brentano}}, \bibinfo {author} {\bibfnamefont {C.}~\bibnamefont {Fransen}},
  \bibinfo {author} {\bibfnamefont {U.}~\bibnamefont {Kneissl}}, \bibinfo
  {author} {\bibfnamefont {N.}~\bibnamefont {Pietralla}}, \ and\ \bibinfo
  {author} {\bibfnamefont {H.~H.}\ \bibnamefont {Pitz}},\ }\href@noop {}
  {\bibfield  {journal} {\bibinfo  {journal} {Phys. Lett. B}\ }\textbf
  {\bibinfo {volume} {506}},\ \bibinfo {pages} {239} (\bibinfo {year}
  {2001})}\BibitemShut {NoStop}%
\bibitem [{\citenamefont {Hennig}\ \emph {et~al.}(2015)\citenamefont {Hennig},
  \citenamefont {Derya}, \citenamefont {Mineva}, \citenamefont {Petkov},
  \citenamefont {Pickstone}, \citenamefont {Spieker},\ and\ \citenamefont
  {Zilges}}]{Henn15}%
  \BibitemOpen
  \bibfield  {author} {\bibinfo {author} {\bibfnamefont {A.}~\bibnamefont
  {Hennig}}, \bibinfo {author} {\bibfnamefont {V.}~\bibnamefont {Derya}},
  \bibinfo {author} {\bibfnamefont {M.~N.}\ \bibnamefont {Mineva}}, \bibinfo
  {author} {\bibfnamefont {P.}~\bibnamefont {Petkov}}, \bibinfo {author}
  {\bibfnamefont {S.~G.}\ \bibnamefont {Pickstone}}, \bibinfo {author}
  {\bibfnamefont {M.}~\bibnamefont {Spieker}}, \ and\ \bibinfo {author}
  {\bibfnamefont {A.}~\bibnamefont {Zilges}},\ }\href@noop {} {\bibfield
  {journal} {\bibinfo  {journal} {Nucl. Instr. and Meth. A}\ }\textbf {\bibinfo
  {volume} {794}},\ \bibinfo {pages} {171} (\bibinfo {year}
  {2015})}\BibitemShut {NoStop}%
\bibitem [{\citenamefont {Belgya}\ \emph {et~al.}(1993)\citenamefont {Belgya},
  \citenamefont {Seckel}, \citenamefont {Johnson}, \citenamefont {Baum},
  \citenamefont {DiPrete}, \citenamefont {Wang},\ and\ \citenamefont
  {Yates}}]{Belg93}%
  \BibitemOpen
  \bibfield  {author} {\bibinfo {author} {\bibfnamefont {T.}~\bibnamefont
  {Belgya}}, \bibinfo {author} {\bibfnamefont {D.}~\bibnamefont {Seckel}},
  \bibinfo {author} {\bibfnamefont {E.~L.}\ \bibnamefont {Johnson}}, \bibinfo
  {author} {\bibfnamefont {E.~M.}\ \bibnamefont {Baum}}, \bibinfo {author}
  {\bibfnamefont {D.~P.}\ \bibnamefont {DiPrete}}, \bibinfo {author}
  {\bibfnamefont {D.}~\bibnamefont {Wang}}, \ and\ \bibinfo {author}
  {\bibfnamefont {S.~W.}\ \bibnamefont {Yates}},\ }\href@noop {} {\bibfield
  {journal} {\bibinfo  {journal} {Phys. Rev. C}\ }\textbf {\bibinfo {volume}
  {47}},\ \bibinfo {pages} {392} (\bibinfo {year} {1993})}\BibitemShut
  {NoStop}%
\bibitem [{\citenamefont {Belgya}\ \emph {et~al.}(1996)\citenamefont {Belgya},
  \citenamefont {Moln\'{a}r},\ and\ \citenamefont {Yates}}]{Belg96}%
  \BibitemOpen
  \bibfield  {author} {\bibinfo {author} {\bibfnamefont {T.}~\bibnamefont
  {Belgya}}, \bibinfo {author} {\bibfnamefont {G.}~\bibnamefont {Moln\'{a}r}},
  \ and\ \bibinfo {author} {\bibfnamefont {S.~W.}\ \bibnamefont {Yates}},\
  }\href@noop {} {\bibfield  {journal} {\bibinfo  {journal} {Nucl. Phys. A}\
  }\textbf {\bibinfo {volume} {607}},\ \bibinfo {pages} {43} (\bibinfo {year}
  {1996})}\BibitemShut {NoStop}%
\bibitem [{\citenamefont {Romig}\ \emph {et~al.}(2015)\citenamefont {Romig},
  \citenamefont {Savran}, \citenamefont {Beller}, \citenamefont {Birkhan},
  \citenamefont {Endres}, \citenamefont {Fritzsche}, \citenamefont {Glorius},
  \citenamefont {Isaak}, \citenamefont {Pietralla}, \citenamefont {Scheck},
  \citenamefont {Schnorrenberger}, \citenamefont {Sonnabend},\ and\
  \citenamefont {Zweidinger}}]{Romi15}%
  \BibitemOpen
  \bibfield  {author} {\bibinfo {author} {\bibfnamefont {C.}~\bibnamefont
  {Romig}}, \bibinfo {author} {\bibfnamefont {D.}~\bibnamefont {Savran}},
  \bibinfo {author} {\bibfnamefont {J.}~\bibnamefont {Beller}}, \bibinfo
  {author} {\bibfnamefont {J.}~\bibnamefont {Birkhan}}, \bibinfo {author}
  {\bibfnamefont {A.}~\bibnamefont {Endres}}, \bibinfo {author} {\bibfnamefont
  {M.}~\bibnamefont {Fritzsche}}, \bibinfo {author} {\bibfnamefont
  {J.}~\bibnamefont {Glorius}}, \bibinfo {author} {\bibfnamefont
  {J.}~\bibnamefont {Isaak}}, \bibinfo {author} {\bibfnamefont
  {N.}~\bibnamefont {Pietralla}}, \bibinfo {author} {\bibfnamefont
  {M.}~\bibnamefont {Scheck}}, \bibinfo {author} {\bibfnamefont
  {L.}~\bibnamefont {Schnorrenberger}}, \bibinfo {author} {\bibfnamefont
  {K.}~\bibnamefont {Sonnabend}}, \ and\ \bibinfo {author} {\bibfnamefont
  {M.}~\bibnamefont {Zweidinger}},\ }\href@noop {} {\bibfield  {journal}
  {\bibinfo  {journal} {Phys. Lett. B}\ }\textbf {\bibinfo {volume} {744}},\
  \bibinfo {pages} {369} (\bibinfo {year} {2015})}\BibitemShut {NoStop}%
\bibitem [{\citenamefont {Tamii}\ \emph {et~al.}(2011)\citenamefont {Tamii},
  \citenamefont {Poltoratska}, \citenamefont {von Neumann-Cosel}, \citenamefont
  {Fujita}, \citenamefont {Adachi}, \citenamefont {Bertulani}, \citenamefont
  {Carter}, \citenamefont {Dozono}, \citenamefont {Fujita}, \citenamefont
  {Fujita}, \citenamefont {Hatanaka}, \citenamefont {Ishikawa}, \citenamefont
  {Itoh}, \citenamefont {Kawabata}, \citenamefont {Kalmykov}, \citenamefont
  {Krumbholz}, \citenamefont {Litvinova}, \citenamefont {Matsubara},
  \citenamefont {Nakanishi}, \citenamefont {Neveling}, \citenamefont {Okamura},
  \citenamefont {Ong}, \citenamefont {\"Ozel-Tashenov}, \citenamefont
  {Ponomarev}, \citenamefont {Richter}, \citenamefont {Rubio}, \citenamefont
  {Sakaguchi}, \citenamefont {Sakemi}, \citenamefont {Sasamoto}, \citenamefont
  {Shimbara}, \citenamefont {Shimizu}, \citenamefont {Smit}, \citenamefont
  {Suzuki}, \citenamefont {Tameshige}, \citenamefont {Wambach}, \citenamefont
  {Yamada}, \citenamefont {Yosoi},\ and\ \citenamefont {Zenihiro}}]{Tami11}%
  \BibitemOpen
  \bibfield  {author} {\bibinfo {author} {\bibfnamefont {A.}~\bibnamefont
  {Tamii}}, \bibinfo {author} {\bibfnamefont {I.}~\bibnamefont {Poltoratska}},
  \bibinfo {author} {\bibfnamefont {P.}~\bibnamefont {von Neumann-Cosel}},
  \bibinfo {author} {\bibfnamefont {Y.}~\bibnamefont {Fujita}}, \bibinfo
  {author} {\bibfnamefont {T.}~\bibnamefont {Adachi}}, \bibinfo {author}
  {\bibfnamefont {C.~A.}\ \bibnamefont {Bertulani}}, \bibinfo {author}
  {\bibfnamefont {J.}~\bibnamefont {Carter}}, \bibinfo {author} {\bibfnamefont
  {M.}~\bibnamefont {Dozono}}, \bibinfo {author} {\bibfnamefont
  {H.}~\bibnamefont {Fujita}}, \bibinfo {author} {\bibfnamefont
  {K.}~\bibnamefont {Fujita}}, \bibinfo {author} {\bibfnamefont
  {K.}~\bibnamefont {Hatanaka}}, \bibinfo {author} {\bibfnamefont
  {D.}~\bibnamefont {Ishikawa}}, \bibinfo {author} {\bibfnamefont
  {M.}~\bibnamefont {Itoh}}, \bibinfo {author} {\bibfnamefont {T.}~\bibnamefont
  {Kawabata}}, \bibinfo {author} {\bibfnamefont {Y.}~\bibnamefont {Kalmykov}},
  \bibinfo {author} {\bibfnamefont {A.~M.}\ \bibnamefont {Krumbholz}}, \bibinfo
  {author} {\bibfnamefont {E.}~\bibnamefont {Litvinova}}, \bibinfo {author}
  {\bibfnamefont {H.}~\bibnamefont {Matsubara}}, \bibinfo {author}
  {\bibfnamefont {K.}~\bibnamefont {Nakanishi}}, \bibinfo {author}
  {\bibfnamefont {R.}~\bibnamefont {Neveling}}, \bibinfo {author}
  {\bibfnamefont {H.}~\bibnamefont {Okamura}}, \bibinfo {author} {\bibfnamefont
  {H.~J.}\ \bibnamefont {Ong}}, \bibinfo {author} {\bibfnamefont
  {B.}~\bibnamefont {\"Ozel-Tashenov}}, \bibinfo {author} {\bibfnamefont
  {V.~{\BIBYu}.}\ \bibnamefont {Ponomarev}}, \bibinfo {author} {\bibfnamefont
  {A.}~\bibnamefont {Richter}}, \bibinfo {author} {\bibfnamefont
  {B.}~\bibnamefont {Rubio}}, \bibinfo {author} {\bibfnamefont
  {H.}~\bibnamefont {Sakaguchi}}, \bibinfo {author} {\bibfnamefont
  {Y.}~\bibnamefont {Sakemi}}, \bibinfo {author} {\bibfnamefont
  {Y.}~\bibnamefont {Sasamoto}}, \bibinfo {author} {\bibfnamefont
  {Y.}~\bibnamefont {Shimbara}}, \bibinfo {author} {\bibfnamefont
  {Y.}~\bibnamefont {Shimizu}}, \bibinfo {author} {\bibfnamefont {F.~D.}\
  \bibnamefont {Smit}}, \bibinfo {author} {\bibfnamefont {T.}~\bibnamefont
  {Suzuki}}, \bibinfo {author} {\bibfnamefont {Y.}~\bibnamefont {Tameshige}},
  \bibinfo {author} {\bibfnamefont {J.}~\bibnamefont {Wambach}}, \bibinfo
  {author} {\bibfnamefont {R.}~\bibnamefont {Yamada}}, \bibinfo {author}
  {\bibfnamefont {M.}~\bibnamefont {Yosoi}}, \ and\ \bibinfo {author}
  {\bibfnamefont {J.}~\bibnamefont {Zenihiro}},\ }\href@noop {} {\bibfield
  {journal} {\bibinfo  {journal} {Phys. Rev. Lett.}\ }\textbf {\bibinfo
  {volume} {107}},\ \bibinfo {pages} {062502} (\bibinfo {year}
  {2011})}\BibitemShut {NoStop}%
\bibitem [{\citenamefont {Pietralla}(1999)}]{Piet99}%
  \BibitemOpen
  \bibfield  {author} {\bibinfo {author} {\bibfnamefont {N.}~\bibnamefont
  {Pietralla}},\ }\href@noop {} {\bibfield  {journal} {\bibinfo  {journal}
  {Phys. Rev. C}\ }\textbf {\bibinfo {volume} {59}},\ \bibinfo {pages} {2941}
  (\bibinfo {year} {1999})}\BibitemShut {NoStop}%
\bibitem [{\citenamefont {Jolos}\ \emph {et~al.}(2004)\citenamefont {Jolos},
  \citenamefont {Shirikova},\ and\ \citenamefont {Voronov}}]{Jolo04}%
  \BibitemOpen
  \bibfield  {author} {\bibinfo {author} {\bibfnamefont {R.~V.}\ \bibnamefont
  {Jolos}}, \bibinfo {author} {\bibfnamefont {N.~Y.}\ \bibnamefont
  {Shirikova}}, \ and\ \bibinfo {author} {\bibfnamefont {V.~V.}\ \bibnamefont
  {Voronov}},\ }\href@noop {} {\bibfield  {journal} {\bibinfo  {journal} {Phys.
  Rev. C}\ }\textbf {\bibinfo {volume} {70}},\ \bibinfo {pages} {054303}
  (\bibinfo {year} {2004})}\BibitemShut {NoStop}%
\bibitem [{\citenamefont {Wilhelm}\ \emph {et~al.}(1996)\citenamefont
  {Wilhelm}, \citenamefont {Radermacher}, \citenamefont {Zilges},\ and\
  \citenamefont {von Brentano}}]{Wilh96}%
  \BibitemOpen
  \bibfield  {author} {\bibinfo {author} {\bibfnamefont {M.}~\bibnamefont
  {Wilhelm}}, \bibinfo {author} {\bibfnamefont {E.}~\bibnamefont
  {Radermacher}}, \bibinfo {author} {\bibfnamefont {A.}~\bibnamefont {Zilges}},
  \ and\ \bibinfo {author} {\bibfnamefont {P.}~\bibnamefont {von Brentano}},\
  }\href@noop {} {\bibfield  {journal} {\bibinfo  {journal} {Phys. Rev. C}\
  }\textbf {\bibinfo {volume} {54}},\ \bibinfo {pages} {R449} (\bibinfo {year}
  {1996})}\BibitemShut {NoStop}%
\bibitem [{\citenamefont {Wilhelm}\ \emph {et~al.}(1998)\citenamefont
  {Wilhelm}, \citenamefont {Kasemann}, \citenamefont {Pascovici}, \citenamefont
  {Radermacher}, \citenamefont {von Brentano},\ and\ \citenamefont
  {Zilges}}]{Wilh98}%
  \BibitemOpen
  \bibfield  {author} {\bibinfo {author} {\bibfnamefont {M.}~\bibnamefont
  {Wilhelm}}, \bibinfo {author} {\bibfnamefont {S.}~\bibnamefont {Kasemann}},
  \bibinfo {author} {\bibfnamefont {G.}~\bibnamefont {Pascovici}}, \bibinfo
  {author} {\bibfnamefont {E.}~\bibnamefont {Radermacher}}, \bibinfo {author}
  {\bibfnamefont {P.}~\bibnamefont {von Brentano}}, \ and\ \bibinfo {author}
  {\bibfnamefont {A.}~\bibnamefont {Zilges}},\ }\href@noop {} {\bibfield
  {journal} {\bibinfo  {journal} {Phys. Rev. C}\ }\textbf {\bibinfo {volume}
  {57}},\ \bibinfo {pages} {577} (\bibinfo {year} {1998})}\BibitemShut
  {NoStop}%
\bibitem [{\citenamefont {Robinson}\ \emph {et~al.}(1994)\citenamefont
  {Robinson}, \citenamefont {Jolie}, \citenamefont {B\"orner}, \citenamefont
  {Schillebeeckx}, \citenamefont {Ulbig},\ and\ \citenamefont {Lieb}}]{Robi94}%
  \BibitemOpen
  \bibfield  {author} {\bibinfo {author} {\bibfnamefont {S.~J.}\ \bibnamefont
  {Robinson}}, \bibinfo {author} {\bibfnamefont {J.}~\bibnamefont {Jolie}},
  \bibinfo {author} {\bibfnamefont {H.~G.}\ \bibnamefont {B\"orner}}, \bibinfo
  {author} {\bibfnamefont {P.}~\bibnamefont {Schillebeeckx}}, \bibinfo {author}
  {\bibfnamefont {S.}~\bibnamefont {Ulbig}}, \ and\ \bibinfo {author}
  {\bibfnamefont {K.~P.}\ \bibnamefont {Lieb}},\ }\href@noop {} {\bibfield
  {journal} {\bibinfo  {journal} {Phys. Rev. Lett.}\ }\textbf {\bibinfo
  {volume} {73}},\ \bibinfo {pages} {412} (\bibinfo {year} {1994})}\BibitemShut
  {NoStop}%
\bibitem [{\citenamefont {Carman}\ \emph {et~al.}(1996)\citenamefont {Carman},
  \citenamefont {Litveninko}, \citenamefont {Madey}, \citenamefont {Neuman},
  \citenamefont {Norum}, \citenamefont {O'Shea}, \citenamefont {Roberson},
  \citenamefont {Scarlett}, \citenamefont {Schreiber},\ and\ \citenamefont
  {Weller}}]{Carm95}%
  \BibitemOpen
  \bibfield  {author} {\bibinfo {author} {\bibfnamefont {T.~S.}\ \bibnamefont
  {Carman}}, \bibinfo {author} {\bibfnamefont {V.}~\bibnamefont {Litveninko}},
  \bibinfo {author} {\bibfnamefont {J.}~\bibnamefont {Madey}}, \bibinfo
  {author} {\bibfnamefont {C.}~\bibnamefont {Neuman}}, \bibinfo {author}
  {\bibfnamefont {B.}~\bibnamefont {Norum}}, \bibinfo {author} {\bibfnamefont
  {P.~G.}\ \bibnamefont {O'Shea}}, \bibinfo {author} {\bibfnamefont {N.~R.}\
  \bibnamefont {Roberson}}, \bibinfo {author} {\bibfnamefont {C.~Y.}\
  \bibnamefont {Scarlett}}, \bibinfo {author} {\bibfnamefont {E.}~\bibnamefont
  {Schreiber}}, \ and\ \bibinfo {author} {\bibfnamefont {H.~R.}\ \bibnamefont
  {Weller}},\ }\href@noop {} {\bibfield  {journal} {\bibinfo  {journal} {Nucl.
  Instr. and Meth. A}\ }\textbf {\bibinfo {volume} {378}},\ \bibinfo {pages}
  {1} (\bibinfo {year} {1996})}\BibitemShut {NoStop}%
\bibitem [{\citenamefont {Weller}\ \emph {et~al.}(2009)\citenamefont {Weller},
  \citenamefont {Ahmed}, \citenamefont {Gao}, \citenamefont {Tornow},
  \citenamefont {Wu}, \citenamefont {Gai},\ and\ \citenamefont
  {Miskimen}}]{Well09}%
  \BibitemOpen
  \bibfield  {author} {\bibinfo {author} {\bibfnamefont {H.~R.}\ \bibnamefont
  {Weller}}, \bibinfo {author} {\bibfnamefont {M.~W.}\ \bibnamefont {Ahmed}},
  \bibinfo {author} {\bibfnamefont {H.}~\bibnamefont {Gao}}, \bibinfo {author}
  {\bibfnamefont {W.}~\bibnamefont {Tornow}}, \bibinfo {author} {\bibfnamefont
  {Y.~K.}\ \bibnamefont {Wu}}, \bibinfo {author} {\bibfnamefont
  {M.}~\bibnamefont {Gai}}, \ and\ \bibinfo {author} {\bibfnamefont
  {R.}~\bibnamefont {Miskimen}},\ }\href@noop {} {\bibfield  {journal}
  {\bibinfo  {journal} {Prog. Part. Nucl. Phys.}\ }\textbf {\bibinfo {volume}
  {62}},\ \bibinfo {pages} {257} (\bibinfo {year} {2009})}\BibitemShut
  {NoStop}%
\bibitem [{\citenamefont {L\"oher}\ \emph {et~al.}(2013)\citenamefont
  {L\"oher}, \citenamefont {Derya}, \citenamefont {Aumann}, \citenamefont
  {Beller}, \citenamefont {Cooper}, \citenamefont {Duch\^{e}ne}, \citenamefont
  {Endres}, \citenamefont {Fiori}, \citenamefont {Isaak}, \citenamefont
  {Kelley}, \citenamefont {Kn\"orzer}, \citenamefont {Pietralla}, \citenamefont
  {Romig}, \citenamefont {Savran}, \citenamefont {Scheck}, \citenamefont
  {Scheit}, \citenamefont {Silva}, \citenamefont {Tonchev}, \citenamefont
  {Tornow}, \citenamefont {Weller}, \citenamefont {Werner},\ and\ \citenamefont
  {Zilges}}]{Loeh13}%
  \BibitemOpen
  \bibfield  {author} {\bibinfo {author} {\bibfnamefont {B.}~\bibnamefont
  {L\"oher}}, \bibinfo {author} {\bibfnamefont {V.}~\bibnamefont {Derya}},
  \bibinfo {author} {\bibfnamefont {T.}~\bibnamefont {Aumann}}, \bibinfo
  {author} {\bibfnamefont {J.}~\bibnamefont {Beller}}, \bibinfo {author}
  {\bibfnamefont {N.}~\bibnamefont {Cooper}}, \bibinfo {author} {\bibfnamefont
  {M.}~\bibnamefont {Duch\^{e}ne}}, \bibinfo {author} {\bibfnamefont
  {J.}~\bibnamefont {Endres}}, \bibinfo {author} {\bibfnamefont
  {E.}~\bibnamefont {Fiori}}, \bibinfo {author} {\bibfnamefont
  {J.}~\bibnamefont {Isaak}}, \bibinfo {author} {\bibfnamefont {J.~H.}\
  \bibnamefont {Kelley}}, \bibinfo {author} {\bibfnamefont {M.}~\bibnamefont
  {Kn\"orzer}}, \bibinfo {author} {\bibfnamefont {N.}~\bibnamefont
  {Pietralla}}, \bibinfo {author} {\bibfnamefont {C.}~\bibnamefont {Romig}},
  \bibinfo {author} {\bibfnamefont {D.}~\bibnamefont {Savran}}, \bibinfo
  {author} {\bibfnamefont {M.}~\bibnamefont {Scheck}}, \bibinfo {author}
  {\bibfnamefont {H.}~\bibnamefont {Scheit}}, \bibinfo {author} {\bibfnamefont
  {J.}~\bibnamefont {Silva}}, \bibinfo {author} {\bibfnamefont {A.~P.}\
  \bibnamefont {Tonchev}}, \bibinfo {author} {\bibfnamefont {W.}~\bibnamefont
  {Tornow}}, \bibinfo {author} {\bibfnamefont {H.~R.}\ \bibnamefont {Weller}},
  \bibinfo {author} {\bibfnamefont {V.}~\bibnamefont {Werner}}, \ and\ \bibinfo
  {author} {\bibfnamefont {A.}~\bibnamefont {Zilges}},\ }\href@noop {}
  {\bibfield  {journal} {\bibinfo  {journal} {Nucl. Instr. and Meth. A}\
  }\textbf {\bibinfo {volume} {723}},\ \bibinfo {pages} {136} (\bibinfo {year}
  {2013})}\BibitemShut {NoStop}%
\bibitem [{\citenamefont {Hartmann}\ \emph {et~al.}(2002)\citenamefont
  {Hartmann}, \citenamefont {Enders}, \citenamefont {Mohr}, \citenamefont
  {Vogt}, \citenamefont {Volz},\ and\ \citenamefont {Zilges}}]{Hart02}%
  \BibitemOpen
  \bibfield  {author} {\bibinfo {author} {\bibfnamefont {T.}~\bibnamefont
  {Hartmann}}, \bibinfo {author} {\bibfnamefont {J.}~\bibnamefont {Enders}},
  \bibinfo {author} {\bibfnamefont {P.}~\bibnamefont {Mohr}}, \bibinfo {author}
  {\bibfnamefont {K.}~\bibnamefont {Vogt}}, \bibinfo {author} {\bibfnamefont
  {S.}~\bibnamefont {Volz}}, \ and\ \bibinfo {author} {\bibfnamefont
  {A.}~\bibnamefont {Zilges}},\ }\href@noop {} {\bibfield  {journal} {\bibinfo
  {journal} {Phys. Rev. C}\ }\textbf {\bibinfo {volume} {65}},\ \bibinfo
  {pages} {034301} (\bibinfo {year} {2002})}\BibitemShut {NoStop}%
\bibitem [{\citenamefont {Pietralla}\ \emph {et~al.}(2002)\citenamefont
  {Pietralla}, \citenamefont {Berant}, \citenamefont {Litvinenko},
  \citenamefont {Hartman}, \citenamefont {Mikhailov}, \citenamefont {Pinayev},
  \citenamefont {Swift}, \citenamefont {Ahmed}, \citenamefont {Kelley},
  \citenamefont {Nelson}, \citenamefont {Prior}, \citenamefont {Sabourov},
  \citenamefont {Tonchev},\ and\ \citenamefont {Weller}}]{Piet02}%
  \BibitemOpen
  \bibfield  {author} {\bibinfo {author} {\bibfnamefont {N.}~\bibnamefont
  {Pietralla}}, \bibinfo {author} {\bibfnamefont {Z.}~\bibnamefont {Berant}},
  \bibinfo {author} {\bibfnamefont {V.~N.}\ \bibnamefont {Litvinenko}},
  \bibinfo {author} {\bibfnamefont {S.}~\bibnamefont {Hartman}}, \bibinfo
  {author} {\bibfnamefont {F.~F.}\ \bibnamefont {Mikhailov}}, \bibinfo {author}
  {\bibfnamefont {I.~V.}\ \bibnamefont {Pinayev}}, \bibinfo {author}
  {\bibfnamefont {G.}~\bibnamefont {Swift}}, \bibinfo {author} {\bibfnamefont
  {M.~W.}\ \bibnamefont {Ahmed}}, \bibinfo {author} {\bibfnamefont {J.~H.}\
  \bibnamefont {Kelley}}, \bibinfo {author} {\bibfnamefont {S.~O.}\
  \bibnamefont {Nelson}}, \bibinfo {author} {\bibfnamefont {R.}~\bibnamefont
  {Prior}}, \bibinfo {author} {\bibfnamefont {K.}~\bibnamefont {Sabourov}},
  \bibinfo {author} {\bibfnamefont {A.~P.}\ \bibnamefont {Tonchev}}, \ and\
  \bibinfo {author} {\bibfnamefont {H.~R.}\ \bibnamefont {Weller}},\
  }\href@noop {} {\bibfield  {journal} {\bibinfo  {journal} {Phys. Rev. Lett.}\
  }\textbf {\bibinfo {volume} {88}},\ \bibinfo {pages} {012502} (\bibinfo
  {year} {2001})}\BibitemShut {NoStop}%
\bibitem [{\citenamefont {Rusev}\ \emph {et~al.}(2013)\citenamefont {Rusev},
  \citenamefont {Tsoneva}, \citenamefont {D\"onau}, \citenamefont {Frauendorf},
  \citenamefont {Schwengner}, \citenamefont {Tonchev}, \citenamefont {Adekola},
  \citenamefont {Hammond}, \citenamefont {Kelley}, \citenamefont {Kwan},
  \citenamefont {Lenske}, \citenamefont {Tornow},\ and\ \citenamefont
  {Wagner}}]{Rus13}%
  \BibitemOpen
  \bibfield  {author} {\bibinfo {author} {\bibfnamefont {G.}~\bibnamefont
  {Rusev}}, \bibinfo {author} {\bibfnamefont {N.}~\bibnamefont {Tsoneva}},
  \bibinfo {author} {\bibfnamefont {F.}~\bibnamefont {D\"onau}}, \bibinfo
  {author} {\bibfnamefont {S.}~\bibnamefont {Frauendorf}}, \bibinfo {author}
  {\bibfnamefont {R.}~\bibnamefont {Schwengner}}, \bibinfo {author}
  {\bibfnamefont {A.~P.}\ \bibnamefont {Tonchev}}, \bibinfo {author}
  {\bibfnamefont {A.~S.}\ \bibnamefont {Adekola}}, \bibinfo {author}
  {\bibfnamefont {S.~L.}\ \bibnamefont {Hammond}}, \bibinfo {author}
  {\bibfnamefont {J.~H.}\ \bibnamefont {Kelley}}, \bibinfo {author}
  {\bibfnamefont {E.}~\bibnamefont {Kwan}}, \bibinfo {author} {\bibfnamefont
  {H.}~\bibnamefont {Lenske}}, \bibinfo {author} {\bibfnamefont
  {W.}~\bibnamefont {Tornow}}, \ and\ \bibinfo {author} {\bibfnamefont
  {A.}~\bibnamefont {Wagner}},\ }\href@noop {} {\bibfield  {journal} {\bibinfo
  {journal} {Phys. Rev. Lett.}\ }\textbf {\bibinfo {volume} {110}},\ \bibinfo
  {pages} {022503} (\bibinfo {year} {2013})}\BibitemShut {NoStop}%
\bibitem [{\citenamefont {Tsoneva}\ and\ \citenamefont {Lenske}(2011)}]{Tso11}%
  \BibitemOpen
  \bibfield  {author} {\bibinfo {author} {\bibfnamefont {N.}~\bibnamefont
  {Tsoneva}}\ and\ \bibinfo {author} {\bibfnamefont {H.}~\bibnamefont
  {Lenske}},\ }\href@noop {} {\bibfield  {journal} {\bibinfo  {journal} {Phys.
  Lett. B}\ }\textbf {\bibinfo {volume} {695}},\ \bibinfo {pages} {174}
  (\bibinfo {year} {2011})}\BibitemShut {NoStop}%
\bibitem [{\citenamefont {Pellegri}\ \emph {et~al.}(2015)\citenamefont
  {Pellegri}, \citenamefont {Bracco}, \citenamefont {Tsoneva}, \citenamefont
  {Avigo}, \citenamefont {Benzoni}, \citenamefont {Blasi}, \citenamefont
  {Bottoni}, \citenamefont {Camera}, \citenamefont {Ceruti}, \citenamefont
  {Crespi}, \citenamefont {Giaz}, \citenamefont {Leoni}, \citenamefont
  {Lenske}, \citenamefont {Million}, \citenamefont {Morales}, \citenamefont
  {Nicolini}, \citenamefont {Wieland}, \citenamefont {Bazzacco}, \citenamefont
  {Bednarczyk}, \citenamefont {Birkenbach}, \citenamefont {Ciema\l{}a},
  \citenamefont {de~Angelis}, \citenamefont {Farnea}, \citenamefont {Gadea},
  \citenamefont {G\"orgen}, \citenamefont {Gottardo}, \citenamefont {Grebosz},
  \citenamefont {Isocrate}, \citenamefont {Kmiecik}, \citenamefont {Krzysiek},
  \citenamefont {Lunardi}, \citenamefont {Maj}, \citenamefont {Mazurek},
  \citenamefont {Mengoni}, \citenamefont {Michelagnoli}, \citenamefont
  {Napoli}, \citenamefont {Recchia}, \citenamefont {Siebeck}, \citenamefont
  {Siem}, \citenamefont {Ur},\ and\ \citenamefont {Valiente-Dob\'on}}]{Pel15}%
  \BibitemOpen
  \bibfield  {author} {\bibinfo {author} {\bibfnamefont {L.}~\bibnamefont
  {Pellegri}}, \bibinfo {author} {\bibfnamefont {A.}~\bibnamefont {Bracco}},
  \bibinfo {author} {\bibfnamefont {N.}~\bibnamefont {Tsoneva}}, \bibinfo
  {author} {\bibfnamefont {R.}~\bibnamefont {Avigo}}, \bibinfo {author}
  {\bibfnamefont {G.}~\bibnamefont {Benzoni}}, \bibinfo {author} {\bibfnamefont
  {N.}~\bibnamefont {Blasi}}, \bibinfo {author} {\bibfnamefont
  {S.}~\bibnamefont {Bottoni}}, \bibinfo {author} {\bibfnamefont
  {F.}~\bibnamefont {Camera}}, \bibinfo {author} {\bibfnamefont
  {S.}~\bibnamefont {Ceruti}}, \bibinfo {author} {\bibfnamefont {F.~C.~L.}\
  \bibnamefont {Crespi}}, \bibinfo {author} {\bibfnamefont {A.}~\bibnamefont
  {Giaz}}, \bibinfo {author} {\bibfnamefont {S.}~\bibnamefont {Leoni}},
  \bibinfo {author} {\bibfnamefont {H.}~\bibnamefont {Lenske}}, \bibinfo
  {author} {\bibfnamefont {B.}~\bibnamefont {Million}}, \bibinfo {author}
  {\bibfnamefont {A.~I.}\ \bibnamefont {Morales}}, \bibinfo {author}
  {\bibfnamefont {R.}~\bibnamefont {Nicolini}}, \bibinfo {author}
  {\bibfnamefont {O.}~\bibnamefont {Wieland}}, \bibinfo {author} {\bibfnamefont
  {D.}~\bibnamefont {Bazzacco}}, \bibinfo {author} {\bibfnamefont
  {P.}~\bibnamefont {Bednarczyk}}, \bibinfo {author} {\bibfnamefont
  {B.}~\bibnamefont {Birkenbach}}, \bibinfo {author} {\bibfnamefont
  {M.}~\bibnamefont {Ciema\l{}a}}, \bibinfo {author} {\bibfnamefont
  {G.}~\bibnamefont {de~Angelis}}, \bibinfo {author} {\bibfnamefont
  {E.}~\bibnamefont {Farnea}}, \bibinfo {author} {\bibfnamefont
  {A.}~\bibnamefont {Gadea}}, \bibinfo {author} {\bibfnamefont
  {A.}~\bibnamefont {G\"orgen}}, \bibinfo {author} {\bibfnamefont
  {A.}~\bibnamefont {Gottardo}}, \bibinfo {author} {\bibfnamefont
  {J.}~\bibnamefont {Grebosz}}, \bibinfo {author} {\bibfnamefont
  {R.}~\bibnamefont {Isocrate}}, \bibinfo {author} {\bibfnamefont
  {M.}~\bibnamefont {Kmiecik}}, \bibinfo {author} {\bibfnamefont
  {M.}~\bibnamefont {Krzysiek}}, \bibinfo {author} {\bibfnamefont
  {S.}~\bibnamefont {Lunardi}}, \bibinfo {author} {\bibfnamefont
  {A.}~\bibnamefont {Maj}}, \bibinfo {author} {\bibfnamefont {K.}~\bibnamefont
  {Mazurek}}, \bibinfo {author} {\bibfnamefont {D.}~\bibnamefont {Mengoni}},
  \bibinfo {author} {\bibfnamefont {C.}~\bibnamefont {Michelagnoli}}, \bibinfo
  {author} {\bibfnamefont {D.~R.}\ \bibnamefont {Napoli}}, \bibinfo {author}
  {\bibfnamefont {F.}~\bibnamefont {Recchia}}, \bibinfo {author} {\bibfnamefont
  {B.}~\bibnamefont {Siebeck}}, \bibinfo {author} {\bibfnamefont
  {S.}~\bibnamefont {Siem}}, \bibinfo {author} {\bibfnamefont {C.}~\bibnamefont
  {Ur}}, \ and\ \bibinfo {author} {\bibfnamefont {J.~J.}\ \bibnamefont
  {Valiente-Dob\'on}},\ }\href@noop {} {\bibfield  {journal} {\bibinfo
  {journal} {Phys. Rev. C}\ }\textbf {\bibinfo {volume} {92}},\ \bibinfo
  {pages} {014330} (\bibinfo {year} {2015})}\BibitemShut {NoStop}%
\bibitem [{\citenamefont {Spieker}\ \emph {et~al.}(2016)\citenamefont
  {Spieker}, \citenamefont {Tsoneva}, \citenamefont {Derya}, \citenamefont
  {Endres}, \citenamefont {Savran}, \citenamefont {Harakeh}, \citenamefont
  {Harissopulos}, \citenamefont {Herzberg}, \citenamefont {Lagoyannis},
  \citenamefont {Lenske}, \citenamefont {Pietralla}, \citenamefont {Popescu},
  \citenamefont {Scheck}, \citenamefont {Schl\"uter}, \citenamefont
  {Sonnabend}, \citenamefont {Stoica}, \citenamefont {W\"ortche},\ and\
  \citenamefont {Zilges}}]{Spie16}%
  \BibitemOpen
  \bibfield  {author} {\bibinfo {author} {\bibfnamefont {M.}~\bibnamefont
  {Spieker}}, \bibinfo {author} {\bibfnamefont {N.}~\bibnamefont {Tsoneva}},
  \bibinfo {author} {\bibfnamefont {V.}~\bibnamefont {Derya}}, \bibinfo
  {author} {\bibfnamefont {J.}~\bibnamefont {Endres}}, \bibinfo {author}
  {\bibfnamefont {D.}~\bibnamefont {Savran}}, \bibinfo {author} {\bibfnamefont
  {M.~N.}\ \bibnamefont {Harakeh}}, \bibinfo {author} {\bibfnamefont
  {S.}~\bibnamefont {Harissopulos}}, \bibinfo {author} {\bibfnamefont {R.-D.}\
  \bibnamefont {Herzberg}}, \bibinfo {author} {\bibfnamefont {A.}~\bibnamefont
  {Lagoyannis}}, \bibinfo {author} {\bibfnamefont {H.}~\bibnamefont {Lenske}},
  \bibinfo {author} {\bibfnamefont {N.}~\bibnamefont {Pietralla}}, \bibinfo
  {author} {\bibfnamefont {L.}~\bibnamefont {Popescu}}, \bibinfo {author}
  {\bibfnamefont {M.}~\bibnamefont {Scheck}}, \bibinfo {author} {\bibfnamefont
  {F.}~\bibnamefont {Schl\"uter}}, \bibinfo {author} {\bibfnamefont
  {K.}~\bibnamefont {Sonnabend}}, \bibinfo {author} {\bibfnamefont {V.~I.}\
  \bibnamefont {Stoica}}, \bibinfo {author} {\bibfnamefont {H.~J.}\
  \bibnamefont {W\"ortche}}, \ and\ \bibinfo {author} {\bibfnamefont
  {A.}~\bibnamefont {Zilges}},\ }\href@noop {} {\bibfield  {journal} {\bibinfo
  {journal} {Phys. Lett. B}\ }\textbf {\bibinfo {volume} {752}},\ \bibinfo
  {pages} {102} (\bibinfo {year} {2016})}\BibitemShut {NoStop}%
\bibitem [{\citenamefont {Herzberg}\ \emph {et~al.}(1999)\citenamefont
  {Herzberg}, \citenamefont {Fransen}, \citenamefont {von Brentano},
  \citenamefont {Eberth}, \citenamefont {Enders}, \citenamefont {Fitzler},
  \citenamefont {K\"aubler}, \citenamefont {Kaiser}, \citenamefont {von
  Neumann-Cosel}, \citenamefont {Pietralla}, \citenamefont {Ponomarev},
  \citenamefont {Prade}, \citenamefont {Richter}, \citenamefont {Schnare},
  \citenamefont {Schwengner}, \citenamefont {Skoda}, \citenamefont {Thomas},
  \citenamefont {Tiesler}, \citenamefont {Weisshaar},\ and\ \citenamefont
  {Wiedenh\"over}}]{Herz99}%
  \BibitemOpen
  \bibfield  {author} {\bibinfo {author} {\bibfnamefont {R.-D.}\ \bibnamefont
  {Herzberg}}, \bibinfo {author} {\bibfnamefont {C.}~\bibnamefont {Fransen}},
  \bibinfo {author} {\bibfnamefont {P.}~\bibnamefont {von Brentano}}, \bibinfo
  {author} {\bibfnamefont {J.}~\bibnamefont {Eberth}}, \bibinfo {author}
  {\bibfnamefont {J.}~\bibnamefont {Enders}}, \bibinfo {author} {\bibfnamefont
  {A.}~\bibnamefont {Fitzler}}, \bibinfo {author} {\bibfnamefont
  {L.}~\bibnamefont {K\"aubler}}, \bibinfo {author} {\bibfnamefont
  {H.}~\bibnamefont {Kaiser}}, \bibinfo {author} {\bibfnamefont
  {P.}~\bibnamefont {von Neumann-Cosel}}, \bibinfo {author} {\bibfnamefont
  {N.}~\bibnamefont {Pietralla}}, \bibinfo {author} {\bibfnamefont
  {V.~{\BIBYu}.}\ \bibnamefont {Ponomarev}}, \bibinfo {author} {\bibfnamefont
  {H.}~\bibnamefont {Prade}}, \bibinfo {author} {\bibfnamefont
  {A.}~\bibnamefont {Richter}}, \bibinfo {author} {\bibfnamefont
  {H.}~\bibnamefont {Schnare}}, \bibinfo {author} {\bibfnamefont
  {R.}~\bibnamefont {Schwengner}}, \bibinfo {author} {\bibfnamefont
  {S.}~\bibnamefont {Skoda}}, \bibinfo {author} {\bibfnamefont {H.~G.}\
  \bibnamefont {Thomas}}, \bibinfo {author} {\bibfnamefont {H.}~\bibnamefont
  {Tiesler}}, \bibinfo {author} {\bibfnamefont {D.}~\bibnamefont {Weisshaar}},
  \ and\ \bibinfo {author} {\bibfnamefont {I.}~\bibnamefont {Wiedenh\"over}},\
  }\href@noop {} {\bibfield  {journal} {\bibinfo  {journal} {Phys. Rev. C}\
  }\textbf {\bibinfo {volume} {60}},\ \bibinfo {pages} {051307} (\bibinfo
  {year} {1999})}\BibitemShut {NoStop}%
\bibitem [{\citenamefont {Raman}\ \emph {et~al.}(1987)\citenamefont {Raman},
  \citenamefont {Malarkey}, \citenamefont {Milner}, \citenamefont {Nestor},\
  and\ \citenamefont {Stelson}}]{Rama87}%
  \BibitemOpen
  \bibfield  {author} {\bibinfo {author} {\bibfnamefont {S.}~\bibnamefont
  {Raman}}, \bibinfo {author} {\bibfnamefont {C.~H.}\ \bibnamefont {Malarkey}},
  \bibinfo {author} {\bibfnamefont {W.~T.}\ \bibnamefont {Milner}}, \bibinfo
  {author} {\bibfnamefont {C.~W.}\ \bibnamefont {Nestor}}, \ and\ \bibinfo
  {author} {\bibfnamefont {P.~H.}\ \bibnamefont {Stelson}},\ }\href@noop {}
  {\bibfield  {journal} {\bibinfo  {journal} {Atomic Data and Nuclear Data
  Tables}\ }\textbf {\bibinfo {volume} {36}},\ \bibinfo {pages} {1} (\bibinfo
  {year} {1987})}\BibitemShut {NoStop}%
\bibitem [{\citenamefont {Burnett}\ \emph {et~al.}(1985)\citenamefont
  {Burnett}, \citenamefont {Baxter}, \citenamefont {Hinds}, \citenamefont
  {Pribac}, \citenamefont {Spear},\ and\ \citenamefont {Vermeer}}]{Burn85}%
  \BibitemOpen
  \bibfield  {author} {\bibinfo {author} {\bibfnamefont {S.~M.}\ \bibnamefont
  {Burnett}}, \bibinfo {author} {\bibfnamefont {A.~M.}\ \bibnamefont {Baxter}},
  \bibinfo {author} {\bibfnamefont {S.}~\bibnamefont {Hinds}}, \bibinfo
  {author} {\bibfnamefont {F.}~\bibnamefont {Pribac}}, \bibinfo {author}
  {\bibfnamefont {R.~H.}\ \bibnamefont {Spear}}, \ and\ \bibinfo {author}
  {\bibfnamefont {W.~J.}\ \bibnamefont {Vermeer}},\ }\href@noop {} {\bibfield
  {journal} {\bibinfo  {journal} {Nucl. Phys. A}\ }\textbf {\bibinfo {volume}
  {432}},\ \bibinfo {pages} {514} (\bibinfo {year} {1985})}\BibitemShut
  {NoStop}%
\bibitem [{\citenamefont {Pitthan}(1970)}]{Pitt70}%
  \BibitemOpen
  \bibfield  {author} {\bibinfo {author} {\bibfnamefont {R.}~\bibnamefont
  {Pitthan}},\ }\href@noop {} {\bibfield  {journal} {\bibinfo  {journal} {Z.
  Naturforsch.}\ }\textbf {\bibinfo {volume} {25}},\ \bibinfo {pages} {1358}
  (\bibinfo {year} {1970})}\BibitemShut {NoStop}%
\bibitem [{\citenamefont {Madsen}\ \emph {et~al.}(1971)\citenamefont {Madsen},
  \citenamefont {Cardman}, \citenamefont {Legg},\ and\ \citenamefont
  {Bockelman}}]{Mads71}%
  \BibitemOpen
  \bibfield  {author} {\bibinfo {author} {\bibfnamefont {D.~W.}\ \bibnamefont
  {Madsen}}, \bibinfo {author} {\bibfnamefont {L.~S.}\ \bibnamefont {Cardman}},
  \bibinfo {author} {\bibfnamefont {J.~R.}\ \bibnamefont {Legg}}, \ and\
  \bibinfo {author} {\bibfnamefont {C.~K.}\ \bibnamefont {Bockelman}},\
  }\href@noop {} {\bibfield  {journal} {\bibinfo  {journal} {Nucl. Phys. A}\
  }\textbf {\bibinfo {volume} {168}},\ \bibinfo {pages} {97} (\bibinfo {year}
  {1971})}\BibitemShut {NoStop}%
\bibitem [{\citenamefont {Barfield}\ \emph {et~al.}(1989)\citenamefont
  {Barfield}, \citenamefont {von Brentano}, \citenamefont {Dewald},
  \citenamefont {Zell}, \citenamefont {Zamfir}, \citenamefont {Bucurescu},
  \citenamefont {Iva\c{s}cu},\ and\ \citenamefont {Scholten}}]{Barf89}%
  \BibitemOpen
  \bibfield  {author} {\bibinfo {author} {\bibfnamefont {A.~F.}\ \bibnamefont
  {Barfield}}, \bibinfo {author} {\bibfnamefont {P.}~\bibnamefont {von
  Brentano}}, \bibinfo {author} {\bibfnamefont {A.}~\bibnamefont {Dewald}},
  \bibinfo {author} {\bibfnamefont {K.~O.}\ \bibnamefont {Zell}}, \bibinfo
  {author} {\bibfnamefont {N.~V.}\ \bibnamefont {Zamfir}}, \bibinfo {author}
  {\bibfnamefont {D.}~\bibnamefont {Bucurescu}}, \bibinfo {author}
  {\bibfnamefont {M.}~\bibnamefont {Iva\c{s}cu}}, \ and\ \bibinfo {author}
  {\bibfnamefont {O.}~\bibnamefont {Scholten}},\ }\href@noop {} {\bibfield
  {journal} {\bibinfo  {journal} {Z. Phys. A}\ }\textbf {\bibinfo {volume}
  {332}},\ \bibinfo {pages} {29} (\bibinfo {year} {1989})}\BibitemShut
  {NoStop}%
\bibitem [{\citenamefont {Hartmann}\ \emph {et~al.}(2000)\citenamefont
  {Hartmann}, \citenamefont {Enders}, \citenamefont {Mohr}, \citenamefont
  {Vogt}, \citenamefont {Volz},\ and\ \citenamefont {Zilges}}]{Hart00}%
  \BibitemOpen
  \bibfield  {author} {\bibinfo {author} {\bibfnamefont {T.}~\bibnamefont
  {Hartmann}}, \bibinfo {author} {\bibfnamefont {J.}~\bibnamefont {Enders}},
  \bibinfo {author} {\bibfnamefont {P.}~\bibnamefont {Mohr}}, \bibinfo {author}
  {\bibfnamefont {K.}~\bibnamefont {Vogt}}, \bibinfo {author} {\bibfnamefont
  {S.}~\bibnamefont {Volz}}, \ and\ \bibinfo {author} {\bibfnamefont
  {A.}~\bibnamefont {Zilges}},\ }\href@noop {} {\bibfield  {journal} {\bibinfo
  {journal} {Phys. Rev. Lett.}\ }\textbf {\bibinfo {volume} {85}},\ \bibinfo
  {pages} {274} (\bibinfo {year} {2000})}\BibitemShut {NoStop}%
\bibitem [{\citenamefont {Hartmann}\ \emph {et~al.}(2004)\citenamefont
  {Hartmann}, \citenamefont {Babilon}, \citenamefont {Kamerdzhiev},
  \citenamefont {Litvinova}, \citenamefont {Savran}, \citenamefont {Volz},\
  and\ \citenamefont {Zilges}}]{Hart04}%
  \BibitemOpen
  \bibfield  {author} {\bibinfo {author} {\bibfnamefont {T.}~\bibnamefont
  {Hartmann}}, \bibinfo {author} {\bibfnamefont {M.}~\bibnamefont {Babilon}},
  \bibinfo {author} {\bibfnamefont {S.}~\bibnamefont {Kamerdzhiev}}, \bibinfo
  {author} {\bibfnamefont {E.}~\bibnamefont {Litvinova}}, \bibinfo {author}
  {\bibfnamefont {D.}~\bibnamefont {Savran}}, \bibinfo {author} {\bibfnamefont
  {S.}~\bibnamefont {Volz}}, \ and\ \bibinfo {author} {\bibfnamefont
  {A.}~\bibnamefont {Zilges}},\ }\href@noop {} {\bibfield  {journal} {\bibinfo
  {journal} {Phys. Rev. Lett.}\ }\textbf {\bibinfo {volume} {93}},\ \bibinfo
  {pages} {192501} (\bibinfo {year} {2004})}\BibitemShut {NoStop}%
\bibitem [{\citenamefont {Isaak}\ \emph {et~al.}(2011)\citenamefont {Isaak},
  \citenamefont {Savran}, \citenamefont {Fritzsche}, \citenamefont {Galaviz},
  \citenamefont {Hartmann}, \citenamefont {Kamerdzhiev}, \citenamefont
  {Kelley}, \citenamefont {Kwan}, \citenamefont {Pietralla}, \citenamefont
  {Romig}, \citenamefont {Rusev}, \citenamefont {Sonnabend}, \citenamefont
  {Tonchev}, \citenamefont {Tornow},\ and\ \citenamefont {Zilges}}]{Isaa11}%
  \BibitemOpen
  \bibfield  {author} {\bibinfo {author} {\bibfnamefont {J.}~\bibnamefont
  {Isaak}}, \bibinfo {author} {\bibfnamefont {D.}~\bibnamefont {Savran}},
  \bibinfo {author} {\bibfnamefont {M.}~\bibnamefont {Fritzsche}}, \bibinfo
  {author} {\bibfnamefont {D.}~\bibnamefont {Galaviz}}, \bibinfo {author}
  {\bibfnamefont {T.}~\bibnamefont {Hartmann}}, \bibinfo {author}
  {\bibfnamefont {S.}~\bibnamefont {Kamerdzhiev}}, \bibinfo {author}
  {\bibfnamefont {J.~H.}\ \bibnamefont {Kelley}}, \bibinfo {author}
  {\bibfnamefont {E.}~\bibnamefont {Kwan}}, \bibinfo {author} {\bibfnamefont
  {N.}~\bibnamefont {Pietralla}}, \bibinfo {author} {\bibfnamefont
  {C.}~\bibnamefont {Romig}}, \bibinfo {author} {\bibfnamefont
  {G.}~\bibnamefont {Rusev}}, \bibinfo {author} {\bibfnamefont
  {K.}~\bibnamefont {Sonnabend}}, \bibinfo {author} {\bibfnamefont {A.~P.}\
  \bibnamefont {Tonchev}}, \bibinfo {author} {\bibfnamefont {W.}~\bibnamefont
  {Tornow}}, \ and\ \bibinfo {author} {\bibfnamefont {A.}~\bibnamefont
  {Zilges}},\ }\href@noop {} {\bibfield  {journal} {\bibinfo  {journal} {Phys.
  Rev. C}\ }\textbf {\bibinfo {volume} {83}},\ \bibinfo {pages} {034304}
  (\bibinfo {year} {2011})}\BibitemShut {NoStop}%
\bibitem [{\citenamefont {Poelhekken}\ \emph {et~al.}(1992)\citenamefont
  {Poelhekken}, \citenamefont {Hesmondhalgh}, \citenamefont {Hofmann},
  \citenamefont {van~der Woude},\ and\ \citenamefont {Harakeh}}]{Poel92}%
  \BibitemOpen
  \bibfield  {author} {\bibinfo {author} {\bibfnamefont {T.~D.}\ \bibnamefont
  {Poelhekken}}, \bibinfo {author} {\bibfnamefont {S.~K.~B.}\ \bibnamefont
  {Hesmondhalgh}}, \bibinfo {author} {\bibfnamefont {H.~J.}\ \bibnamefont
  {Hofmann}}, \bibinfo {author} {\bibfnamefont {A.}~\bibnamefont {van~der
  Woude}}, \ and\ \bibinfo {author} {\bibfnamefont {M.~N.}\ \bibnamefont
  {Harakeh}},\ }\href@noop {} {\bibfield  {journal} {\bibinfo  {journal} {Phys.
  Lett. B}\ }\textbf {\bibinfo {volume} {278}},\ \bibinfo {pages} {423}
  (\bibinfo {year} {1992})}\BibitemShut {NoStop}%
\bibitem [{\citenamefont {Papakonstantinou}\ \emph {et~al.}(2012)\citenamefont
  {Papakonstantinou}, \citenamefont {Hergert}, \citenamefont {Ponomarev},\ and\
  \citenamefont {Roth}}]{Papa12}%
  \BibitemOpen
  \bibfield  {author} {\bibinfo {author} {\bibfnamefont {P.}~\bibnamefont
  {Papakonstantinou}}, \bibinfo {author} {\bibfnamefont {H.}~\bibnamefont
  {Hergert}}, \bibinfo {author} {\bibfnamefont {V.~{\BIBYu}.}\ \bibnamefont
  {Ponomarev}}, \ and\ \bibinfo {author} {\bibfnamefont {R.}~\bibnamefont
  {Roth}},\ }\href@noop {} {\bibfield  {journal} {\bibinfo  {journal} {Phys.
  Lett. B}\ }\textbf {\bibinfo {volume} {709}},\ \bibinfo {pages} {270}
  (\bibinfo {year} {2012})}\BibitemShut {NoStop}%
\end{thebibliography}

\providecommand{\BIBYu}{Yu} \providecommand{\BIBCh}{Ch}
  \providecommand{\BIBTh}{Th}

\end{document}